\documentclass{aa}  

\usepackage{graphicx}
\usepackage{txfonts}
\usepackage[colorlinks=true,citecolor=blue,linkcolor=blue,urlcolor=blue,breaklinks=true]{hyperref}
\usepackage{xcolor,float,placeins}
\usepackage[ruled,vlined]{algorithm2e}
\usepackage{multirow}
\usepackage{siunitx}
\usepackage{wasysym}
\usepackage{rotating}
\usepackage{stfloats}
\usepackage[version=4]{mhchem}
\usepackage{natbib}
\bibpunct{(}{)}{;}{a}{}{,}

\newcommand{\RIII}{FRB\,20180916B}

\newcommand{\Msol}{M$_{\astrosun}$}
\newcommand{\jyms}{\,Jy\,ms} % Fluence units
\newcommand{\radsqm}{\,rad\,m$^{-2}$\xspace} % Rotation measure units
\newcommand{\pccm}{\,pc\,cm$^{-3}$} % Dispersion measure units
\newcommand{\tscat}{\ensuremath{\tau_{\text{sc}}}}
\newcommand{\scbw}{$\Delta\nu_{\text{sc}}$}
\newcommand{\pckm}{pc$^{-2/3}$\,km$^{-1/3}$}
\newcommand{\skyday}{sky$^{-1}$\,day$^{-1}$}
\newcommand{\dmeg}{\ensuremath{\text{DM}_{\text{EG}}}}
\newcommand{\dmobs}{\ensuremath{\text{DM}_{\text{obs}}}}
\newcommand{\dmmw}{\ensuremath{\text{DM}_{\text{MW}}}}
\newcommand{\dmhalo}{\ensuremath{\text{DM}_{\text{halo}}}}
\newcommand{\dmhost}{\ensuremath{\text{DM}_{\text{host}}}}
\newcommand{\zmacquart}{\ensuremath{z_{\text{Macquart}}}}
\newcommand{\rmmw}{\ensuremath{\text{RM}_{\text{MW}}}}
\newcommand{\rmobs}{\ensuremath{\text{RM}_{\text{obs}}}}
\newcommand{\rmhost}{\ensuremath{\text{RM}_{\text{host}}}}
\newcommand{\zmax}{\ensuremath{z_{\text{max}}}}
\newcommand{\HII}{H$_{\text{II}}$}

\def\ipm#1{\ignorespaces}
\def\ep#1{\ignorespaces}
\def\jvl#1{\ignorespaces}

\def\refbf#1{#1}

\institute{
ASTRON, the Netherlands Institute for Radio Astronomy, Oude Hoogeveensedijk 4,7991 PD Dwingeloo, The Netherlands\label{astron}
  \and
Anton Pannekoek Institute, University of Amsterdam, Postbus 94249, 1090 GE Amsterdam, The Netherlands\label{uva}
  \and
Jodrell Bank Centre for Astrophysics, Department of Physics and Astronomy, The University of Manchester, Manchester, M13 9PL, UK\label{jbca}
\email{ines.pastor.marazuela@gmail.com}
  \and
Rubicon Fellow\label{rubicon}
  \and
Cahill Center for Astronomy, California Institute of Technology, Pasadena, CA, USA\label{caltech}
  \and
National Centre for Radio Astrophysics, Tata Institute of Fundamental Research, Pune 411007, Maharashtra, India\label{ncra}
  \and
Netherlands eScience Center, Science Park 402, 1098 XH, Amsterdam, The Netherlands\label{escience}
  \and
Perimeter Institute for Theoretical Physics, Waterloo ON N2L 2Y5, Canada\label{pi}
  \and
Department of Space, Earth and Environment, Chalmers University of Technology, Onsala Space Observatory, 439 92, Onsala, Sweden\label{oso}
}

\def\orcid#1{}
\author{%
     In\'es~Pastor-Marazuela    \orcid{0000-0002-4357-8027}\ \inst{\ref{astron} \and \ref{uva} \and \ref{jbca} \and \ref{rubicon}}
\and Joeri~van~Leeuwen          \orcid{0000-0001-8503-6958}\ \inst{\ref{astron}}
\and Anna~Bilous                \orcid{0000-0002-7177-6987}\ \inst{\ref{astron}}
\and Liam~Connor                \orcid{0000-0002-7587-6352}\ \inst{\ref{caltech} \and \ref{uva}}
\and Yogesh~Maan                \orcid{0000-0002-0862-6062}\ \inst{\ref{ncra} \and \ref{astron}}
\and Leon~Oostrum               \orcid{0000-0001-8724-8372}\ \inst{\ref{astron} \and \ref{uva} \and \ref{escience}}
\and Emily~Petroff              \orcid{0000-0002-9822-8008}\ \inst{\ref{uva} \and \ref{pi}}
\and Dany~Vohl                  \orcid{0000-0003-1779-4532}\ \inst{\ref{uva} \and \ref{astron}}
\and Kelley~M.~Hess             \orcid{0000-0001-9662-9089}\ \inst{\ref{oso} \and \ref{astron}}
\and Emanuela~Orr\`u              \orcid{0009-0001-2882-7195}\ \inst{\ref{astron}}
\and Alessio~Sclocco            \orcid{0000-0003-3278-0518}\ \inst{\ref{escience}}           
\and Yuyang~Wang                \orcid{0000-0002-3822-0389}\ \inst{\ref{uva}}
} 
\authorrunning{In{\'e}s~Pastor-Marazuela~et~al.}

\begin{document}

% \title{The Apertif Fast Radio Burst sample:\\ Frequency evolution of multi-component burst fraction}
% \title{Morphological evolution with frequency in Fast Radio Bursts \\ as revealed %exposed 
% by the  Apertif survey}
% \ipm{Final results? Insights? Include the word Survey?}
% \title{Comprehensive analysis of the Apertif FRB sample}
%
%\title{Results from the Apertif Radio Transient System: \\Comprehensive analysis of the Fast Radio Burst sample}
%
%\title{The morphology, polarisation and local environments of the Fast Radio Bursts in the final, high-resolution ALERT sample strongly resemble those seen in magnetars}
%
%\title{The ALERT survey: \\at high resolution, Fast Radio Bursts strongly resemble magnetars}
%
%\title{At high resolution, the ALERT survey shows: \\Fast Radio Bursts strongly resemble energetic neutron stars}
%
% \title{The final, high resolution ALERT sample shows: \\Fast Radio Bursts strongly resemble energetic neutron stars}

\title{Comprehensive analysis of the Apertif Fast Radio Burst sample:\\
similarities with young, energetic neutron stars}

% \title{The Fast Radio Bursts in the final Apertif sample resemble young, energetic neutron stars}

\titlerunning{The Apertif fast radio burst sample}

   % \author{I. Pastor-Marazuela\inst{1,2,3,4}
   %        \and
   %        J. van Leeuwen\inst{1}
   %        \and
   %        ARTS Team
   %        \and
   %        Apertif Team Contributors
   %        }

   % \institute{
   %      ASTRON, Netherlands Institute for Radio Astronomy, Oude Hoogeveensedijk 4, 7991 PD Dwingeloo, The Netherlands
   %      \and 
   %      Anton Pannekoek Institute for Astronomy, University of Amsterdam, Science Park 904, 1098 XH Amsterdam, The Netherlands
   %      \and
   %      Jodrell Bank Centre for Astrophysics, University of Manchester, Oxford Road, Manchester M13 9PL, UK
   %      \email{ines.pastor.marazuela@gmail.com}
   %      \and
   %      Rubicon Fellow
   %           }

  % \input{incl-authors.tex}
  % \input{incl-affil.tex}

   \date{}
 
   \abstract{
Understanding the origin of the energetic fast radio bursts (FRBs) has become the main science driver of recent dedicated FRB surveys powered by real-time instrumentation.
Between July 2019 and February 2022, we carried out ALERT, an FRB survey at 1370\,MHz using the Apertif Radio Transient
System (ARTS) installed at the Westerbork Synthesis Radio Telescope (WSRT).
Here we report the detection of 18 new FRBs, and we study the properties of the entire 24 burst sample that were detected during the survey. 
For five bursts, we identify host galaxy candidates within their error regions with >50\% probability association. 
We observe an average linear polarisation fraction of $\sim43\%$ and an average circular polarisation fraction
consistent with 0\%. A third of the FRBs display multiple components. 
These burst structures and the polarisation fractions are strikingly similar to those observed in young, energetic pulsars and magnetars.
The Apertif FRBs next reveal a population of highly scattered bursts. Given the observing frequency and time resolution, the scattering of most FRBs is likely to have been produced in the immediate circumburst environment.
Furthermore, two FRBs show evidence for high rotation measure values, \refbf{which could reach} $|\text{RM}|>10^3$\,\radsqm\ in the source reference frames. This corroborates that some source environments are dominated by  magneto-ionic effects.
Together, the scattering and rotation measures ALERT finds prove that a large fraction of FRBs are embedded in complex media such as star forming regions or supernova remnants.
Through the discovery of FRB\,20200719A, the third most dispersed FRB so far, we further show that one-off FRBs emit at frequencies in excess of  6\,GHz, the highest known to date.
We compare this to the radio-bright, high-frequency emission seen in magnetars.
Finally, we determine an FRB all-sky rate of $459^{+208}_{-155}$\,\skyday\ above a fluence limit of 4.1\,\jyms, and a fluence cumulative distribution with a power law index ${\gamma=-1.23\pm0.06\pm0.2}$, which is roughly consistent with the Euclidean Universe predictions.
Through the high resolution in time, frequency, polarisation and localisation that ALERT featured, we were able to determine the morphological complexity, polarisation, local scattering and magnetic environment, and high-frequency luminosity of FRBs.
We find all these   strongly resemble those seen in young, energetic, highly magnetised neutron stars.
   }


   \maketitle
%
%------------------------------------------------------
\section{Introduction}

The field of Fast Radio Bursts (FRBs) -- extragalactic radio flashes of millisecond duration with extreme luminosities
\refbf{\citep[$>10^{46}$\,erg\,s$^{-1}$][]{lorimer_bright_2007, petroff_fast_2019, cordes_fast_2019,
    ryder_luminous_2023}} -- has been rapidly evolving in recent years. The number of published FRBs is now in the
hundreds \citep{petroff_fast_2022}, more than forty have been localised to their host galaxies\footnote{FRB Hosts
Catalog: \url{https://www.frb-hosts.org/}}, and about fifty are known to repeat
\citep{chimefrb_collaboration_chimefrb_2023}. Although these discoveries have not yet fully unveiled the origin of FRBs,
the detection of a bright radio burst from the Galactic magnetar SGR\,1935+2154 demonstrated that at least some FRBs may
be produced by magnetars \citep{bochenek_fast_2020, chimefrb_collaboration_bright_2020}, \refbf{although a repeating FRB also exists in a globular cluster
% suggests there might be different channels through which FRBs are born
%% Joeri changed this because it's just one counterexample  and I felt the previous phrasing gave it too much weight.
%% We already wrote that only "some FRBs .. etc" -- JVL
\citep{kirsten_repeating_2022}}. These findings have been possible thanks to the increased number of observations and surveys dedicated to FRB searching in recent years, enabled by real-time search hardware and algorithms \citep[see, e.g.,][for a review]{2024Univ...10..158R}.

The current published sample of both repeating and seemingly one-off FRBs is dominated by sources discovered in the CHIME/FRB project, which searches in the 400 -- 800\,MHz band \citep{chimefrb_collaboration_first_2021}. Given the large sample size, population studies have been possible with the CHIME/FRB data to look at properties such as the bulk burst morphology of a large FRB sample. In studies of this sample, \citet{pleunis_fast_2021} find four burst archetypes:  single component FRBs, classified as either narrow or broadband; multi-component bursts with each component spanning a similar frequency extent; or multi-component bursts with `sad-trombone-like' downward drifting structure.
A population study of burst properties in the first CHIME/FRB Catalog by \citet{chawla_modeling_2022} also reports the overabundance of scattering detected in this sample. Additionally, injections performed for the catalog analysis confirm CHIME/FRB detections are biased against highly scattered events, hinting at the presence of a wider FRB population with large scattering timescales to which CHIME/FRB is less sensitive
 \citep{chimefrb_collaboration_first_2021}.

Several observed properties of FRBs, including dispersion measure (DM), scattering, scintillation, and polarisation, are highly frequency dependent. While the CHIME/FRB sample is by far the largest, studies at higher (and lower) radio frequencies are essential to probe the full extent of the FRB population across all parameters, including DM and scattering. One-off FRBs discovered at frequencies of $\sim$1 GHz by the Murriyang Telescope at the Parkes Observatory and the Australian Square Kilometre Array Pathfinder (ASKAP) also show evidence of scatter broadening and multiple components \refbf{\citep{day_high_2020,champion_five_2016, qiu_population_2020}}, some of which might be beyond the width detection threshold at CHIME/FRB frequencies.

The best-studied FRBs by far have been the small but productive sample of repeating FRB sources. Both the low \citep[110\,MHz;][]{pastor-marazuela_chromatic_2021, pleunis_lofar_2021} and high \citep[8\,GHz;][]{gajjar_highest-frequency_2018} frequency detections of FRBs have been made through targeted observations of known prolific repeaters. Comparing activity of the repeating FRB 20180916B at 150\,MHz and 1.4\,GHz simultaneously has shown frequency-dependent activity of this particular source \citep{pastor-marazuela_chromatic_2021}. For one-off FRBs, direct comparisons of behaviour at high and low frequencies is not yet possible. As such, assembling large samples of one-off FRBs at different radio frequencies may prove the most fruitful in uncovering frequency-dependent properties. Comparing the observed distributions of DM, scattering, flux, fluence, and scintillation of FRB samples from different instruments will provide insight into the underlying FRB population distribution -- either directly, or after correcting for the survey selection effects, as in e.g.~\citet{gardenier_multi-dimensional_2021} -- as well as the properties of the burst environment and host galaxy.

In this paper we present the sample of 24 one-off FRBs discovered with the Apertif system on the Westerbork Synthesis Radio Telescope (WSRT),  during the Apertif-LOFAR Exploration of the Radio Transient sky survey \citep[ALERT;][]{van_leeuwen_apertif_2023}.
Its high spectro-temporal resolution search has yielded a self-contained sample of FRBs for which we report DM, burst morphology, frequency structure, scattering, scintillation, and polarisation. 
In Section~\ref{sec:observations} we present the observing strategy for Apertif, while Section~\ref{sec:data_release} presents the data release. In Section~\ref{sec:data_analysis} we present the data analysis method of the detected bursts. In Section~\ref{sec:results} we present the detected FRB sample, the burst properties and the results of population analysis across DM, propagation effects, morphology, and polarisation; we discuss further in Section~\ref{sec:discussion} and conclude in Section~\ref{sec:conclusions}.

%------------------------------------------------------
\section{Observations}\label{sec:observations}

Apertif, the APERture Tile in Focus, is a front-end instrument installed at the Westerbork Synthesis Radio Telescope (WSRT), in twelve of fourteen 25\,m dishes of the interferometer \citep{van_cappellen_apertif_2022}, located in the Netherlands.
Apertif consists of phased array feeds (PAFs), with each dish forming 40 compound beams (CBs) on the sky and thus increasing the original Field of View (FoV) of the WSRT to 8.2\,deg$^2$ \citep{adams_radio_2019}. 
Apertif has carried out an imaging and a time domain survey between July 2019 and February 2022. The Apertif Radio Transient System (ARTS) was designed to carry out the time domain survey, as described in \cite{van_leeuwen_apertif_2023}, with supporting detail in \citet{van_leeuwen_apertif_2022}. \refbf{The system for commensally observing in imaging and time-domain mode was not yet operational in the period presented here. The ARTS observations were thus carried out separately from the imaging observations.}
The CBs from each dish are coherently beamformed into 12 tied-array beams \citep[TABs, see][]{maan_real-time_2017}, and these are next recombined in frequency to form 71 synthesised beams (SBs) per compound beam \citep{van_leeuwen_apertif_2023}.
The SBs of all CBs generate a total of 2840 Stokes I, Q, U, and V data-streams at a central frequency of 1370\,MHz and a bandwidth of 300\,MHz, with a time and frequency resolution of 81.92\,$\mu$s and 195\,kHz respectively.
The Stokes I data-streams are then searched for single-pulses with the software \texttt{AMBER}\footnote{\texttt{AMBER}: \url{https://github.com/TRASAL/AMBER}} \citep{sclocco_real-time_2014, sclocco_real-time_2016,   sclocco_real-time_2019}. The data post-processing is implemented with the Data Analysis of Real-time Candidates from the Apertif Radio Transient System \citep[\texttt{DARC ARTS}\footnote{\texttt{DARC}: \url{https://github.com/TRASAL/darc}},][]{oostrum_fast_2020}, and includes real-time candidate classification through a neural network that is public \citep{connor_applying_2018, connor_2024_10656614}.
The single pulse searches and data post-processing are run on a 40-node graphics processing unit (GPU) cluster at the WSRT. The observations were scheduled with \texttt{apersched} \citep{hess_kmhessapersched_2022}.

ARTS has  proven its FRB searching capabilities at high time and frequency resolution through the follow-up and detection of known repeating FRBs \citep{oostrum_repeating_2020, pastor-marazuela_chromatic_2021}, as well as through the discovery of new one-off FRBs \citep{connor_bright_2020, pastor-marazuela_fast_2023, van_leeuwen_apertif_2023}.
Here, we present the discovery of a new population of, as of yet, one-off FRBs that have been detected in the ALERT time domain survey.

% \ep{Some brief description of Apertif and cite overview paper?}

\subsection{Pointings and sky exposure}

% python /home/arts/pastor/scripts/arts-analysis/sky_exposure.py -g -f
\begin{figure*}
    \centering
    \includegraphics[width=17cm]{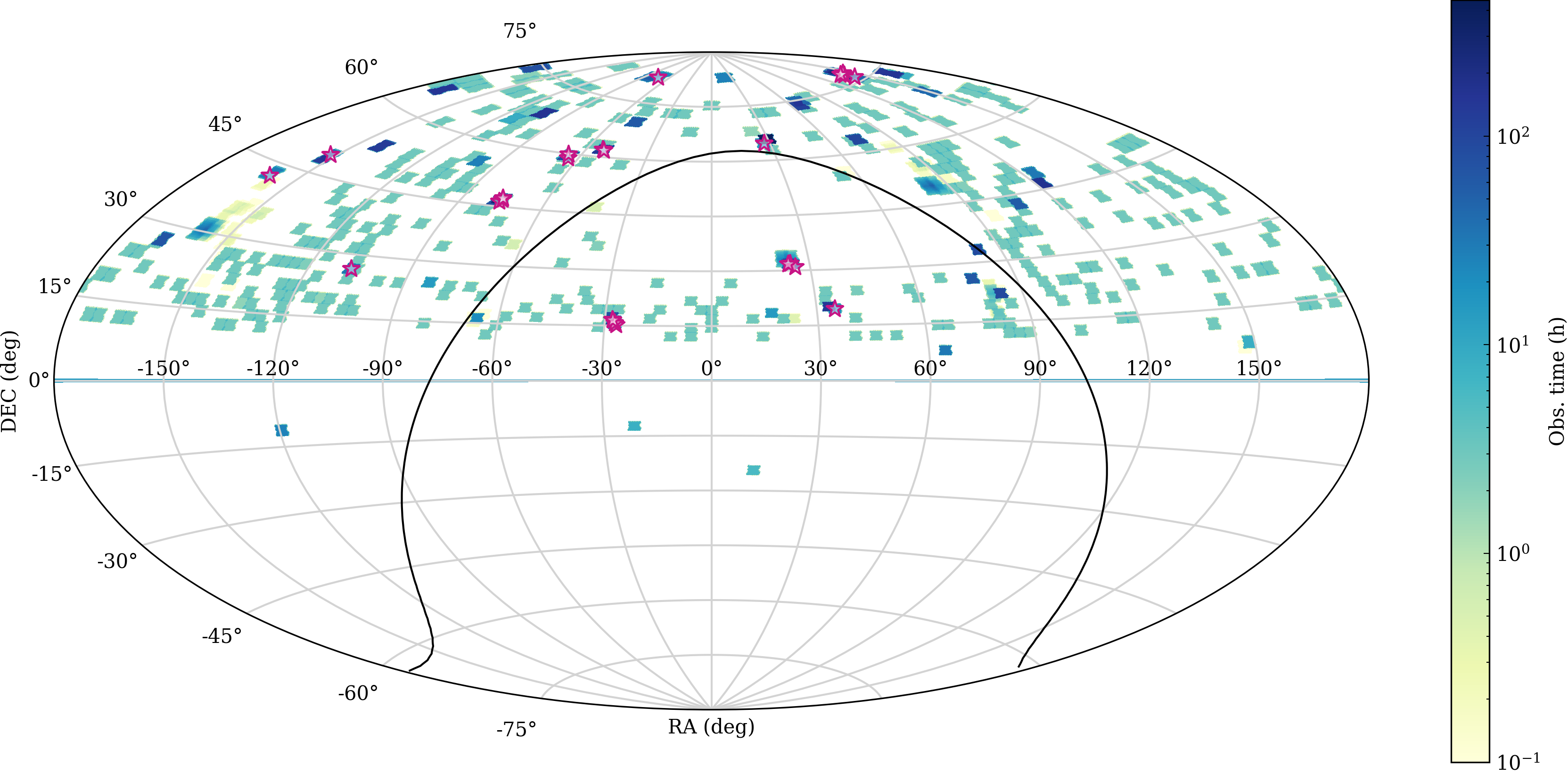}
    \caption{Exposure time per sky area in equatorial coordinates. Dark blue regions correspond to larger exposure times, and white regions have not been observed. The position of the detected FRBs is marked by magenta stars. The Galactic plane is indicated by the black line.}
    \label{fig:sky_exposure}
\end{figure*}

The priority source list and thus pointing definition evolved during the ALERT survey, in order to adapt to the rapidly evolving FRB discoveries \citep{bailes_discovery_2022}. While only two repeating sources were known at the beginning of the survey in July 2019 \citep{spitler_repeating_2016, chime/frb_collaboration_second_2019}, several new repeaters were reported soon after \citep{chime/frb_collaboration_chime/frb_2019, fonseca_nine_2020}. Simultaneously, the number of Apertif detections was increasing. From 2020 onward, the observing shifted away from fields with no known FRBs in order to prioritise the follow up of repeaters and Apertif-discovered FRBs. Given the isotropic sky distribution of FRBs \citep{bhandari_survey_2018}, one-off FRBs should be detected blindly at the same rate in pointings with and without known FRBs.

\begin{figure}
    \centering
    \includegraphics[width=\hsize]{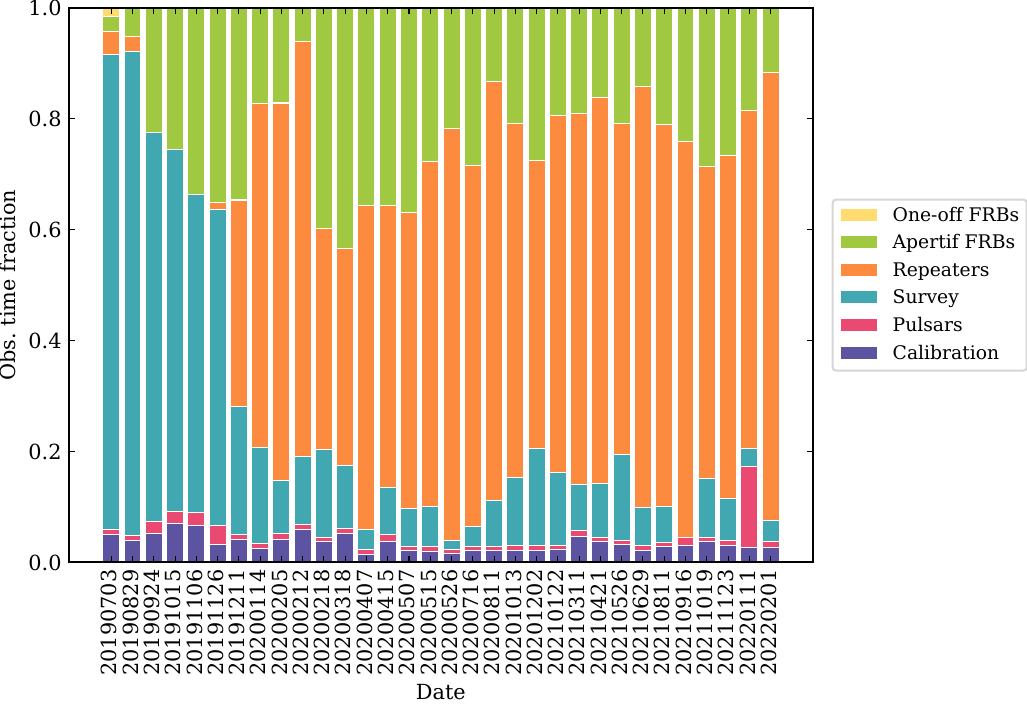}
    \caption{Evolution of the target class fractions with time. Blue represents fields without known FRBs, orange are
      fields with known repeating FRBs. Green revisits  Apertif-discovered FRBs, and yellow
      observations target one-off FRBs discovered by other instruments.}
    \label{fig:source_type_exposure}
\end{figure}

Figure~\ref{fig:sky_exposure} shows the exposure time per sky region in equatorial coordinates and the location of the newly discovered FRBs, while  Fig.~\ref{fig:source_type_exposure} shows the fraction of time spent on survey pointings, repeating FRBs, one-off FRBs, new FRBs discovered with Apertif, pulsars and calibration observations. During the 2019 observations, $\sim$70\% of the time was spent on the Apertif survey pointings, while the remaining $\sim$30\% was divided between follow up of known and newly discovered one-off FRBs and calibration observations. The evolution in pointing strategy in 2020 and 2021 to prioritise the follow up of known repeaters and newly discovered Apertif FRBs is reflected in Fig.~\ref{fig:source_type_exposure}, where the changes implemented around Jan 1 of each calendar year are visible. Roughly 60\% of the time was dedicated to repeater follow up, 20\% to the follow up of Apertif FRBs and the remaining 20\% in survey pointings and calibration observations.

% python /home/arts/pastor/scripts/arts-analysis/survey_pointings.py

%------------------------------------------------------
\section{Data release}\label{sec:data_release}

The observations described above were archived in 
the down-sampled, Stokes-I, PSRFITS format described in \citet{van_leeuwen_apertif_2023}, which are now public.
The 2019 data was released before,
as Apertif Time-Domain Data Release 1, together with \citet{van_leeuwen_apertif_2023}.
Accompanying the current paper
% \footnote{At publication date}
we make public also the 2020$-$2022 data.
Together, the 2019-2022 data comprise 
Apertif Time-Domain Data Release 2,
containing 1666 multi-hr pointings that
form the whole Apertif Time-Domain legacy data set,
since Apertif observations ended in 2022.
This complete set
\refbf{occupies a data volume of 0.8\,PB and}
is accessible at ASTRON, through the
Data
Explorer\footnote{\url{https://science.astron.nl/sdc/astron-data-explorer/data-releases/}; ``Apertif Time-Domain DR2''}, the 
Virtual Observatory (VO) Interface\footnote{\url{https://vo.astron.nl/}; ``Apertif Time Domain FRB search (DR2)''} and the  Apertif Long Term Archive
(ALTA)\footnote{\url{https://alta.astron.nl/science/dataproducts/release__release_id=APERTIF_DR2_TimeDomain}}.
The release includes the subset of 
pointings with FRB 
detections\footnote{\url{https://hdl.handle.net/21.12136/383f3c18-9c2c-495e-9d4c-d3b4192a5b7d}}, that can be directly downloaded; 
and the set of all pointings\footnote{\url{https://hdl.handle.net/21.12136/03a6775b-e768-4212-bd06-027267d21c0a}},
which require staging from tape by the
ASTRON helpdesk.
% \ipm{Joeri: write total size/volume of the data archive.-- DONE} 

% Includes the repeater fields

%------------------------------------------------------
\section{Data analysis}\label{sec:data_analysis}

In this section, we detail the post-processing data analysis performed on the bursts that resulted from the
\texttt{AMBER} and \texttt{DARC} searches and candidate selection \citep[as described in][]{van_leeuwen_apertif_2023}. Here we describe the methods to determine the burst properties, including dispersion measure, scattering, scintillation, flux calibration, morphology, frequency structure, polarisation, and localisation. The results of these analyses are presented in Section~\ref{sec:results}. Although for some discoveries the  analysis and the results are intertwined to some extent, separating the analysis into its own Section makes it easier to reference later when discussing the results.

\subsection{Dispersion measure and redshift estimation} \label{sec:DM}

Each FRB candidate detected by the \texttt{AMBER} pipeline has an associated DM that maximises the signal-to-noise ratio
(S/N) at the given downsampling factor. Some of the detected FRBs present multiple components, and we thus used an
algorithm based on \cite{hessels_frb_2019}\footnote{\texttt{DM\_phase}:
\url{https://github.com/danielemichilli/DM_phase}}, that was already put to test in \cite{pastor-marazuela_chromatic_2021}
and \citet{Bilous_24A}, to find the DM maximising the structure of the burst. For faint and/or scattered bursts with no signs of multi-component structure, we used \texttt{pdmp}\footnote{\texttt{pdmp}: \url{http://psrchive.sourceforge.net/manuals/pdmp/}} instead, since it maximises S/N and in these cases this method is more robust at determining the correct DM. The measured DMs of all FRBs presented in this paper are given in Table~\ref{tab:frb_table} of Appendix~\ref{app:frb_table}.

In order to determine the redshift upper limit for each FRB, we first estimate the extragalactic DM (\dmeg) from the observed DM$_{\text{obs}}$. We predict the Milky Way (MW) contribution to the DM (DM$_{\text{MW}}$) from the NE2001 \citep{cordes_ne2001i_2003} and YMW16 \citep{yao_new_2017} Galactic electron density models, and take the average of the two. Since the Galactic halo can also significantly contribute to the DM, we adopt the model from \cite{yamasaki_galactic_2020} to compute the MW halo DM (\dmhalo) in the direction of each FRB. The extragalactic DM will thus be \dmeg\,=\,\dmobs\,-\dmmw\,-\,\dmhalo.
Next we apply the DM--$z$ relation from \cite{macquart_census_2020} assuming the cosmological parameters from \cite{planck_collaboration_planck_2020} to obtain the mean redshift and 95\% errors. For this, we use the python package \texttt{FRB}\footnote{\texttt{FRB}: \url{https://github.com/FRBs/FRB/tree/main}}. We assume a host galaxy contribution to the DM of 100\,\pccm. The measured DMs and estimated redshifts of our FRB sample are detailed in Section~\ref{sec:result_dm}.

% Next we apply the DM--$z$ relation from \cite{zhang_fast_2018} and assume the cosmological parameters from \cite{planck_collaboration_planck_2020} to obtain the redshift upper limit. For this, we use the python package \texttt{fruitbat}\footnote{\texttt{fruitbat}: \url{https://fruitbat.readthedocs.io/en/latest/}} \citep{batten_fruitbat_2019}. We assume a negligible host galaxy contribution to the DM. The measured DMs and estimated redshifts of our FRB sample are detailed in Section~\ref{sec:result_dm}.

\subsection{Flux calibration}

To perform the flux calibration of all FRBs, we scheduled drift scan observations of the bright calibrator sources
3C147, 3C286, and/or 3C48 at the beginning and the end of each observing run. As the flux densities of these sources are known \citep{perley_accurate_2017},  they can be used as calibrators to obtain the system-equivalent flux density (SEFD). 
For each FRB, we used the drift scan taken during the same observing run that was the least affected by radio frequency interference (RFI). 
To convert the pulse profile into flux units, we applied the radiometer equation using the obtained SEFD. We define the peak flux as the maximum flux value at the instrument time resolution (0.08192\,ms), and it is thus a lower limit. Finally we integrated over a time window covering the whole burst duration to obtain the FRB fluences in units of Jy\,ms. Based on the measured stability %% instability sounds a bit too harsh
of the system, we assume 20\% errors on the fluence. The resulting fluxes and fluences are given in Table~\ref{tab:frb_table} and detailed in Section~\ref{sec:event_rate}.

\subsection{Localisation and host candidate identification} \label{sec:localisation_data_analysis}

To determine the localisation of the detected FRBs, we implement the localisation method described in \citet{oostrum_fast_2020} and \cite{van_leeuwen_apertif_2023}.
The method consists of creating a model of the telescope response in sky coordinates and comparing it to the observed response pattern as follows: first, a model of the Compound Beams (CBs) is created based on drift-scan data. From this we construct the TABs and SBs, consequently obtaining a model of the SB sensitivity in Right Ascension (RA) and Declination (Dec).
To localise a burst, we next compare its S/N per SB detection pattern against the predicted SB model.
We define the best position of each burst as the resulting 99\% confidence regions, which have narrow elliptical shapes
since the WSRT is an East-West array. The size of the confidence region shrinks with both higher detection S/N and with
a larger number of CBs in which the burst was detected. The orientation of the ellipse depends on the hour angle of the detection, due to the Earth rotation.

The localisation regions of all detected FRBs were covered by the Pan-STARRS1\footnote{Pan-STARRS1: \url{https://outerspace.stsci.edu/display/PANSTARRS}} survey \citep{chambers_pan-starrs1_2019}, thus giving us access to deep images and a source catalogue with a median photometric depth of 23.2 in the \textit{g}-band. The project Pan-STARRS1 Source Types and Redshifts with Machine Learning\footnote{PS1-STRM: \url{https://archive.stsci.edu/hlsp/ps1-strm}} \citep[PS1-STRM,][]{beck_ps1-strm_2021} provides source classifications and photometric redshifts for all the sources contained in the PanSTARRS1 3$\pi$ DR1, computed through a neural network. Hence, for each FRB, we searched for all the PS1-STRM sources classified as ``galaxies'', ``quasi-stellar objects'' (QSOs), or ``unsure'' contained within or near the localisation region, and within the expected redshift limits.
When the number of known galaxies within an FRB error region was $\leq5$, we performed a Probabilistic Association of Transients to their Hosts (\texttt{PATH}) analysis \citep{aggarwal_probabilistic_2021}, to determine the probability of the FRB being associated to each host galaxy candidate.

% For each FRB, we looked for known galaxies within the localisation error region and below the redshift upper limit (Section~\ref{sec:DM}) using two galaxy catalogues: the NASA/IPAC extragalactic database (NED) and the GLADE v2.3 catalogue \citep{dalya_glade_2018}. 
In most cases, the total FRB error region is of order 5\,arcmin$^2$, which is too large to unambiguously identify a
unique FRB host galaxy candidate \citep[see, e.g.,][]{eftekhari_associating_2017}.
Additionally, although some FRB error regions might contain more than one known host galaxy candidate, there are probably more that are too faint to be detected. 
FRBs have been localised to galaxies of different types spanning a broad range of masses \citep[][]{bhandari_characterizing_2022, gordon_demographics_2023}, from the dwarf galaxy hosts of FRB\,20121102A \citep{chatterjee_direct_2017} and FRB\,20190520B \citep{niu_repeating_2022}, to massive galaxies reaching close to $10^{11}$\,\Msol\ in the case of FRB\,20200120E, which has been localised to a globular cluster of M81 \citep{bhardwaj_nearby_2021, kirsten_repeating_2022}.
Following \citet{petroff_fast_2018} and \citet{van_leeuwen_apertif_2023}, we estimate the expected number of galaxies within the comoving volume determined by the error region and the redshift upper limit of each burst. 
We adopt a dwarf galaxy number density of $n=(0.02-0.06)$\,Mpc$^{-3}$ for galaxy masses $4\times10^7$\,\Msol$<$\,M$_{\text{stellar}}<10^{10}$\,\Msol\ \citep{baldry_galaxy_2012, haynes_arecibo_2011}, and a massive galaxy number density of $n=(1.5-2.0)\times 10^{-3}$\,Mpc$^{-3}$ for galaxy masses M$_{\text{stellar}}>10^{11}$\,\Msol\ \citep{faber_galaxy_2007}. The expected number of galaxies within the comoving volume $V_{\text{co}}$ is simply $N_{\text{gal}}=n V_{\text{co}}$.
The results of this analysis are given in Section~\ref{sec:result_localisation}.

%\ep{Here maybe it would be better to say `In most cases the total error region is of order XX arcmin$^2$, which is too large to unambiguously identify a unique FRB host galaxy \citep{Eftekhari_2017_ApJ}'.}

\subsection{Burst morphologies} \label{sec:morphologies}

We characterise the morphology of all FRBs by fitting their dedispersed pulse profiles to a single or multi-component
model through minimisation of residuals.
A human expert determines the number of components,
 guided in edge cases by the  Bayesian information criterion (BIC) values for the fits.
% For 20200719A we checked the BIC
Each burst is fitted to a single or multi-component Gaussian model given by Eq.~\ref{eq:gaussian}, with and without a
convolution with an exponential decay given by Eq.~\ref{eq:exp_decay} to represent scattering, thus assuming the
scattering timescale to be the same for all components. After fitting the scattered and unscattered models, the model
with the lowest
%Bayesian information criterion (BIC)
BIC
is selected, with BIC$_{\text{g}}$ for the Gaussian, unscattered model, and BIC$_{\text{sc}}$ for the scattered model. The resulting expression for the fitted pulse profile $I(t)$ is given by Eq.~\ref{eq:pulse_profile}.

\begin{equation}\label{eq:gaussian}
    G_i(t) = A_i \exp\left(-\frac{(t-t_i)^2}{2\sigma_i^2}\right)
\end{equation}

\begin{equation}\label{eq:exp_decay}
    F(t) = 
    \begin{cases}
        e^{-t/\tau_{\text{sc}}},& \text{if } t\geq 0\\
        0,              & \text{otherwise}
    \end{cases}
\end{equation}

\begin{equation}\label{eq:pulse_profile}
    I(t) = 
    \begin{cases}
        \sum_{i=0}^n G_i(t), & \text{if BIC$_{\text{sc}}$>BIC$_{\text{g}}$}\\
        F(t') \circledast \sum_{i=0}^n G_i(t'), & \text{otherwise}
    \end{cases}
\end{equation}

Scattering is not the only explanation for the exponential broadening of the burst; intra-channel dispersive smearing is an instrumental effect that can also produce such broadening. While scattering is roughly proportional to $\nu^{-4}$, intra-channel smearing is proportional to $\nu^{-3}$, and it becomes significant when the burst width is not resolved at the time-frequency resolution of the instrument. 
For each burst, we compute the expected intra-channel smearing $\Delta t_{\text{DM}}$ with the following equation \citep[][Section 4.1.2]{petroff_fast_2019}:
\begin{equation} \label{eq:smearing}
    \Delta t_{\text{DM}} = 8.3\times10^6\, \text{DM}\, \Delta\nu_{\text{ch}}\, \nu^{-3}\, \text{ms,}
\end{equation}
where $\Delta\nu_{\text{ch}}$ is the frequency resolution, and $\nu$ is the observing frequency, both in MHz. For Apertif, we have $\Delta\nu_{\text{ch}}=0.195$\,MHz and $\nu=1370$\,MHz.
For FRBs where BIC$_{\text{sc}}$<BIC$_{\text{g}}$, we compare the resulting scattering timescale to the expected intra-channel smearing. If we find that \tscat$<\Delta t_{\text{DM}}$, we will consider that the burst scattering is not resolved.

For bursts where we determine the exponential broadening to be produced by scattering, we compute the frequency-dependent exponential broadening of the form \tscat$\propto\nu^{-\alpha}$, in order to get the dependence on frequency and determine the scattering index $\alpha$ when possible. To do this, we use \texttt{scatfit}\footnote{\texttt{scatfit}: \url{https://github.com/fjankowsk/scatfit}} \citep{jankowski_scatfit_2022}. %Since bursts with low S/N or narrowband bursts might not allow for such analysis, we only use frequency subbands where the S/N>3.5.
\refbf{We divide the full bandwidth into several subbands so that the signal in at least two subbands has S/N>3.5, ideally four or five. We specify the subband number for the scattering analysis of each FRB in Section~\ref{sec:results}.}

We define the width of each burst component as the full width at tenth maximum (FWTM) of the fitted Gaussian for consistency with the First CHIME/FRB Catalog \citep{chimefrb_collaboration_first_2021}, plus a factor \tscat$\ln 10$ to take into account the scatter broadening. 
The total width of the burst is defined as follows in the general case of a multi-component burst:
\begin{equation} \label{eq:width_mc}
    \mathcal{W} \text{ (ms)} = t_f -t_0 + (\text{FWTM}_0 + \text{FWTM}_f)/2 + \tau_{sc}\ln{10},
\end{equation}
where $t_0$ and $t_f$ are respectively the arrival time of the first and last subcomponents of the burst, and $\text{FWTM}_0$ and $\text{FWTM}_f$ the full width at tenth maximum of the first and last components, respectively. \refbf{In this way, the total burst width includes all of its subcomponents.}
For a single component burst and the independent burst subcomponents, the total width is defined as:
\begin{equation} \label{eq:width_sc}
    \mathcal{W} \text{ (ms)} = \text{FWTM} + \tau_{sc}\ln{10}.
\end{equation}

If the scattering timescale of the FRB is unresolved, the term depending on \tscat\ in Eq.~\ref{eq:width_mc} and \ref{eq:width_sc} equals zero.
Section~\ref{sec:multi-component} details the results of this analysis.

\subsection{Frequency structure}
\label{sec:fstruct}

% python /home/arts/pastor/scripts/arts-analysis/plot_frbs.py -all -fluxcal -dm
\begin{figure*}
    \centering
    \includegraphics[width=17cm]{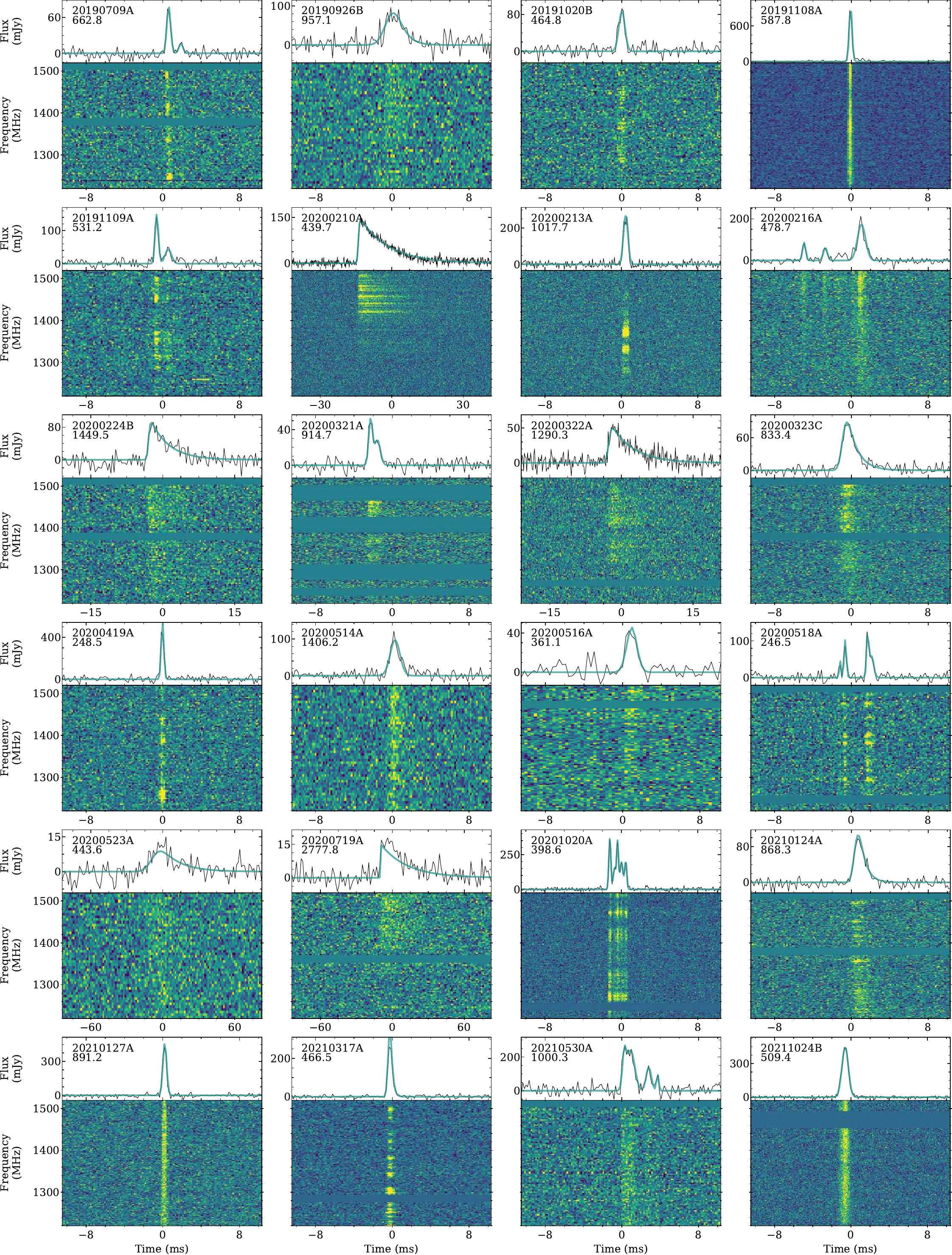}
    \caption{Dynamic spectra of all FRBs detected with Apertif. Top subpanels: averaged pulse profile in black, and
      fitted profile in teal. Each top left corner contains the name from the
      TNS (the Transient Name Server; \url{www.wis-tns.org}), and the applied DM in \pccm. Bottom subpanels: the dynamic spectrum, rebinned in time and frequency to optimise burst visibility.}
    \label{fig:frbs_dynspec}
\end{figure*}

The frequency structure of the detected FRBs provides information about the intrinsic burst spectrum and bandwidth, as well as phase modulations that could be intrinsic or produced by the propagation of the radio waves through the turbulent interstellar medium (ISM), known as scintillation.
We obtain the FRB spectra $S(\nu)$ by averaging their frequency structure over the total burst duration $\mathcal{W}$ defined in Section~\ref{sec:morphologies}.
In the case of bursts with a frequency extent narrower than the observing bandwidth, 
we fit the averaged spectrum to a Gaussian and a power law, and select the function with the lowest BIC. In the case of
a Gaussian fit, we define the burst peak frequency and the burst bandwidth respectively as the centre and the FWTM of
the fitted Gaussian.
For the power law spectral fits,
we derive the resulting spectral index, $\Gamma$.
When $\Gamma$ is positive, the burst is brighter at the top of the band than at the bottom.
\refbf{The index we report is corrected for the fact that the  primary-beam gain steadily increases from the bottom to the top of the band \citep{van_cappellen_apertif_2022}, which would otherwise lead  to an overestimation of $\Gamma$ by $\sim$1. }

To determine the scintillation bandwidth, we compute the auto-correlation function (ACF) of all burst spectra, removing the zero-lag frequency value, and fit the central peak to a Lorentzian. We define the scintillation bandwidth $\Delta\nu_{\text{sc}}$ as the half width at half maximum (HWHM) of the fitted Lorentzian. The ACF is defined as follows \citep[see Section 4.2.2 from][and references therein]{lorimer_handbook_2004}:

\begin{equation}\label{eq:acf}
    \text{ACF}(\Delta\nu) = \dfrac{\displaystyle \sum_{\nu}(S(\nu))(S(\nu+\Delta\nu)) }{\displaystyle\sqrt{\sum_{\nu}(S(\nu))^2  \sum_{\nu}(S(\nu+\Delta\nu))^2}},
\end{equation}
where $S(\nu)$ is the burst averaged spectrum at frequency $\nu$ and $\Delta\nu$ the frequency lag.
In the case of multi-component bursts, we assume the scintillation to be the same for all subcomponents, since the subcomponent separation is small compared to the typical scintillation timescales of a few minutes observed in Galactic pulsars \citep{narayan_physics_1992, bhat_pulsar_1998}, and observed differences in frequency structure do not appear to change between subcomponents, as might be the case from an intrinsic structure.
The scintillation analysis results are summarised in Section~\ref{sec:result_scintillation}.

For every FRB that is detected away from boresight, 
the spectrum we analyse is provided by an SB that is composed from the bands of several TABs (cf.~Sect.~\ref{sec:observations}). 
The number of combined TABs ranges from 0 (the central SB) to 8 (an outer SB). 
These TABs overlap but have some roll-off~\citep[see][]{van_leeuwen_apertif_2023}.
Variations in S/N with frequency of order 10\% may be introduced throughout the band, between the edge and peak of each subsequent TABs. 

\subsection{Polarisation} \label{sec:polcal}

For any ALERT observation, the Stokes $I$ data, or total intensity, are always saved as filterbank files. However, the Stokes $Q$, $U$, and $V$ data are only saved if \texttt{AMBER} identifies a candidate with S/N>10, a duration <10\,ms, and a DM greater than 1.2 times the predicted Milky Way contribution in the direction of the FRB according to the YMW16 electron density model \cite{yao_new_2017}.
Such a detection triggers data dumps of the four Stokes parameters \citep[see][]{van_leeuwen_apertif_2023}.

The Stokes parameters allow us to carry out polarisation analyses, and thus are a powerful tool for understanding the properties intrinsic to the FRBs and the environment where they live. This includes estimating the Faraday rotation measure (RM, \radsqm), the linear ($L=\sqrt{Q^2+U^2}$) and circular ($V$) polarisation intensities, and studying the polarisation position angle (PPA, $\psi$) evolution.

Once we obtain the observed source RM (\rmobs) using the analysis detailed in Appendix~\ref{app:RM_synth}, we compare this to the expected MW contribution in the direction of each burst (\rmmw) using the Faraday rotation map from \cite{hutschenreuter_galactic_2022}. We then convert the resulting RM to the FRB redshift expected from the Macquart relation ($z$):

\begin{equation} \label{eq:rm_redshift}
    \rmhost = (\rmobs - \rmmw)\times(1+z)^2 
\end{equation}

The Stokes data must be calibrated for leakage between the different Stokes parameters before applying any polarisation analysis. \refbf{It has been shown that although at the Apertif beam centres the fractional leakage between Stokes I/Q and U/V is low ($\sim0.001$), at the edge beams it can be up to 0.1 \citep{van_cappellen_apertif_2022}.}
The calibration is performed using a linearly polarised calibrator (we used 3C286 or 3C138), and an unpolarised calibrator (3C147). A phase difference between the $x$ and $y$ complex gains will result in leakage between $V$ and $U$, and this phase can be solved with a source with linear but no circular polarisation. On the other hand, an unpolarised source will determine the gain amplitude difference between the $x$ and $y$ feeds, and thus the leakage of $I$ into $Q$.
We only started adding the unpolarised source to the calibration observations from April 2020, so all FRBs from earlier dates have no $I/Q$ leakage correction. %Some FRBs were detected on the last day of the observing run; in those cases no calibrator observations are available. 
After any new FRB detection, scheduling constraints permitting, we carried out observations on and off the linearly
polarised and unpolarised calibrators. The calibrators were placed at the centre of the CBs where the FRB had been
discovered. This centre corresponds to SB 35, though the FRBs were often detected in different SBs. We thus made the assumption that there is a negligible leakage difference between the central and surrounding SBs.

Upon detailed analysis and calibration of the Stokes data, after the survey completion, we concluded that this assumption did not always hold true. In some cases, after calibrating the $U/V$ leakage, a residual $V$ signal oscillating with frequency at the same rate as $Q$ was still observed, which is not consistent with expected physical phenomena (e.g. Faraday conversion is expected to be much smaller than Faraday rotation at our observing frequencies, \citet{gruzinov_conversion_2019}). 
In those cases, we applied a technique in which we identified a frequency dependent phase minimising the $V$ oscillations, and rotated Stokes $U/V$ by the resulting phase. 
To test this technique, we applied it to FRB\,20191108A \citep{connor_bright_2020}, which was detected in SB\,37 but calibrated with the linearly polarised calibrator 3C286 observed in SB\,35. 
Using this technique to minimise the circular polarisation, we find a linear polarisation fraction $L=86\pm2\%$, higher
than the original $50\%$, and a circular polarisation fraction consistent with 0 ($V=5\pm7\%$), stricter than the
originally reported limit of $\lesssim13\%$. The rotation measure computed through RM synthesis is $+473.1\pm2.2$\,\radsqm, consistent with the $+474\pm3$\,\radsqm\ obtained through a least square fit of the position angle as a function of $\lambda^2$ in \cite{connor_bright_2020}. In this way we validated the technique that we subsequently applied to several FRBs.

% Many FRBs, both repeaters and one-offs, have now been observed to have high RM values originating either in the local environments or in the host galaxies of the FRBs \citep[e.g.][]{michilli_extreme_2018, anna-thomas_magnetic_2023, connor_bright_2020, mckinven_polarization_2021}.  
% \newpage
% ~\\
% \newpage % at start: p. 8-28
%-------------------------------------------------------
\section{Results}\label{sec:results}

Between July 2019 and February 2022, a total of 24 new FRBs were discovered within the % Apertif FRB 
ALERT survey. This includes  FRB\,20190709A, FRB\,20190926B, FRB20191020B, FRB\,20191108A, FRB\,20191109A, and
FRB\,20201020A that were reported in previous publications \citep{van_leeuwen_apertif_2023, connor_bright_2020, pastor-marazuela_fast_2023}. 
% This is one of the largest samples of one-off FRBs above 1\,GHz. 
The dynamic spectra and fitted pulse profiles of all 24 FRBs are presented in Fig.~\ref{fig:frbs_dynspec}. All FRBs were followed up with Apertif observations for 30\,h up to 450\,h, but none were seen to repeat. The bursts display different morphologies, including broadband and narrowband single components, and a high fraction of bursts with multiple components peaking at the same frequency. These morphologies are typical of the one-off FRBs in the First CHIME/FRB Catalog \citep{pleunis_fast_2021}; the lack of observed repetitions thus reinforces this apparent relation between morphology and repetition.
Additionally, the bursts display a broad range of propagation properties that we will discuss below.
In this Section, we first describe in detail the properties of some of the FRBs in our sample with remarkable features (Section~\ref{sec:special_frbs}), and next the properties of the FRB ensemble from Section~\ref{sec:result_localisation} onwards. We discuss these results in Section~\ref{sec:discussion}.

% Changing figure location for better placement

\subsection{FRBs of special interest} \label{sec:special_frbs}
This section first describes the most interesting FRBs in our sample individually, ordered by detection date. 
Group of FRBs with similar features are presented together next.

\subsubsection{FRB\,20200210A} \label{sec:FRB20200210A}

\begin{figure}[h!]
    \centering
    \includegraphics[width=\hsize]{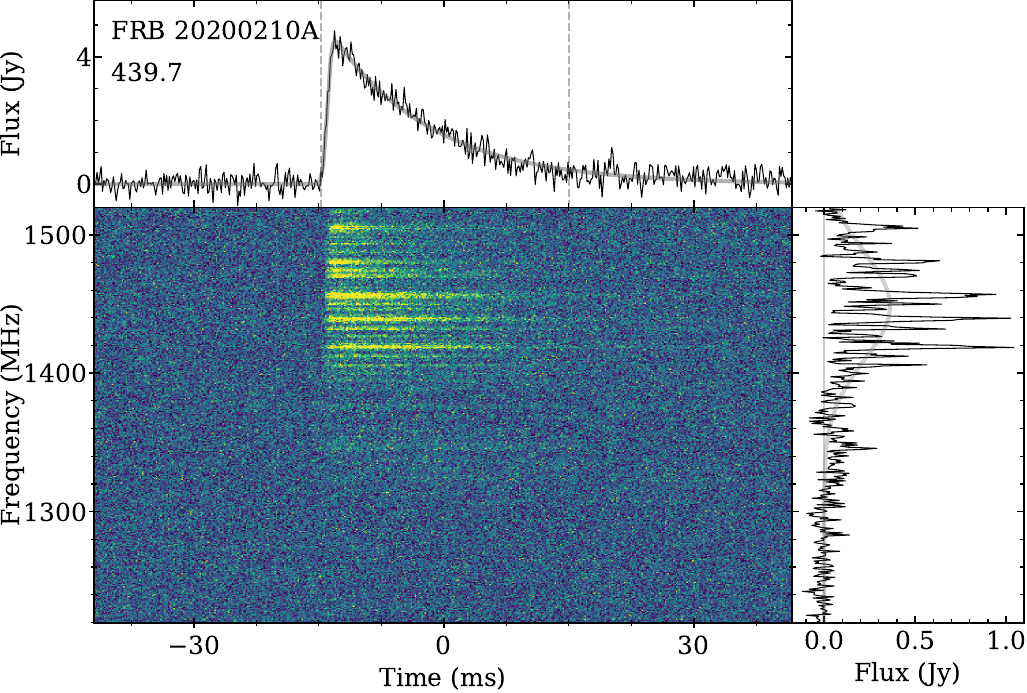}
    \caption{Dynamic spectrum of FRB\,20200210A. We re-use the conventions of
      Fig.~\ref{fig:frbs_dynspec}, but with fit lines now gray. The vertical, dashed lines
      in the  top panel demark the burst section from which the spectrum is extracted.
      This spectrum and its Gaussian fit are
      shown bottom-right. We ascribe the large intensity fluctuations to scintillation in the Milky Way. }
    \label{fig:FRB20200210A}
\end{figure}

This FRB (Fig.~\ref{fig:FRB20200210A}) presents a rare set of properties. It displays both temporal broadening from multi-path propagation, with \tscat$=12.6\pm0.3$\,ms, as well as a scintillation pattern with \scbw$=1.6\pm0.1$\,MHz, both measured at 1370\,MHz, which indicates the burst has traveled through two distinct scattering screens. Furthermore, it is a narrowband burst, with a bandwidth of $\sim170$\,MHz. The scattering timescale is uncommonly large for its DM of 439.7\pccm. Such a large scattering timescale at 1370\,MHz \refbf{is unlikely to be dominated by the Intergalactic Medium \citep[IGM,][]{zhu_scattering_2018}}. %or an intervening galaxy halo; we thus associate the first scattering screen with the host galaxy. 
The scintillation bandwidth falls within the expected ranges from the YMW16 \citep{yao_new_2017} and NE2001 \citep{cordes_ne2001i_2003} electron density models \refbf{for the FRB galactic coordinates ($l\sim76^{\circ}$, $b\sim+19^{\circ}$)}; the scattering screen producing scintillation is thus likely to be located in the Milky Way.

Scintillation can only occur when the scattering diameter by the first scattering screen is unresolved by the second, and this permits us to put constraints on the distance between the FRB and the first scattering screen \citep{masui_dense_2015, cordes_fast_2019}. 
\citet{cordes_fast_2019} determine the source size requirements for scattering and scintillation to be present at the same frequency band:
\begin{equation} \label{eq:scint_src}
    \tau_X \tau_G < \dfrac{1}{(2\pi\nu)^2}\dfrac{d_{so}^2}{L_X L_G} \simeq (0.16 \text{ ms})^2 \left( \dfrac{d_{so}^2}{\nu^2 L_X L_G}\right),
\end{equation}
where $\tau_X$ and $\tau_G$ are respectively the extragalactic and Galactic scattering timescales in ms, $\nu$ is the
observing frequency in GHz, $d_{so}$ the angular diameter distance from source to observer in Gpc, and $L_X$ and $L_G$
the distances of the lenses to the source and the observer  respectively, in kpc. We want to determine the distance upper limit between the source and the first scattering screen it encounters, $L_X$.
Scintillation bandwidths can be converted to scattering timescales with the following equation:
\begin{equation} \label{eq:scint_to_scat}
    \tau_{\text{sc}} = C_1/2\pi\Delta\nu_{\text{sc}},
\end{equation}
where $C_1$ is a constant with a value close to unity that depends on the medium scattering properties, and we assume $C_1=1$ for a thin scattering screen \citep[Eq.~8 from][Section 4.2.3]{cordes_diffractive_1998, lorimer_handbook_2004}.

In the case of FRB\,20200210A at a frequency $\nu=1.37$\,GHz, we have the extragalactic scattering timescale $\tau_X=12.6\pm0.3$\,ms, and the Galactic scintillation bandwidth $\Delta\nu_{\text{sc}}=1.6\pm0.1$\,MHz, which yields $\tau_G=0.1\ \mu$s. Given the Galactic latitude of the FRB, the scintillation is likely produced in the Milky Way thick disk, at $L_G\sim1$\,kpc \citep{ocker_large_2022}. 
From the extragalactic DM of the FRB alone, we estimate a redshift of \zmacquart$=0.36$, and thus an angular diameter distance upper limit of $d_{so} = 1.09$\,Gpc. By using these values in Eq.~\ref{eq:scint_src}, we find an upper limit on the distance between the FRB and the scattering screen at its host galaxy of $L_X\lesssim12$\,kpc. 
However, the presence of scattering allows us to use a joint scattering-dispersion redshift estimator. We do this by applying the method described in \cite{cordes_redshift_2022}, and assume a lognormal probability density function (PDF) for the scattering parameter $\phi_{\tau}\equiv \tilde{F}G$. We find the estimated median redshift to be $z=0.11$, which corresponds to an angular diameter distance of $d_{so} = 0.43$\,Gpc. With this new redshift constraint, we find the distance upper limit between the FRB and its scattering screen to be just $\sim2$\,kpc. This is fully consistent with scattering in the host galaxy, even for a dwarf host, though it is not constraining enough to determine if the scattering originated in the circumburst environment. \refbf{Although such a large scattering timescale could in theory be produced by an intervening galaxy halo \citep{zhu_scattering_2018}, the close distance to the scattering screen we determined rules this out.}

The high S/N of this FRB allowed us to subdivide the burst into several frequency subbands to perform a fit of the scattering index (See Appendix~\ref{app:FRB20200210A_scattering} and Fig.~\ref{fig:FRB20200210A_scattering} for further details). We determined a robust measurement of the scattering index, $\alpha=13.8\pm0.9$. This scattering index is anomalous when compared to the pulsar population, and it will be further discussed in Section~\ref{sec:discussion_scattering_origin}. In the top panel of Fig.~\ref{app:FRB20200210A_scattering}, we notice that the two lower frequency subbands display wider fitted Gaussian components. This could indicate the presence of a second component at those frequencies, unresolved due to scattering. However, it is unclear how such component would affect the measured scattering index, since it will depend on its relative amplitude and frequency extent \citep{oswald_thousand-pulsar-array_2021}. Even when removing those two subbands from the fit, we obtain a similar $\alpha$.

\begin{figure}[h!]
    \centering
    \includegraphics[width=\hsize]{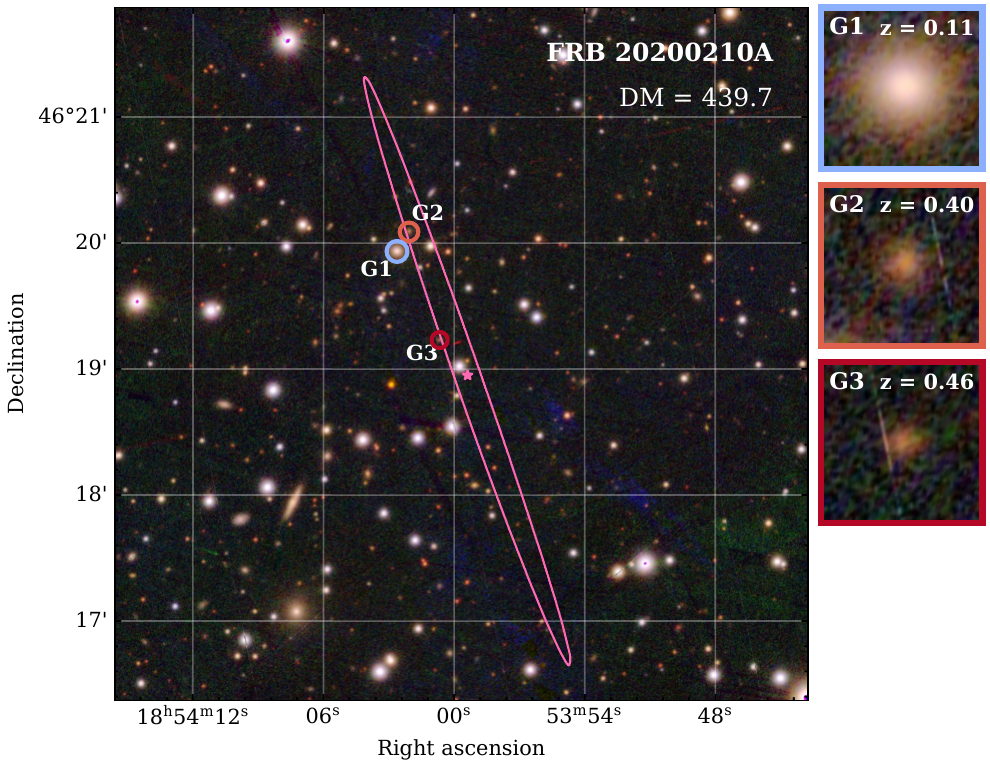}
    \caption{Localisation region (pink contour) of FRB\,20200210A. The three
      galaxies within 7" of the error region are marked with circles,
       coloured ranging from  bluer (lower redshift)
      to redder (higher redshift), with the galaxy ID (same as in Table~\ref{tab:host_galaxies}). The background image is from Pan-STARRS DR1.
      The subplots (right) are zoomed images of these three galaxies, with 12\,$\arcsec$ FoV, and border colour matching the main-panel circles. Each galaxy ID and photometric redshift are indicated at the top.}
    \label{fig:FRB20200210A_loc}
\end{figure}

We localised this FRB to an error region of 0.78\,arcmin$^2$, centred at the coordinates 18:53:59.4	+46:18:57.4 in
RA (hms) and DEC (dms).
\refbf{As detailed in Sect.~\ref{sec:localisation_data_analysis}, we next compare this area to
Pan-STARRS1, down to magnitude 23.2.}
We find one galaxy, G2, contained within the error region, at a photometric redshift of 0.40. This is close to the upper limit set by the Macquart relation assuming \dmhost=100\,\pccm, \zmacquart=0.36$^{+0.10}_{-0.22}$, but much higher than expected from the scattering. However, there are two additional galaxies within 7" of the error region and the \zmacquart\ upper limit. G3 is located 1" from the error region and has $z_{\text{phot}}\sim0.46$, while G1 is 7" away from the error region and has $z_{\text{phot}}\sim0.11$. The latter has a Kron radius \citep{kron_photometry_1980} of 6.1", placing it very close to the FRB localisation region. Since we have performed no astrometric corrections, the galaxy could well be inside the FRB error region. After performing a \texttt{PATH} analysis, assuming an unseen prior $P(U)=0.01$ given the expected low redshift of the galaxy, we find the most likely host to be G1, with $P(G1|x)=0.58$. The host galaxy candidates and \texttt{PATH} results are presented in Table~\ref{tab:host_galaxies}, while the FRB localisation region and the host galaxy candidates are shown in Fig.~\ref{fig:FRB20200210A_loc}.

%  /Users/user/Documents/projects/ARTS/FRBs/FRB200210/localisation/localisation_frb.ipynb

\subsubsection{FRB\,20200213A} \label{sec:FRB20200213A}

\begin{figure}[h!]
    \centering
    \includegraphics[width=\hsize]{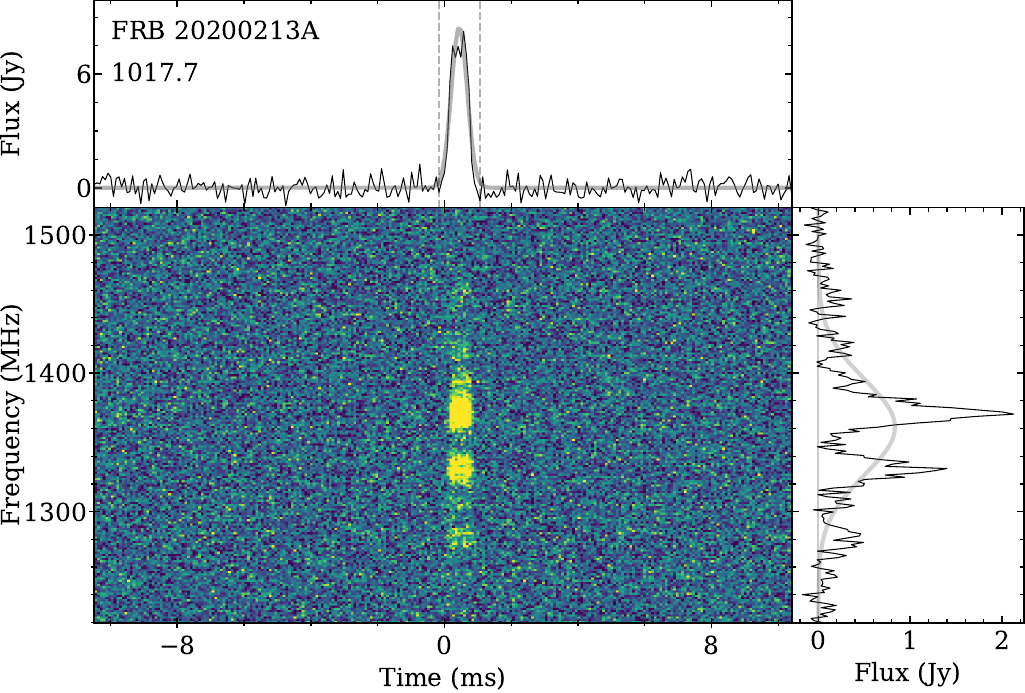}
    \caption{Dynamic spectrum of FRB\,20200213A. The pulse profile and the spectrum are both fitted to a Gaussian.}
    \label{fig:FRB20200213A}
\end{figure}

This FRB (Fig.~\ref{fig:FRB20200213A}) is the most narrow-banded of the sample. It has a bandwidth of 145\,MHz, less than half of the total observing bandwidth. Additionally, it displays a strong frequency modulation, with two main patches of similar intensity and an array of lower intensity patches above and below the central ones. \refbf{This resembles some of the FRBs detected in the ASKAP fly's eye survey \citep{shannon_dispersionbrightness_2018}.} The frequency modulation has a 19\,MHz bandwidth, significantly larger than the 4\,MHz \scbw\ predicted by the NE2001 model (YMW16 predicts 1\,MHz). This suggests that either the Galactic ISM is more uniform than predicted by the electron density models, or that the frequency structure is intrinsic to the source.
The temporal structure of the burst presents a single component with a flat peak. Given the DM of 1017.7\pccm\ and the instrument frequency resolution of 195\,kHz, the FRB width is close to the dispersion broadening  \citep{petroff_fast_2019}. The flat peak could thus be a result of instrumental smearing instead of the intrinsic structure of the burst, or be the signature of a second or even third component indistinguishable from the first.

The detection of this burst in SB\,48 triggered the storage of the Stokes data. Subsequently, we scheduled observations of the linearly polarised calibrator 3C138 in SB\,35. The calibrated Stokes data is presented in Appendix Fig.~\ref{fig:stokes}. Since we observe a faint indication of Q/U oscillations, we applied the RM synthesis algorithm to the frequency channels where the burst is bright enough; we selected a frequency extent contained within the FWTM of the spectrum fitted to a Gaussian, between 1291 and 1436\,MHz. We find a resulting RM of $300.3\pm2.1$\,\radsqm, and after Faraday de-rotating we obtain linear and circular polarisation fractions of $L=10\pm3\%$ and $V=8\pm6\%$ respectively. \refbf{Given the S/N=18 of the burst and the relatively low $L$, the RM versus total linearly polarised flux plot (Fig.~\ref{fig:RM_FRB200213}) shows several peaks of comparable height to the largest one. The RM measurement is thus ambiguous, but we will assume this value hereafter}.
Given the expected MW contribution $\rmmw\sim-17$\,\radsqm in the direction of the FRB, the RM in the host galaxy could be as high as \rmhost$=1461^{+380}_{-576}$\,\radsqm for the expected redshift range \zmacquart$=1.15^{+0.24}_{-0.58}$. Since the polarised fraction is low and the burst is narrowband, we advise caution when interpreting this RM.

We localised this burst to a small area of 0.94\,arcmin$^2$. However, since the FRB could be located at a redshift as
high as $\sim1.4$, we identify nine galaxies in the error region as host galaxy candidates. Dimmer galaxies, too faint
to appear in the Pan-STARRS1 catalogue, could also exist in the area.
These results are included in 
the overview Figure of FRB localisations and host galaxy candidate
positions, Fig.~\ref{fig:localisation}.

\subsubsection{FRB\,20200216A} \label{sec:FRB20200216A}

\begin{figure}[h!]
    \centering
    \includegraphics[width=\hsize]{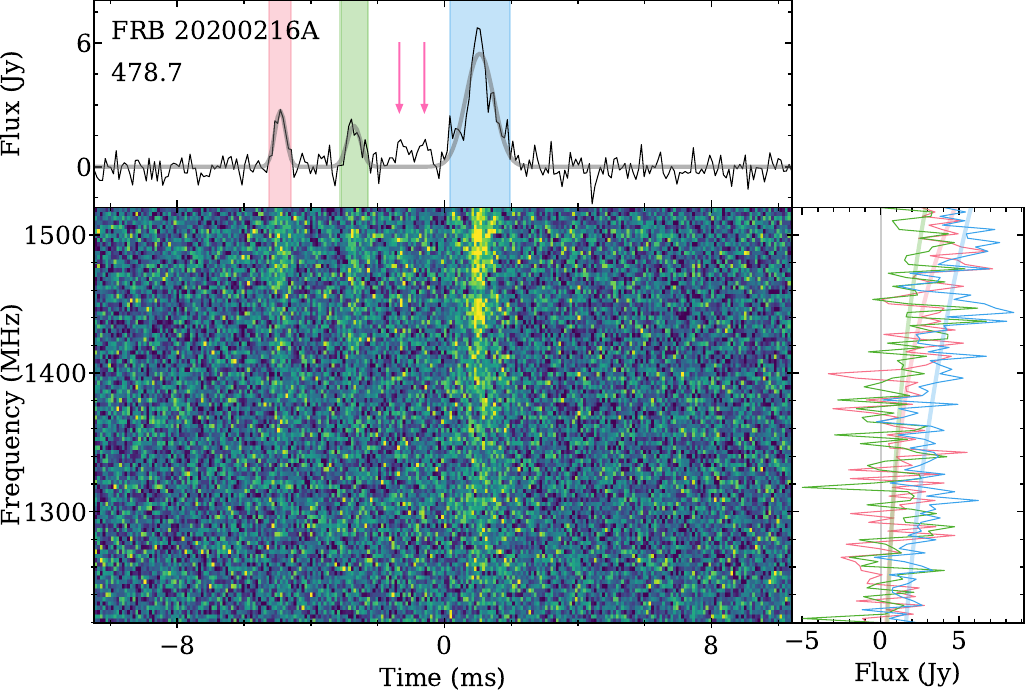}
    \caption{Dynamic spectrum of FRB\,20200216A. On the pulse profile (top panel), each coloured shaded region shows a distinct component and fitted to a Gaussian. Their respective spectra are shown on the bottom right panel with the same colour, and fitted to a power law (transparent solid lines). The two pink arrows on the pulse profile indicate the position of the two potential subcomponents between the precursors and the main component.}
    \label{fig:FRB20200216A}
\end{figure}

This FRB consists of a bright main burst subcomponent with two narrow precursors of about a third of the amplitude of the main burst, as can be seen in Fig.~\ref{fig:FRB20200216A}. The separation between the two precursors is $\sim2.2$\,ms, while the main component arrives $\sim3.8$\,ms after the second precursor. The pulse profile appears to contain two bumps between the second precursor and the main component, but their amplitude is too low to be identified as real subcomponents. We carried out a timing analysis identical to the one described in \citet{pastor-marazuela_fast_2023}, where the presence of a periodicity in  FRB\,20201020A was investigated, including the power spectrum analysis and the study of the time separation between subcomponents, but we find no evidence of periodicity in this burst. 

The spectrum of each of the FRB\,20200216A subcomponents can be well fitted by a power law. The power law spectral indices of the first, second, and third components are respectively 10.6, 7.7 and 4.6. 
The precursors seem to peak at higher frequencies than the main subcomponent,  reminiscent of the downwards drifting effect typically observed in repeating FRBs \citep[e.g.][]{hessels_frb_2019}. However, the main subcomponent is brighter at the top of the band. The lack of visible emission at the bottom of the band of the two precursors could be simply explained by their lower amplitude, which is below the noise level at lower frequencies. The emission of each component is likely to peak at similar frequencies, but above the highest observing frequency. 
We further applied Gaussian function fits to the spectra, but the BIC favours a power law model for all components.
We thus identify the morphology of this FRB as a multi-component burst with components peaking at the same frequency \citep{pleunis_fast_2021}. 

Given the DM, the expected MW and halo contribution to the DM, and assuming a host galaxy contribution of 100\,\pccm\ in the host frame, the expected redshift for this FRB is \mbox{\zmacquart$=0.44^{+0.12}_{-0.24}$}. We localised it to an error region of 2.34\,arcmin$^2$ around the coordinates 22:08:24.7 +16:35:34.6 in RA and DEC, within which we identify one galaxy, with a photometric redshift $z_{\text{phot}}=0.52\pm0.09$, consistent with the expected limits, that we label G3. We find three additional galaxies within 5" of the error region, at similar redshifts of $z\sim0.5$. 
After running a \texttt{PATH} analysis, we find the brightest and nearest of the galaxies, labeled G1, to be the most likely host, with $P(G_1|x)\sim0.42$. This galaxy is located 4" away from the error region, and has a photometric redshift $z_{\text{phot}}\sim0.49$. G3 is the second brightest and second most likely host, with $P(G_3|x)\sim0.37$. Since these posterior probabilities are similar, we cannot confidently identify the host galaxy of FRB\,20200216A. The details on the host galaxy candidates and \texttt{PATH} analysis are presented in Table~\ref{tab:host_galaxies}.

\begin{figure}[h!]
    \centering
    \includegraphics[width=\hsize]{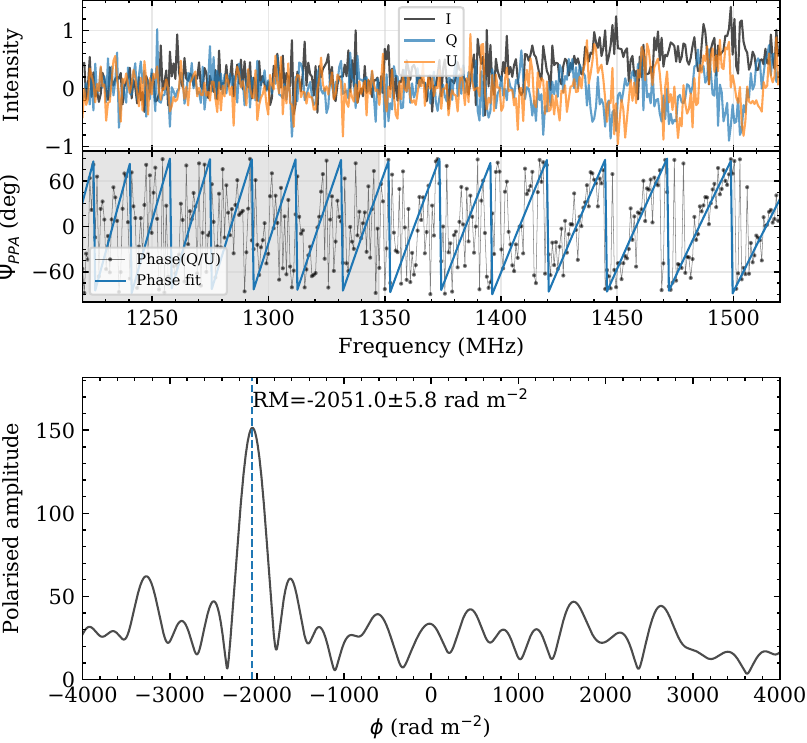}
    \caption{Measured polarisation properties of FRB\,20200216A. Only frequencies above 1347\,MHz were used for RM synthesis, since there is not enough signal below that frequency.}
    \label{fig:pol_FRB20200216A}
\end{figure}

The detection of FRB\,20200216A triggered a full-Stokes data dump. Subsequently we scheduled on/off observations of the linearly polarised source 3C286 to calibrate the UV leakage. We observe quick oscillations in the sign of Stokes Q and U in the main component of the burst after calibration, which we associate with Faraday rotation (See Appendix Fig.~\ref{fig:stokes}). We selected the frequency channels contained within the full width at fifth maximum (FWFM) of a Gaussian fit to the spectrum of the main component, which are all those above 1347\,MHz, and then we performed the RM synthesis technique on the data. We find the best solution to be RM$=-2051\pm6$\radsqm, as shown in Fig.~\ref{fig:pol_FRB20200216A}. 
After Faraday de-rotating, we obtain linear and circular polarisation fractions of $38\pm6\%$ and $11\pm4\%$ respectively for the main component. The polarisation fraction of the two fainter precursors appears to be slightly lower.
Although the linear polarisation fraction is low and the frequency extent of the burst is narrow, the resulting RM and $\psi_{\text{PPA}}$ match well the phase between Stokes $Q$ and $U$.
This is the second largest RM ever measured in a one-off FRB to date \citep{sherman_deep_2024, mckinven_polarization_2021}. The expected \rmmw\ contribution in the direction of the burst is $-36\pm10$\,\radsqm, totalling $\sim-2015$\,\radsqm\ from an extragalactic origin.
Assuming that the extragalactic RM originates from the host galaxy, this would translate to an RM of $\sim-4200^{+1300}_{-800}$\,\radsqm\ in the host reference frame at \zmacquart. Three repeating FRBs with high or extreme RM values have been associated to persistent radio sources \citep[PRSs,][]{marcote_repeating_2017, niu_repeating_2022, bruni_nebular_2023}, and hence finding a galaxy within the error region associated with such a radio source might be a strong indication that the FRB was produced in that galaxy. The field is not covered by an Apertif imaging survey (see Sect.~\ref{sec:result_localisation}). We searched for radio emission within the error region of the FRB in the Rapid ASKAP Continuum Survey (RACS) Mid \citep{duchesne_rapid_2023}, but we found no radio source associated to any of the host galaxy candidates. \refbf{From the RACS-Mid data, we set a $3\sigma$ continuum radio flux upper limit of 0.6\,mJy. This translates to a luminosity upper limit of $4.8^{+3.7}_{-4.0}\times10^{30}$\,erg\,Hz$^{-1}$\,s$^{-1}$. Compared to the typical luminosities often used to define a PRS \citep[$10^{29}$\,erg\,Hz$^{-1}$\,s$^{-1}$]{law_fast_2022}, this upper limit is not highly constraining. Deeper radio observations might reveal any potential continuum emission in the future.}

% %  /Users/user/Documents/projects/ARTS/FRBs/FRB200216/localisation/localisation_frb.ipynb
% \begin{figure}[h!]
%     \centering
%     \includegraphics[width=\hsize]{plots/localisations/FRB20200216A_localisation_zphot.pdf}
%     \caption{Localisation region (pink contour) of FRB\,20200216A and galaxies within 5" of the error region.
%       The subplots on the right shows zoomed images of the galaxies, with 12\,$\arcsec$ FoV, and border colour matching the main-panel circle. The galaxy IDs and photometric redshift are indicated at the top.}
%     \label{fig:FRB20200419A_loc}
% \end{figure}

\subsubsection{FRB\,20200419A}

This FRB with a low DM of 248.5\,\pccm\ consists of a single component with a width of 0.58\,ms and no measurable scattering. The burst is broadband, with an intensity that fluctuates with frequency with a decorrelation bandwidth of 7.5\,MHz, slightly higher than the expected 2.4\,MHz from the NE2001 model.
Upon detection of the burst, the full-Stokes data were saved, and observations of both the linearly polarised calibrator 3C286 and the unpolarised calibrator 3C147 were carried out. The burst is highly polarised, with a linear polarisation fraction of $L=77\pm6\%$ and null circular polarisation fraction $V=4\pm6\%$. All the linear polarisation is observed in Stokes Q, and thus no RM can be measured. 

\begin{figure}[h!]
    \centering
    \includegraphics[width=\hsize]{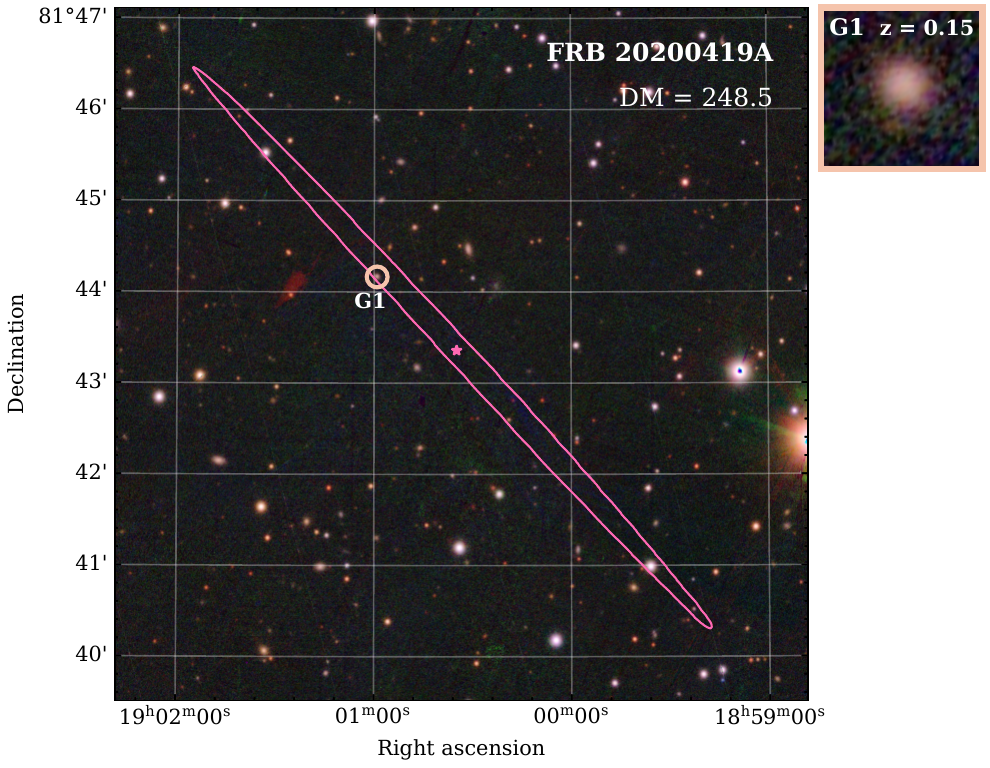}
    \caption{Localisation region and host galaxy candidate of FRB\,20200419A (as in Fig.~\ref{fig:FRB20200210A_loc}).}
    \label{fig:FRB20200419A_loc}
\end{figure}

We localised this FRB to an ellipse centred around 19:00:34.2 +81:43:20.5 in RA and DEC with a 1.29\,arcmin$^2$ error region.
The expected redshift from the Macquart relation is \zmacquart$=0.08^{+0.04}_{-0.06}$, but we find no known galaxies at such low redshift within the FRB error region. The lowest redshift galaxy we identify has a photometric redshift z$=0.15\pm0.03$, consistent with the expected redshift within errors, and we find no other host galaxy candidates within 10" of the error region.
A \texttt{PATH} analysis determines that the galaxy is $\sim70\%$ likely to be associated with the FRB. Given the low DM of the FRB and the depth of the Pan-STARRS catalogue, we assumed a very small unseen prior, $P(U)=0.001$, since even dwarf galaxies would be detected at such low redshift (See Table~\ref{tab:host_galaxies} for details).
The host galaxy candidate would be a good target for optical follow-up to determine its spectroscopic redshift. A confirmation of the galaxy redshift might indicate a lower DM contribution from the MW or the halo, or a host galaxy contribution <100\,\pccm, or a combination of both. The FRB error region and host galaxy candidate are shown in Fig.~\ref{fig:FRB20200419A_loc}.

%  /Users/user/Documents/projects/ARTS/FRBs/FRB200419/localisation/localisation_frb.ipynb

\subsubsection{FRB\,20200514A} \label{sec:frb20200514a}

% \begin{figure}[h!]
%     \centering
%     \includegraphics[width=\hsize]{plots/stokes/FRB200514_RM_phase.pdf}
%     \caption{Measured polarisation properties of FRB\,20200514A. \ipm{Write this section, maybe add dynamic spectrum or Stokes?}}
%     \label{fig:pol_frb20200514a}
% \end{figure}

FRB\,20200514A was detected with a DM of 1406.2\,\pccm\ as a single component burst with a total width of 2.2\,ms. Although the burst is broadband, it becomes brighter at the top of the band. Its detection triggered the dump of the Stokes data, and after calibration with the linearly polarised 3C286 and the unpolarised 3C147, we observe rapid oscillations of Q and U with frequency (See Fig.~\ref{fig:stokes}). After applying RM synthesis, we obtain an RM$=966.1\pm20.5$\,\radsqm, although the FDF shows significant secondary peaks \refbf{with heights similar to the largest one (See Fig.~\ref{fig:RM_FRB200514}). The reported RM results in polarisation fractions of $L=51\pm5\%$ and $V=21\pm9\%$.
Given the large secondary peaks, the reported RM is ambiguous. This likely arises from the relatively low S/N of the linearly polarised intensity, which is further lowered in each frequency channel and contributes to the appearance of multiple peaks of comparable height in the RM synthesis plot. However, since the peak at 966.1\,\radsqm\ is the largest, we will assume this value hereafter.}
The expected MW contribution to the RM in the direction of the burst is $\sim-215$\,\radsqm, \refbf{which would result} in an extragalactic RM of $\sim765$\,\radsqm. Given the high excess DM of the burst, its expected redshift is \zmacquart$=1.35^{+0.30}_{-0.66}$. Hence, if we assume the RM to be produced within the FRB host galaxy, it could be as high as \rmhost$=6500^{+2000}_{-3200}$\,\radsqm. 
The PPA remains roughly flat, with a marginal decrease of $\sim5^{\circ}$ along the burst duration.

\subsubsection{FRB\,20200518A} \label{sec:FRB20200518A}
% arts:~/pastor/scripts/arts-analysis/plot_single_frbs.ipynb
\begin{figure}[h!]
    \centering
    \includegraphics[width=\hsize]{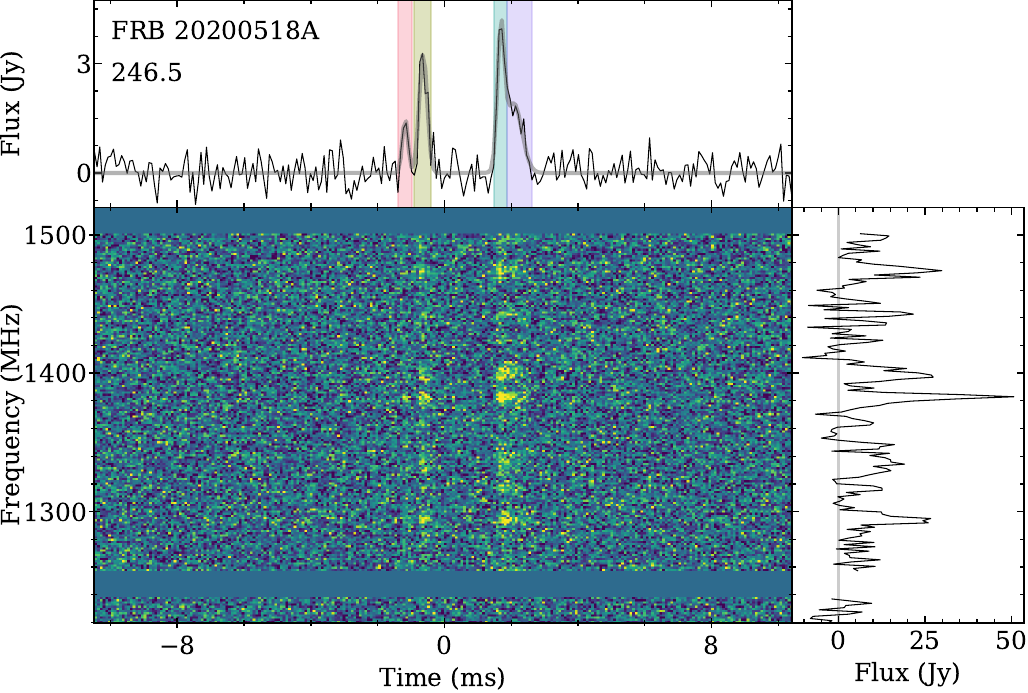}
    \caption{Dynamic spectrum of FRB\,20200518A. On the pulse profile, each of the four components is marked by a coloured shaded region, and fitted to a Gaussian. The spectrum on the bottom right panel is the sum of all components.}
    \label{fig:FRB20200518A}
\end{figure}

This FRB consists of two groups of two narrowly spaced subcomponents each. The space between the two groups is $\sim2.3$\,ms, while the space between the subcomponents of each group is 0.54\,ms on the first and 0.34\,ms on the second. In the first group, the second component has a larger amplitude, while the first component of the second group is the brightest of all four. The power spectrum of the average pulse profile presents several peaks, but each one corresponds to the separation between different components. The timing analysis does not provide evidence for periodicity.

All four subcomponents present a similar frequency extent. The peak frequency of the emission cannot be easily determined since the burst presents strong frequency modulations that we associate with scintillation, with a decorrelation bandwidth of \scbw$=5\pm2$\,MHz. This matches the expected Milky Way contribution of $\sim4$\,MHz from NE2001 \citep{cordes_ne2001i_2003}. 
\refbf{We tested whether the spectral modulation is consistent between all four subcomponents by cross-correlating the spectra between them. For all six subcomponent pairs, we find peaks at the zero-lag frequency, with large correlation coefficients. We compute the standard scores $z = (x-\mu)/\sigma$ of the zero-lag peaks, where $x$ is the correlation value at the peak, and $\mu$ and $\sigma$ are respectively the mean and the standard deviation of the correlation values for all frequency lags. For each subcomponent pairs, we find $z$ between three and seven, indicating that the spectral modulation is consistent between all four subcomponents.}
The burst shows no evidence of scatter broadening at the Apertif resolution.

\begin{figure}[h!]
    \centering
    \includegraphics[width=\hsize]{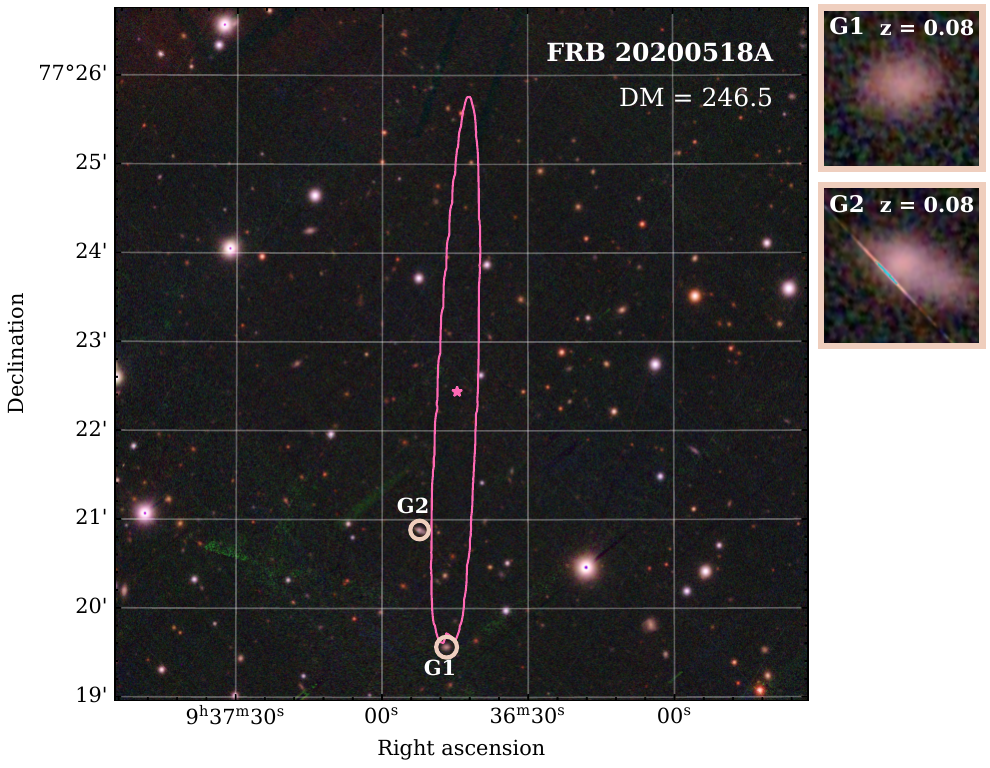}
    \caption{Localisation region of FRB\,20200518A and two host galaxy candidates.}
    \label{fig:FRB20200518A_loc}
\end{figure}

With a DM of 246.5\,\pccm, it is the least dispersed burst of our sample.
We localised this FRB to a narrow ellipse with a localisation area of 1.67\,arcmin$^2$ centred at the coordinates 09:36:45.3 +77:22:36.8 in RA and DEC. Given the source redshift and expected MW and halo contributions, we derive \zmacquart$=0.10^{+0.04}_{-0.08}$. Although we found no galaxies within the error region and redshift limit, we identified two galaxies with photometric redshifts $\sim0.08$ within 3.5" and 8" of the nominal error region edge, which given the astrometric uncertainty, could well be inside the actual error ellipse. The FRB localisation region and the two host galaxy candidates are shown in Fig.~\ref{fig:FRB20200518A_loc}.
We ran a \texttt{PATH} analysis on the two candidates assuming a small unseen prior, $P(U)=0.01$, given the low expected redshift. We found that both galaxies have a similar likelihood of being the host, with $P(G_1|x)\sim0.42$ and $P(G_2|x)=0.56$, as detailed in Table~\ref{tab:host_galaxies}.

%  /Users/user/Documents/projects/ARTS/FRBs/FRB200518/localisation/localisation_frb.ipynb

The Stokes data were saved after the detection of this burst, but unfortunately no calibration observations were carried out. The raw Stokes data show signal in I, Q, and V. By assuming the circular polarisation to be \refbf{low}, we apply the method to find the phase that would minimise $V$. As a result, we obtain a linear polarisation fraction $L=72\pm12\%$, and a residual circular polarisation fraction $V=29\pm17\%$. This would represent a highly linearly polarised burst if our assumptions are correct. \refbf{The residual $V$ appears mostly in the second group of components, which might indicate a difference in the polarisation between the two groups. However, $V$ remains low and with large errors, so we cannot confirm whether it is intrinsic to the burst or a result of the corrections applied without calibrators.} The burst does not display any significant Q/U modulations with frequency, and thus no RM can be estimated. The first group of components appears to show a higher linear polarisation fraction than the second one, although the latter also displays a peak in $V$ that could be an inaccuracy of the calibration procedure. The resulting PPA remains constant between the two component groups.
The Stokes data calibrated through the circular polarisation minimisation technique are shown in Fig.~\ref{fig:stokes}.

\subsubsection{FRB\,20200719A} \label{sec:FRB20200719A}

\begin{figure}[h!]
    \centering
    \includegraphics[width=\hsize]{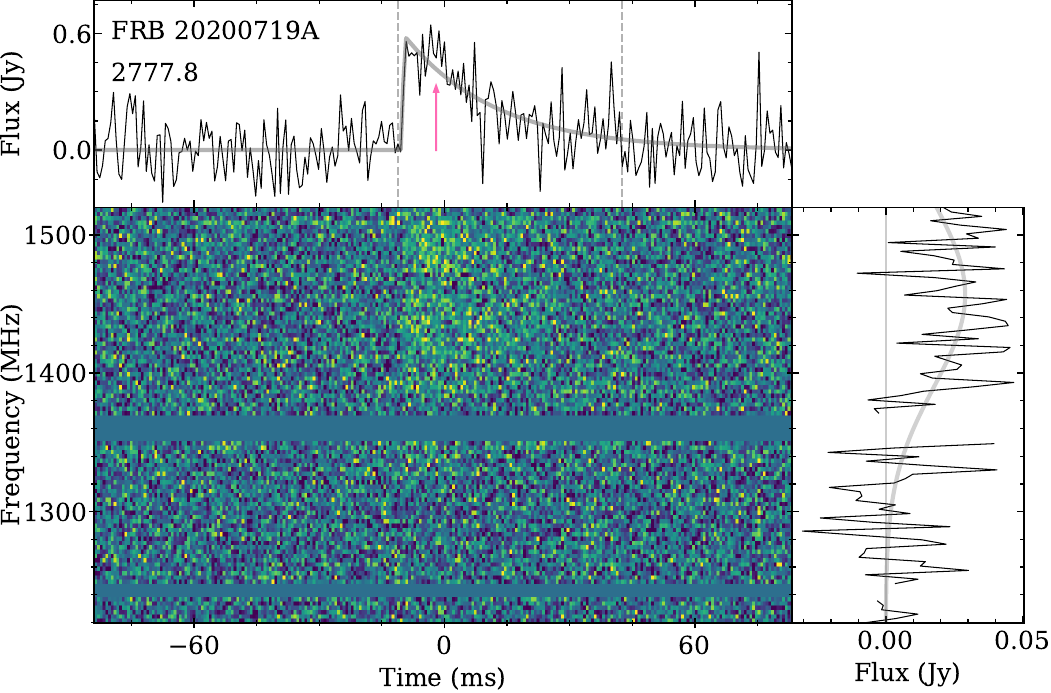}
    \caption{Dynamic spectrum of FRB\,20200719A. The pulse profile is fitted to a scattered Gaussian. The pink arrow indicates the position of an excess emission that might be explained by a second component merged to the first by scattering. The spectrum is fitted to a Gaussian.}
    \label{fig:FRB20200719A}
\end{figure}

FRB\,20200719A, with a DM of 2778\,\pccm, is the most dispersed FRB of our sample, as well as the most scattered, with \tscat=21\,ms.
It differs by more than 1000\,\pccm\  from the Apertif FRB with the second-highest DM.
Compared to the FRBs in the TNS\footnote{Accessed 2023 Nov 01}, it is the FRB with the third-largest DM known to date,
after
the Parkes 70-cm FRB 19920913A \citep{2022MNRAS.515.3698C} with a DM of 3338\,\pccm\ and the  
CHIME/FRB source FRB\,20180906B with a DM of 3038\,\pccm.
The inferred redshift of FRB\,20200719A is \zmacquart$\sim3.26^{+0.62}_{-1.35}$ if we assume a \dmhost\ contribution of 100\,\pccm. The large scattering timescale might however be an indication of a significant contribution to the DM from the host galaxy and the circumburst environment \citep{cordes_redshift_2022, ocker_large_2022}, which would place the FRB at a lower redshift. Nonetheless, since the host galaxy contribution to the observed DM evolves as DM${_{\text{host}}= \text{DM}_{\text{host,loc}}/(1+z)}$ \citep{deng_cosmological_2014}, even a large host local DM contribution would be diluted at high redshift.
We can estimate a  redshift lower limit assuming the host galaxy has a DM contribution as large as that originally found for the repeating FRB\,20190520B \citep{niu_repeating_2022}. 
Even though foreground studies of the FRB\,20190520B field
have since identified two intervening galaxy cluster halos that
reduce the required host DM contribution by as much as 50$-$70\% \citep{2023ApJ...954L...7L},
we use the original value here, to be  conservative in our limits.
% 20210117A  arXiv:2211.16790   -- arXiv:2303.07387 find DM_host=665 (rest frame) ; similar or a little higher.
The FRB\,20190520B host galaxy contributes DM$_{\text{host}}=902$\,\pccm\ to the observed DM. Since its host is located at $z=0.241$, the local DM contribution in the host frame is DM$_{\text{host,loc}}\sim1119$\,\pccm. 
If we now assume the host galaxy of FRB\,20200719A has a contribution to the DM as large as that originally suggested
for FRB\,20190520B, its redshift would still be $z_{\text{min}}\sim2.8^{+0.6}_{-1.2}$. This highly constraining lower limit still places the FRB at very large cosmological distances.

In principle, further increasing DM$_{\text{host,loc}}$
reduces the required distance, 
hence further overcoming the 1+$z$ dilution in DM$_{\text{host}}$. 
One might wonder if this double action  allows for a reasonable combination of distance and DM$_{\text{host}}$. But even if we place the host at z=1.0, the distance of the currently farthest FRB  \citep{2023Sci...382..294R}, we require a DM$_{\text{host,loc}}$ of over 2600\,{\pccm}, an extreme outlier of known values for DM$_{\text{host,loc}}$.

% log_frb_analysis.txt: Computing intersecting halos 
The probability of intersecting a foreground galactic halo increases with distance. For instance, the most highly dispersed FRB\,20180906B from the CHIME/FRB sample \citep{chimefrb_collaboration_first_2021} was shown to intersect within 1.4\,Mpc of a galaxy cluster \citep{connor_observed_2022}. For FRB\,20200719A, the possibility of intersecting foreground galaxies is not negligible. We follow \cite{prochaska_probing_2019}\footnote{\url{https://github.com/FRBs/FRB}} to determine how likely this FRB is of intersecting an intervening galaxy with a mass greater than the Milky Way within the line of sight (LoS).
We use the Aemulus halo mass function \citep{mcclintock_aemulus_2019} to generate galaxy halos with masses between $10^{12}$\,M$_{\odot}$ (roughly the MW mass) and $10^{16}$\,M$_{\odot}$. Next, we compute the average number of halos expected to occupy the comoving volume at the expected redshift of this FRB, \zmax$=3.26$. If we consider an intersection within the virial radius of the galaxies within the comoving volume, which is the distance at which we expect a foreground galaxy to have a significant contribution to the DM, we find the average number of galaxies to be $N(z)=2.633$. However, in order to have a significant contribution to scattering, the impact parameter must be lower \citep{ocker_constraining_2021}. If we consider 0.15 times the virial radius \citep[roughly 10 times the half mass radius, and between 20 and 40\,kpc depending on the mass,][]{kravtsov_size-virial_2013}, we find $N(z)=0.059$.
Assuming the location of the foreground galaxies within the comoving volume follows a Poisson distribution, the probability that the LoS crosses $k$ halos is given by:

\begin{equation}
    P(k|N(z)) = \dfrac{N^k e^{-N}}{k!}.
\end{equation}

The probability of intersecting at least one foreground halo is thus given by ${P(k\geq1|N(z)) = 1-e^{-N}}$. We find the probability of at least one intersection within the virial radius of the foreground galaxy to be $\sim$93\%, and within 0.15 times virial radius it is $\sim$5.8\%.
Foreground galaxies are thus very likely to contribute to the DM of this FRB, while the contribution to scattering is less likely. 

\refbf{Since foreground galaxy clusters could also contribute to the large observed DM, we searched for galaxy clusters intersecting or around the FRB LoS. Although we find several galaxy clusters around the FRB localisation ellipse from \cite{zou_galaxy_2021}, who identify galaxy clusters in the DESI Legacy Survey, none of these intersect the FRB LoS within their characteristic radius $R_{500}$, where $R_{500}$ is defined as the radius within which the mean density of the cluster is 500 times the critical density of the Universe, $\rho_c$, as shown in Fig.~\ref{fig:FRB20200719A_gal_cluster}. Foreground galaxy clusters are thus unlikely to contribute to a significant fraction of the observed DM.}

% The localisation region contains no known galaxies from the NED and GLADE databases, 
Within the localisation region, we find only two galaxies below a conservative redshift lower limit of 1.6 (at $1.4\pm0.5$ and $0.81\pm0.96$ respectively, see Fig.~\ref{fig:localisation}), but with such a high redshift upper limit, roughly $10^3$ dwarfs and $\sim$50 massive galaxies are expected to be contained within the localisation comoving volume. We thus ran a \texttt{PATH} analysis on the two galaxies assuming a large unseen prior, $P(U)=0.9$, \refbf{to take into account the expected number of massive galaxies within the comoving volume versus the ones we see}. The brightest galaxy, G2, is found to be the most likely host, with $P(G_2|x)\sim0.1$, while $P(G_1|x)\sim2\times10^{-4}$, as detailed in Table~\ref{tab:host_galaxies}. We thus cannot confidently associate FRB\,20200719A to any host galaxy.

The spectrum of the FRB can be fitted to a Gaussian peaking at $\nu_{\text{obs}}=1460$\,MHz and with a bandwidth (FWTM) of 260\,MHz. \refbf{The emission thus extends above the observing band.} The pulse profile fitted to a single scattered component shows excess of emission after the peak. This might be the signature of a second component that is blurred together with the first due to scattering. By fitting a two-component scattered model, we find a potential component separation of 5.95\,ms. However, the BIC of the single component model is marginally lower and hence it is preferred.
Assuming a single component scattered burst, we divided the total bandwidth into four subbands and fitted the scattering tail separately in the top two, where there was enough signal to perform the fit. From the difference in scattering timescale between these two subbands, we obtain a scattering index $\alpha=-11.1\pm4.5$. In spite of the large error bars, this index is still inconsistent with scattering by a thin screen or a turbulent medium. We will discuss this further in Section~\ref{sec:discussion_scattering_origin}.

Given the large distance at which this FRB was emitted, its peak frequency is highly redshifted towards lower frequencies. The observed frequencies evolve as $\nu_{\text{obs}}=\nu_0/(1+z)$, which means the peak frequency in the host galaxy frame would have been between 4.2 and 7.1\,GHz for the expected redshift range. As  will be further discussed in Section~\ref{sec:high_f_lum}, and shown in Fig.~\ref{fig:freq_energy}, this is the highest inferred rest frame frequency of a one-off FRB to date.
The implications of such a high DM FRB are also reviewed later, in Section~\ref{sec:discussion_dm}.

\subsubsection{FRB\,20210124A} \label{sec:FRB20210124A}

This burst, with a DM of 869.2\,\pccm, consists of a scattered single component with \tscat$=0.65$\,ms. If we divide the burst into six subbands, we measure a scattering index $\alpha=4.4\pm3.3$, fully consistent with scattering by a turbulent medium or a thin screen. Additionally, the FRB presents intensity modulations with frequency, with a decorrelation bandwidth \scbw$=1.7$\,MHz. These modulations are likely to be a product of scintillation in the MW, since the expected scintillation bandwidth predicted by NE2001 in the direction of the FRB, $\sim1$\,MHz, agrees with  our measurement well within a factor of two.
If we consider the screen producing the scattering to be closer to the production site of the burst and the one producing the scintillation to be within the MW, we can set an upper limit on the distance between the FRB and the first screen to be $\sim600$\,kpc assuming the host galaxy to be at $z=0.9$. The scattering must thus have been produced within the galactic neighbourhood of the FRB host galaxy.

The Stokes data of this FRB were saved, and we carried out observations of a linearly polarised and an unpolarised calibrator (3C286 and 3C147 respectively). In the Stokes data (See Fig.~\ref{fig:stokes}), we observe signal in Stokes $I$ and $Q$, but not in $U$ and $V$. The linear polarisation fraction adds up to $L=86\pm8\%$, the highest in our sample together with FRB\,20191108A \citep[][see Section~\ref{sec:polcal}]{connor_bright_2020}. The resulting circular polarisation fraction is in turn $V=15\pm12\%$, roughly consistent with 0. The PPA remains constant within errors throughout the burst duration.

Although the FRB was localised to an error region as small as 0.89\,arcmin$^2$, we identified 11 galaxies within the error region and redshift upper limit, \zmax$=1.13$. It is thus not possible to identify the most likely host galaxy. The localisation region and host galaxy candidates are displayed in Fig.~\ref{fig:localisation}.

\subsubsection{FRB\,20210127A} \label{sec:FRB20210127A}

This FRB, detected at a DM of 891.7\,\pccm, consists of a single component, 0.83\,ms wide, with no measurable
scattering. The burst extends over the whole observing bandwidth, and no scintillation is visible at Apertif frequencies. Its full-Stokes data were saved, and subsequent observations of a linearly polarised and and unpolarised calibrators (3C286 and 3C147 respectively) were carried out. The burst was detected in SB\,34 while the calibrators were centred in SB\,35, and although these SBs are adjacent, the calibrated data still shows a residual $V$ signal with oscillating intensity that is unlikely to arise from a physical phenomenon (See Fig.~\ref{fig:stokes}). Stokes $Q$ and $U$ display similar oscillations, and we thus apply a phase correction to minimise the $V$ signal. After implementing RM synthesis, we obtain an RM$=123.5\pm0.4$\,\radsqm, as presented in Fig.~\ref{fig:RM_FRB210127}. This RM is in excess of what we expect from the MW contribution, \rmmw$\sim35$\,\radsqm\ (See Table~\ref{tab:stokes}). If the FRB was produced at \zmacquart$\sim0.98$, this would translate to \rmhost$\sim335$\,\radsqm\ in the source reference frame.
The PPA shows a marginally significant decrease of about 5 degrees.

We localised this FRB to a small error region of 0.75\,arcmin$^2$, but we identified eight galaxies with photometric redshifts below the upper limit \zmax$=1.23$, and it is thus not possible to determine the most probable host. See Fig.~\ref{fig:localisation} for the localisation region and candidate hosts.

\subsubsection{FRB\,20210317A}

This burst has a DM of 466.4\,{\pccm} and it is formed by a single component. The burst displays frequency modulations
with a decorrelation bandwidth of $4.8$\,MHz without measurable scattering. The NE2001 model predicts the MW
scintillation bandwidth in the direction of the FRB to be $\sim1.1$\,MHz, and since these agree within an order of magnitude, we attribute the burst modulation to scintillation in the MW. 

The burst triggered a full-Stokes data dump, but since it was detected during the last observation of the observing run,
no calibration observations could be scheduled. The raw Stokes $V$ data show oscillations resembling those produced by
Faraday rotation in $Q$ and $U$, hence we apply the circular polarisation minimisation technique to rotate the phase of
the $V$ signal into $U$. This produces the calibrated data presented in Fig.~\ref{fig:stokes}. After applying RM
synthesis, we determine an RM of $-252.5\pm1.3$\,{\radsqm} (See Fig.~\ref{fig:RM_FRB210317}), and the resulting linear and circular polarisation fractions are $L=50\pm5\%$ and $V=3\pm4\%$. 
Given the expected contribution $\rmmw\sim26$\,\radsqm, the RM at the expected redshift \zmacquart$=0.38^{+0.10}_{-0.22}$ would be around \rmhost$\sim-530$\,\radsqm.
The PPA shows a slow increase of $\sim7$ degrees throughout the burst duration.

\begin{figure}[h!]
    \centering
    \includegraphics[width=\hsize]{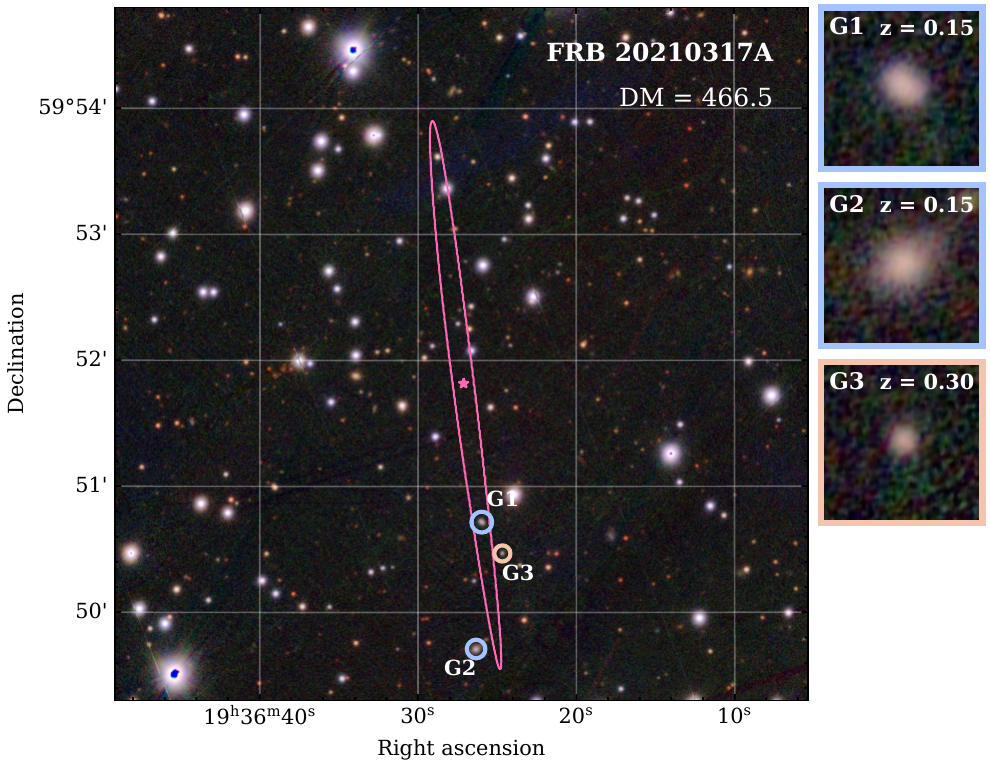}
    \caption{Localisation region (pink contour) of FRB\,20210317A. The three galaxies next to or inside the error region within the redshift limit are indicated by circles and zoomed in on the right.}
    \label{fig:FRB20210317A_loc}
\end{figure}

Since we localised the burst to a relatively small error region of 0.54\,arcmin$^2$ centred at the coordinates 19:36:27.4 +59:51:50.7 in RA and DEC, we searched for host galaxy candidates in and around the error region. We identified one galaxy with photometric redshift $z=0.15\pm0.05$ within the error region, and two additional ones at $z=0.30\pm0.05$ and $z=0.15\pm0.06$ respectively 4.5" and 8.5" away from the error region. Although G1 is the most likely host at 54\%, G2 also has a 35\% probability of being associated to the FRB. Hence, we cannot confidently identify the host of FRB\,20210317A. The localisation region and host galaxy candidates are shown in Fig.~\ref{fig:FRB20210317A_loc}, while the details of the \texttt{PATH} analysis are shown in Table~\ref{app:host_gal}.

%  /Users/user/Documents/projects/ARTS/FRBs/FRB210317/localisation/localisation_frb.ipynb

\subsubsection{FRB\,20210530A} \label{sec:FRB20210530A}

\begin{figure}[h!]
    \centering
    \includegraphics[width=\hsize]{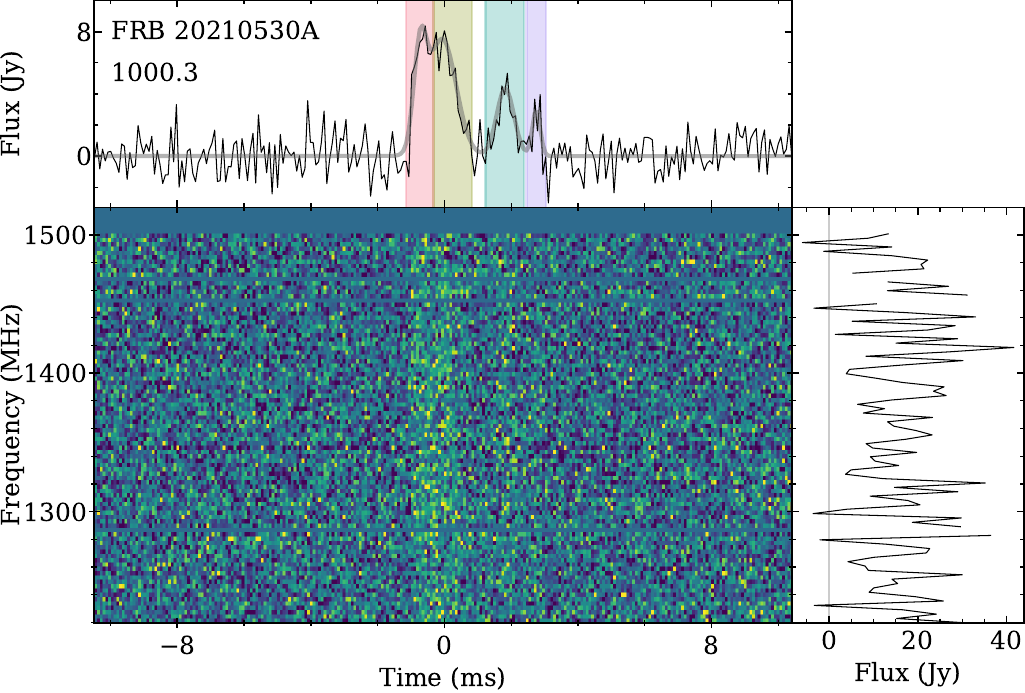}
    \caption{Dynamic spectrum of FRB\,20210530A, displayed as in Fig.~\ref{fig:FRB20200518A}. The spectrum is flat throughout the band.}
    \label{fig:FRB20210530A}
\end{figure}

The pulse profile of this FRB consists of a main, broad component with a flat top followed by two postcursors, and it is well fitted by four Gaussian components, as shown in Fig.~\ref{fig:FRB20210530A}. The first two Gaussians model the profile of the main component, and they have a similar amplitude and a separation of 0.67\,ms. The two postcursors have a separation of 1.89\,ms and 0.93\,ms with respect to their preceding subcomponents each. A timing analysis does not reveal evidence for periodicity.

The burst triggered a full-Stokes data dump, and  calibrators 3C286 and 3C147 were observed. The two main components display an oscillating signal in the calibrated Stokes Q and U, but not in V. After applying RM synthesis, we find an RM$=-125.1\pm4.6$\,\radsqm, as seen in Fig.~\ref{fig:RM_FRB210530}. The resulting polarisation fractions are $L=52\pm4\%$ and $V=0\pm7\%$. The PPA of the two main components appears first to increase by $\sim20$ degrees and then to decrease back to the initial value.

We localised this FRB to an error region of 2.97\,arcmin$^2$. We find 36 galaxies contained within this region and the
Macquart redshift upper limit \zmax$=1.35$, as shown in Fig.~\ref{fig:localisation}, too many to identify the most likely host.

\subsubsection{FRB\,20211024B} \label{sec:FRB20211024B}

\begin{figure}[h!]
    \centering
    \includegraphics[width=\hsize]{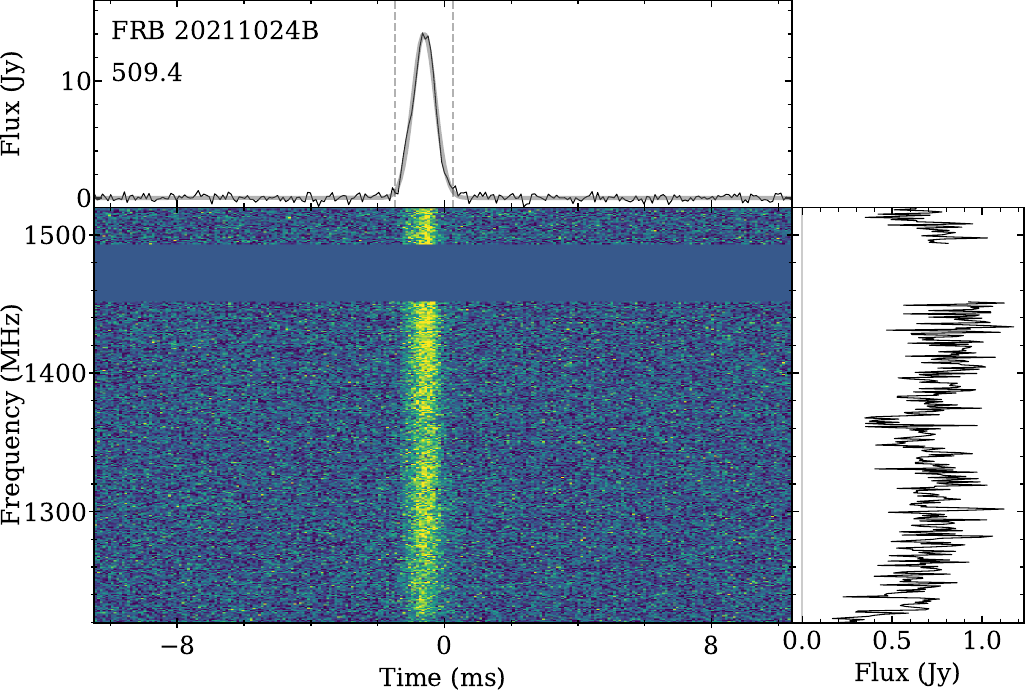}
    \caption{Dynamic spectrum of FRB\,20211024B. The spectrum shows a morphology that cannot be well fitted to a Gaussian nor a power law, but that could be of instrumental origin: the FRB was detected in SB 40 (Table~\ref{tab:frb_table2}), which is composed of 2 TABs (see Sect.~\ref{sec:fstruct}).}
    \label{fig:FRB20211024B}
\end{figure}

This FRB, displayed in Fig.~\ref{fig:FRB20211024B}, consists of a single discernible component with a FWTM of 1.45\,ms. The burst presents a slight asymmetry, with the intensity increasing more slowly than it decreases at later time. This could be an intrinsic property of the burst, or a hint of an unresolved precursor. 
The burst was detected with a DM of 509.4\,\pccm, and after removing the MW and halo contribution, we expect a redshift of \zmacquart$=0.52^{+0.12}_{-0.30}$. 

This FRB has a small localisation region of 0.77\,arcmin$^2$, and within the redshift upper limit \zmax=0.64, we find five host galaxy candidates with photometric redshifts ranging from 0.2 to 0.6. We find the most likely host to be the brightest galaxy, G2, with $P(G_2|x)\sim61\%$, while for the second brightest galaxy we find $P(G_1|x)\sim24\%$. These two host candidates have similar photometric redshifts $z_{\text{phot}}\sim0.24$, on the lower end of what is expected from the Macquart relation for the assumed \dmhost. The details of the host candidates and \texttt{PATH} analysis are presented in Table~\ref{tab:host_galaxies}.

\subsubsection{Bicomponent bursts} \label{sec:bicomponent_frbs}

Three bursts \refbf{exhibit} two components, where the first is brighter than the second. 
Two of the FRBs, FRB\,20190709A and FRB\,20191109A, were originally presented in \cite{van_leeuwen_apertif_2023}. As discussed there, the subcomponents of FRB\,20190709A have a separation of 1.3\,ms, and the amplitude of the first is roughly five times larger than the second. Each has a FWTM of $\sim$0.9\,ms, and no scattering can be resolved. The first component is broadband and shows intensity variations in frequency consistent with the expected scintillation in the MW. The second component is mainly visible at the bottom of the band, coincident in frequency with a bright scintillation `patch' from the first component.

Similarly, the subcomponents of FRB\,20191109A have a separation of 1.2\,ms, the main component has a width of 0.7\,ms and it is $\sim$3.5 times larger in amplitude than the second, with a width of 1.4\,ms. The pulse profile shows a bump about a millisecond after the first component, but its S/N is to low to confidently associate it with a third component. The two components have a similar frequency extent. The emission extends from the top of the band down to 1280\,MHz. There appears to be a gap in emission between 1370 and 1440\,MHz, but we associate it to an instrumental effect (lower sensitivity at those frequencies during the observation) rather than to an intrinsic property of the burst.

FRB\,20200321A is the last burst with two components. The observation where this burst was detected was highly affected by RFI, and nearly half of the observing bandwidth had to be masked. The subcomponent separation is 0.7\,ms, while the widths are 0.9\,ms and 1.3\,ms for the first and the second subcomponents respectively. The subcomponents are thus nearly merged together. From the limited available bandwidth, the two components appear to be narrowband, with a frequency extent of $\sim$230\,MHz at a peak frequency 1435\,MHz, and both subcomponents extending the same range of frequencies.
This FRB triggered the dump of the full-Stokes data, and observations of the linearly polarised calibrator 3C286 were subsequently scheduled. The Stokes Q and U parameters do not display any discernible oscillation where an RM could be estimated, and we measure low polarisation fractions of $L=17\pm5\%$ and $V=13\pm9\%$. The calibrated Stokes parameters are shown in Fig.~\ref{fig:stokes}.

\subsubsection{Low polarisation bursts} \label{sec:unpolarised_frbs}

In this section we include the bursts that triggered the storage of the full-Stokes data which have an average linear polarisation fraction $L<35\%$ and a circular polarisation fraction $V<30\%$. Following the FRB polarisation classification from \cite{sherman_deep_2024}, these bursts would be considered to be unpolarised. Six out of the 16 bursts with Stokes data fall into this category, including FRB\,20200213A presented in Section~\ref{sec:FRB20200213A}, and FRB\,20200321A described in Section~\ref{sec:bicomponent_frbs}. The calibrated Stokes data of these bursts are displayed in Fig.~\ref{fig:stokes}, and the remaining four bursts are described below.

FRB\,20191020B, originally presented in \cite{van_leeuwen_apertif_2023}, was the first FRB of the sample that triggered the storage of the Stokes data. Although no calibrator observations were taken at the time, in the raw data we observe signal in the linear polarisation of $L=31\pm7\%$, and a circular polarisation fraction consistent with 0. There is no sign of $Q$/$U$ oscillations with frequency that could be attributed to Faraday rotation. 

FRB\,20200322A, detected at a DM of 1290.3\,\pccm\ presents a scattering tail with \tscat$=4.2$\,ms.  The burst appears narrowband, with most of the emission observed above 1300\,MHz. By dividing the bandwidth into four subbands, we were able to measure the scattering timescale in the to three subbands and infer a scattering index $\alpha=-4.5\pm2.3$, which is consistent with both sattering in a turbulent medium or by a thin screen. The spectrum can be fitted to a Gaussian peaking at 1406\,MHz, and additionally it shows spectral modulations with a decorrelation bandwidth of $5\pm2$\,MHz, consistent with the expected MW contribution. The combination of scattering and scintillation would place an upper limit between the FRB location and the scattering screen of $\sim300$\,kpc assuming a redshift of \zmacquart$\sim1.46$. 
The Stokes data of the FRB were saved upon its detection, and the $U$/$V$ leakage was calibrated using the linearly polarised source 3C286. The burst appears to be unpolarised, with $L=3\pm6\%$ and $V=14\pm9\%$. The Stokes data are presented in Fig.~\ref{fig:stokes}.

FRB\,20200323C was detected at a DM of 833.4\,\pccm\ and consists of a single component with a scattering timescale of \tscat$=1.3$\,ms. The burst is brighter at the top of the band, and its spectrum can be well fitted by a power law with a spectral index $\Gamma\sim6.3$. By dividing the dynamic spectrum into 8 subbands, we measured the scattering timescale in the top 6, where the burst is bright enough, resulting in a scattering index of $\alpha=-3.5\pm3.0$. 
Its detection triggered a full-Stokes data dump, and 3C286 observations were obtained to calibrate the $U$/$V$ leakage. The burst presents a polarisation fraction consistent with 0, although some residual $Q$/$U$ signal is apparent in Fig.~\ref{fig:stokes}. The signal is however not strong enough to apply the RM synthesis technique.

FRB\,20200516A, with a DM of 361.1\,\pccm, is a single component burst with a temporal width of $\sim2.2$\,ms and no measurable scattering, whose spectrum can be well fitted by a power law with a spectral index $\Gamma\sim+6.9$. Unfortunately, no calibration observations were performed after the detection of this burst. The raw data shows however a low polarisation fraction of $L=17\pm13\%$ and $V=14\pm7\%$. Calibration would have been unlikely to significantly modify this result.

%--------------------------------------------------

\subsection{Localisation} \label{sec:result_localisation}

% Placing figure here to force the position
% /home/ines/Documents/projects/ARTS/scripts/localisation_hips.py
\begin{figure*}
    \centering
    \includegraphics[width=6cm]{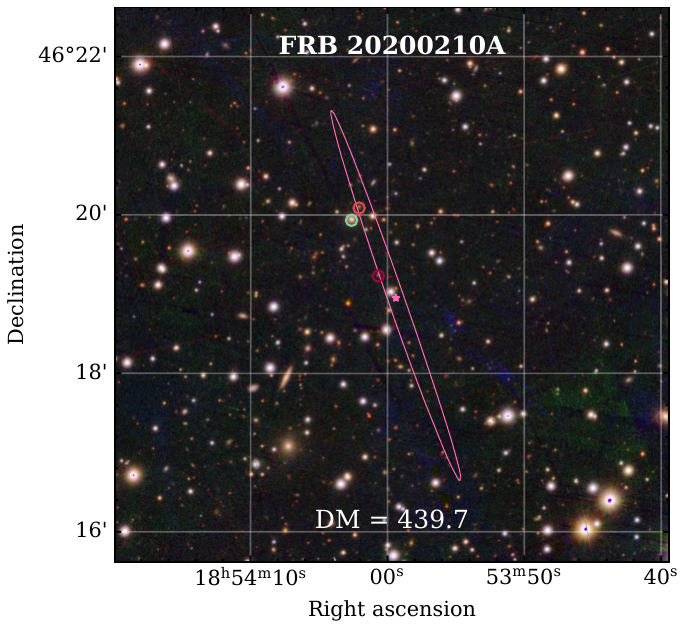}
    \includegraphics[width=6cm]{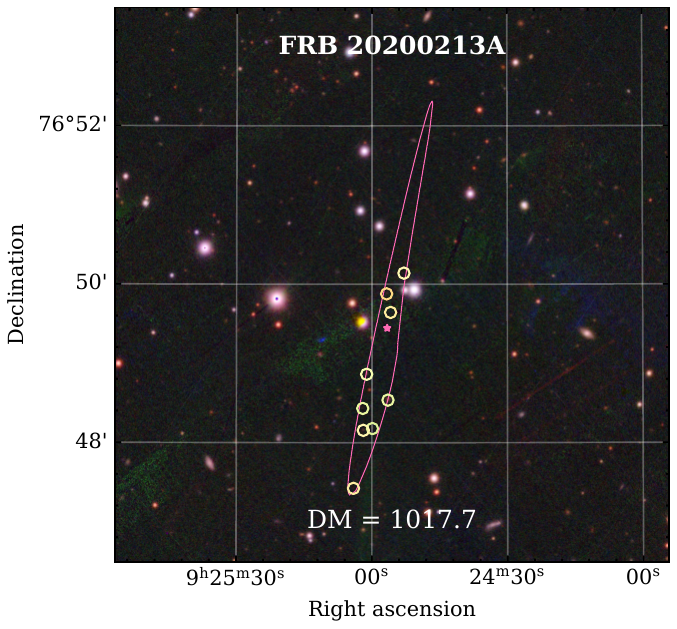}
    \includegraphics[width=6cm]{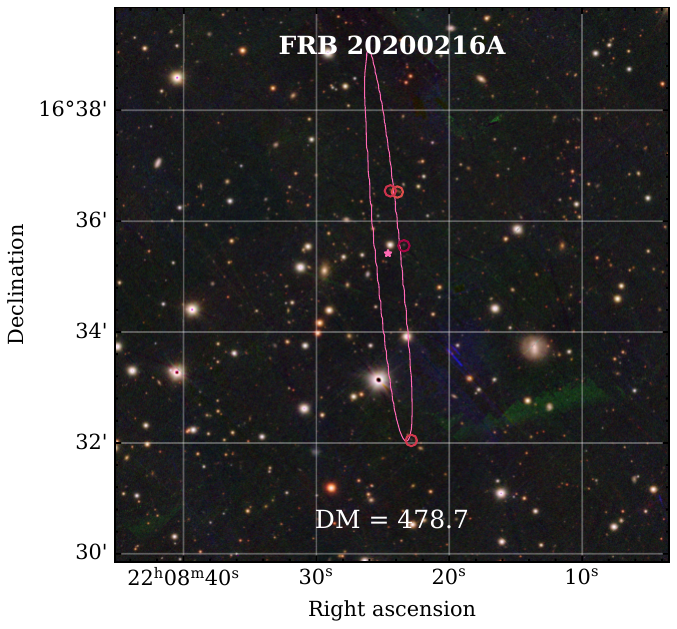}\\
    % FRB200224
    \includegraphics[width=6cm]{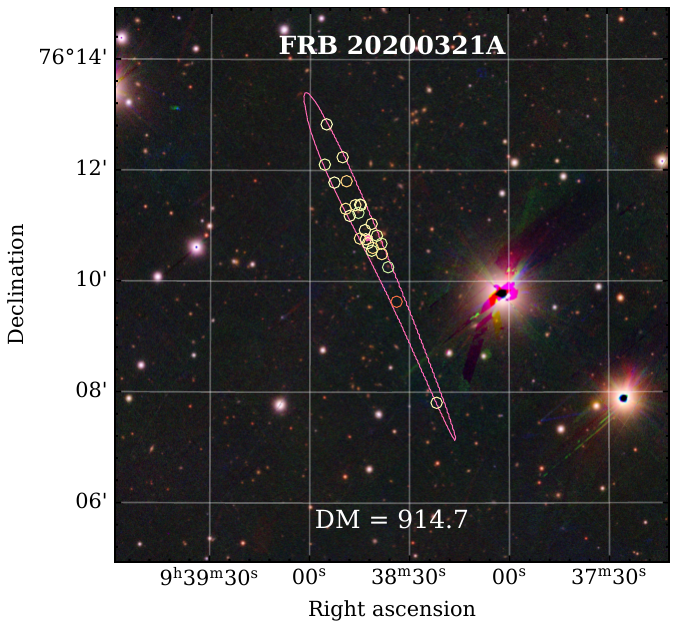}    \includegraphics[width=6cm]{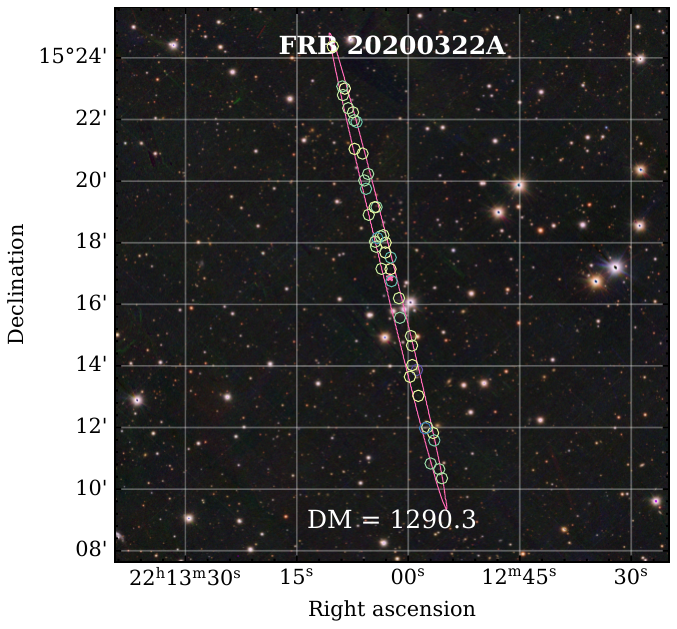}
    \includegraphics[width=6cm]{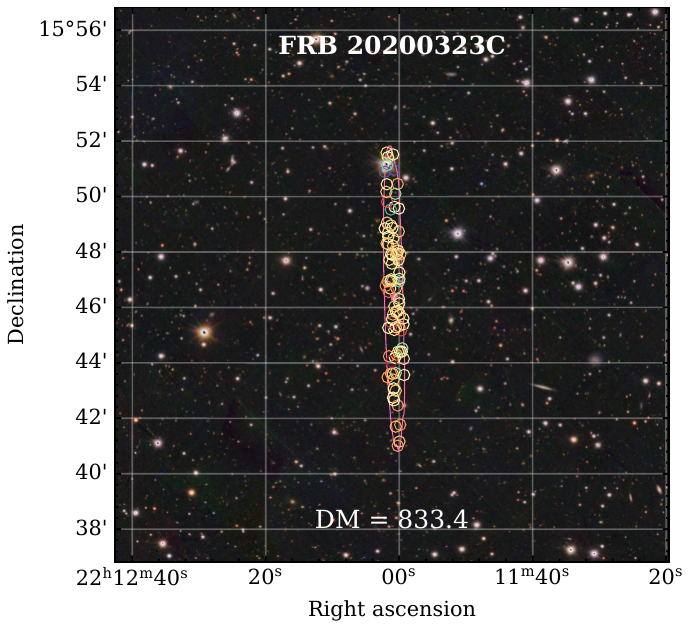}\\ \includegraphics[width=6cm]{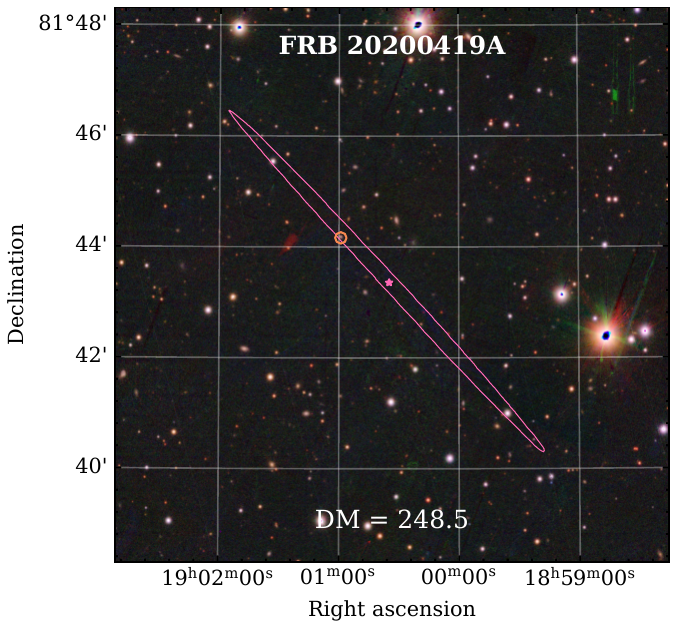}
    % FRB200514\\
    % FRB200516
    \includegraphics[width=6cm]{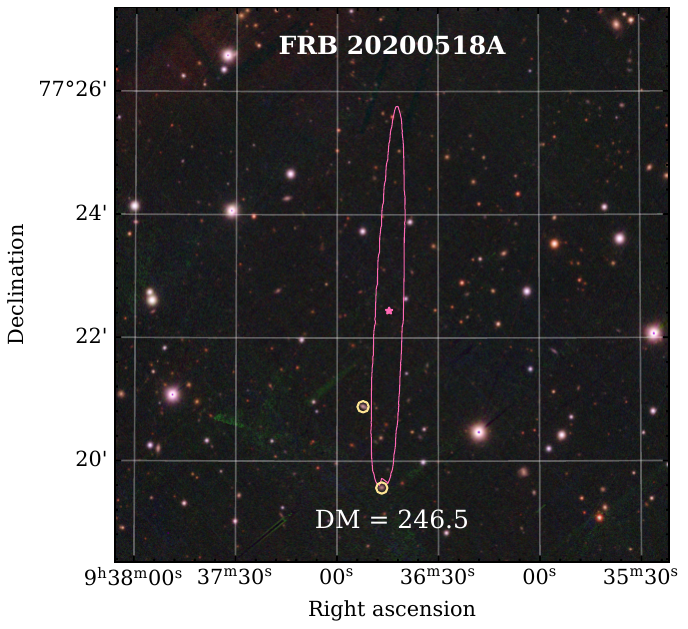}
    % FRB200523\\
    \includegraphics[width=6cm]{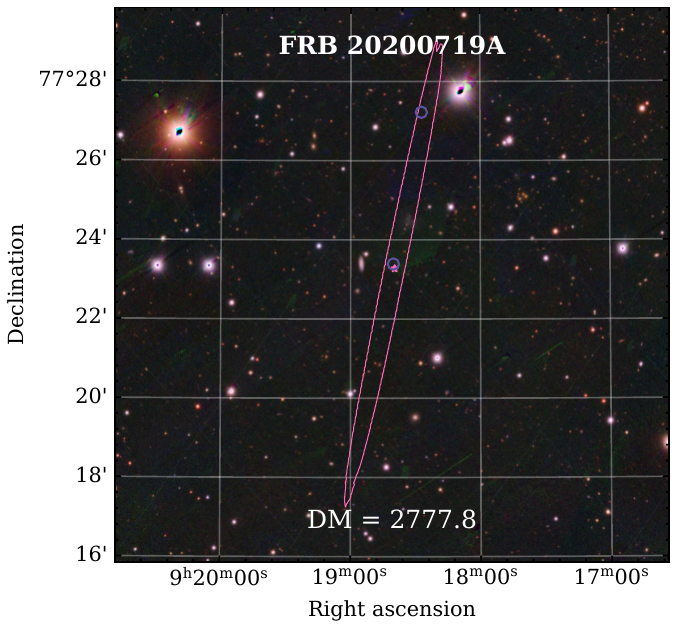}\\
    \includegraphics[width=6cm]{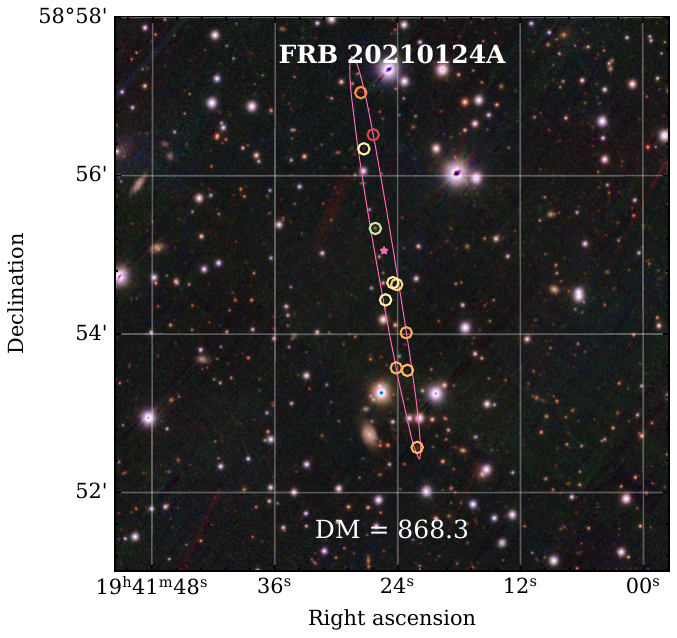}
    \includegraphics[width=6cm]{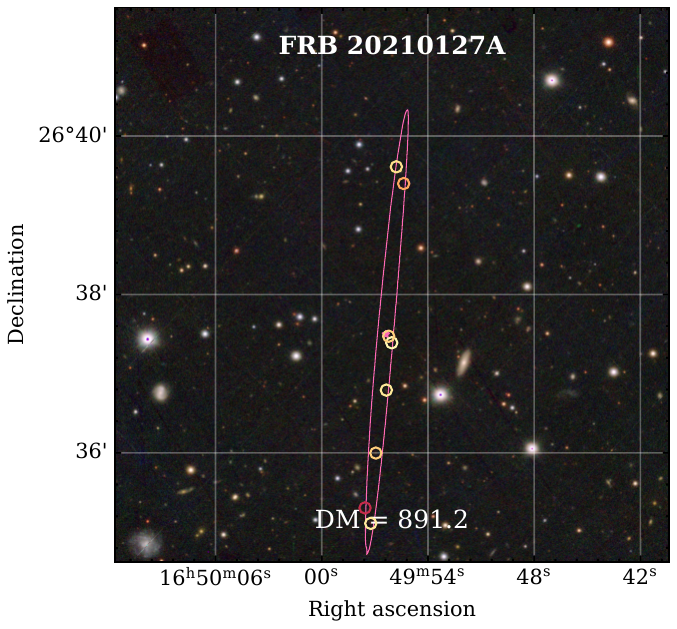}
    \includegraphics[width=6cm]{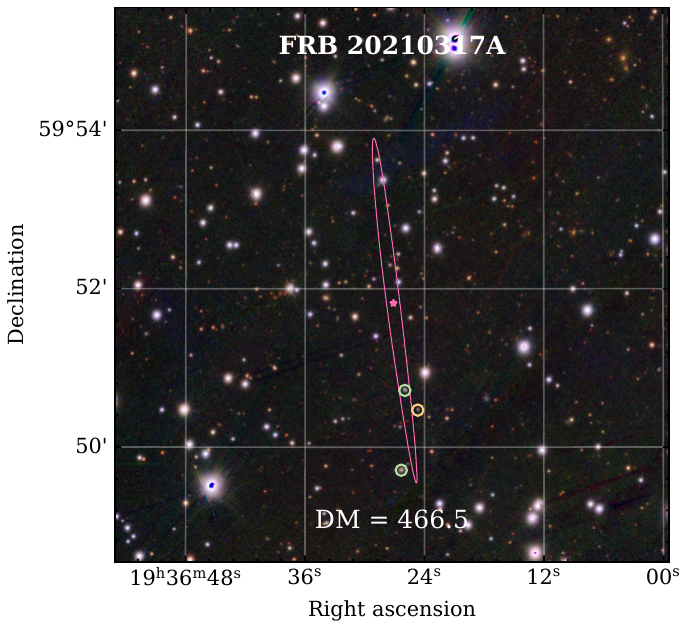}

    % \includegraphics[width=0.32\textwidth]{plots/localisations/FRB20200210A_localisation.pdf}
    % \includegraphics[width=0.32\textwidth]{plots/localisations/FRB20200213A_localisation.pdf}
    % \includegraphics[width=0.32\textwidth]{plots/localisations/FRB20200216A_localisation.pdf}\\
    % % FRB200224
    % \includegraphics[width=0.32\textwidth]{plots/localisations/FRB20200321A_localisation.pdf}    
    % \includegraphics[width=0.32\textwidth]{plots/localisations/FRB20200322A_localisation.pdf}
    % \includegraphics[width=0.32\textwidth]{plots/localisations/FRB20200323C_localisation.pdf}\\ \includegraphics[width=0.32\textwidth]{plots/localisations/FRB20200419A_localisation.pdf}
    % % FRB200514\\
    % % FRB200516
    % \includegraphics[width=0.32\textwidth]{plots/localisations/FRB20200518A_localisation.pdf}
    % % FRB200523\\
    % \includegraphics[width=0.32\textwidth]{plots/localisations/FRB20200719A_localisation.pdf}\\
    % \includegraphics[width=0.32\textwidth]{plots/localisations/FRB20210124A_localisation.pdf}
    % \includegraphics[width=0.32\textwidth]{plots/localisations/FRB20210127A_localisation.pdf}
    % \includegraphics[width=0.32\textwidth]{plots/localisations/FRB20210317A_localisation.pdf}
    
    \caption{Localisation regions of the new Apertif FRBs with an error region <6\,arcmin$^2$. In each subplot, the pink contour represents the 99\% confidence region of the localisation, and the pink star the centroid of the error region. The circles show the PS1-STRM galaxies identified within or close to the error region and redshift range of each FRB, with colours from blue to red as redshift increases. The text on top of each plot gives the TNS identifier of each FRB, and the bottom text the DM in units of \pccm. The background images are from the PanSTARRS DR1 \citep{chambers_pan-starrs1_2019}. In each plot, the grids are spaced by 2\,arcmin in declination.}
    \label{fig:localisation}
\end{figure*}

\addtocounter{figure}{-1}
\begin{figure*}
    \centering
    \includegraphics[width=6cm]{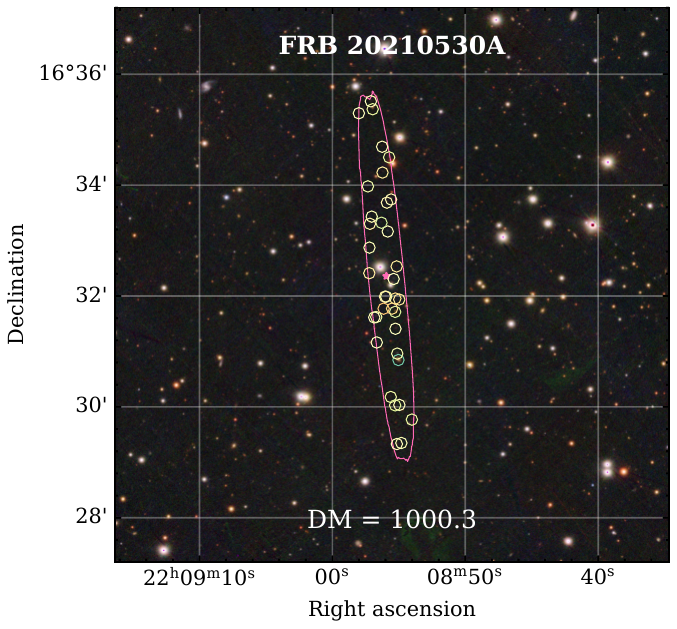}
    \includegraphics[width=6cm]{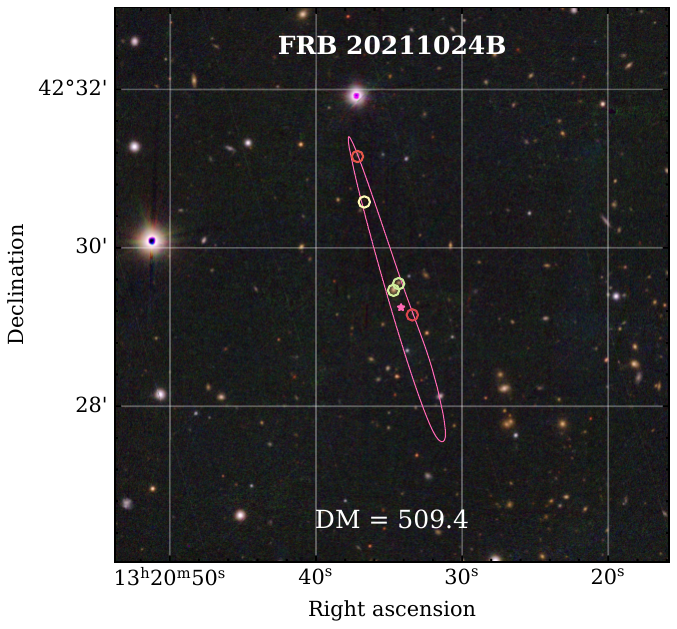}

    \caption{Continued}
\end{figure*}

% /Users/user/Documents/projects/ARTS/scripts/galaxy_density.py
\begin{figure}
    \centering
    \includegraphics[width=\columnwidth]{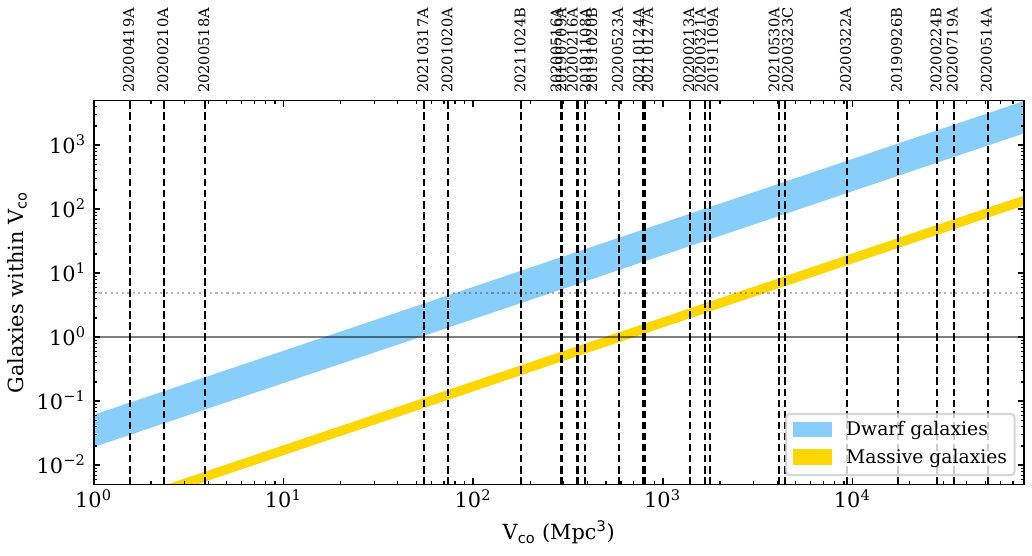}
    \caption{Expected galaxies within the comoving volume of the FRB localisation. The blue and yellow shaded regions give the expected number of dwarf and massive galaxies respectively within the localisation region and redshift upper limits. Each vertical dashed line gives the upper limit on the comoving volume of an FRB, whose TNS identifiers are given on top. The horizontal solid line shows the one galaxy limit, while the dotted line shows the five galaxy limit.}
    \label{fig:exp_galaxies}
\end{figure}

The Apertif localised FRBs have an average error region of $\sim$5\,arcmin$^2$ and a median of
$\sim$2\,arcmin$^2$. Figure~\ref{fig:localisation} displays the 99\% confidence levels on the localisation of the new
Apertif FRBs with a localisation area $<6$\,arcmin$^2$, as well as the galaxies identified within the error
regions. Depending on the mean redshift from the Macquart relation, we estimate the number of dwarf and massive galaxies
expected to be contained in the comoving volume $V_{\text{co}}$ of the localisation region, as shown in
Fig.~\ref{fig:exp_galaxies}. For FRB\,20200210A, we assume the redshift obtained by combining scattering timescale and
DM (See Section~\ref{sec:FRB20200210A}). For the FRBs published in \citet{van_leeuwen_apertif_2023}, we provide updated
error regions after fitting them to an ellipse at the 99\% level.
% Placing figure 4 earlier to force the position on the text

For several of the Apertif FRBs, the expected number of dwarf galaxies within the error region computed as described in Section~\ref{sec:localisation_data_analysis} is $<$5, while the number of expected massive galaxies is $\ll$1, namely FRB\,20200210A, FRB\,20200419A, FRB\,20200518A, FRB\,20210317A, FRB\, and 20201020A.
After searching for known galaxies within the expected redshift limits for the relatively well localised FRBs, we find seven FRBs with $\leq5$ host galaxy candidates, listed in Table~\ref{tab:host_galaxies}. For FRB\,20200419A, we find a single host galaxy candidate with a \texttt{PATH} association probability of $\sim70\%$, while FRB\,20200210A, FRB\,20200518A, FRB\,20210317A, and FRB\,20211024B have host galaxy candidates with association probabilities between 50\% and 60\%.

For searches of associated radio continuum emission at 1.4\,GHz, only FRB\,20190709A, FRB\,20190926B and 
FRB\,20191108A
\citep[][]{van_leeuwen_apertif_2023} lie within the Apertif imaging footprint \citep{adams_first_2022}.
For the latter FRB, the lowest-redshift
one of the set,  \citet{connor_bright_2020} report an imaging non-detection. 

All FRBs were found with Galactic latitudes $|l|>12^{\circ}$ with the exception of FRB\,20200514, with $l=2.48^{\circ}$, which was detected in an observation of the repeater \RIII.

The localisation regions of all Apertif FRBs are reported in Table~\ref{tab:frb_table2}. The regions were fitted to a 2D Gaussian and we provide the ellipse properties that best match the 99\% error region of each FRB, with RA and DEC indicating the ellipse centre, a and b the semi-major and semi-minor axes respectively, and $\theta$ the angle of the ellipse measured from West (lower RA) through North, following the SAOImageDS9\footnote{\texttt{ds9}: \url{https://sites.google.com/cfa.harvard.edu/saoimageds9}} convention.

\subsection{Spectral properties}

About two thirds of the Apertif FRBs are broadband, i.e. they have emission from the bottom to the top of the observing
bandwidth at roughly the same intensity. The spectra of the remaining bursts can be classified in two different
categories. Four have emission at the top of the band, and can be well fitted by a power law with a positive spectral
index $\Gamma$, ranging from 4.4 to 10.6. The remaining five are narrowband and their spectra can be well fitted by a
Gaussian. Although the broadband bursts are likely to have, in reality, a Gaussian or power law spectrum, the fractional
bandwidth of Apertif of $\sim$0.2 and the presence of structure in frequency from scintillation impede a further
characterisation of the wider spectral properties. The burst spectral properties are detailed in Table~\ref{tab:frb_table2}.

%------------------------------------------------------
\section{Discussion}\label{sec:discussion}

Based on the Apertif results presented in the previous Section, we here discuss the implications, along a range of sub
topics. 

\subsection{Propagation effects}

The Apertif FRB sample displays a large variety of propagation effects. In this section we describe the observed
scattering timescales and scintillation bandwidths, and compare them to the FRB samples collected by other instruments
and to the expected Milky Way (MW) contribution \citep[queried from NE2001 and YMW16 at 1370\,MHz using \texttt{pygedm};][]{price_comparison_2021}.

\subsubsection{Scattering and comparison to other surveys} \label{sec:discussion_scattering} \label{sec:scattering}

% % ~/Documents/projects/ARTS/scripts/FRB_analysis/multi-component_analysis.ipynb
% \begin{figure*}[htbp]
%     \centering
%     \includegraphics[width=18cm]{plots/multcomp_analysis.pdf}
%     \caption{Observed scattering and results of the multi-component analysis.
%     \textbf{Left:} Histogram of observed scattering timescales in CHIME/FRB and Apertif FRBs. The dashed orange line represents the fiducial scattering timescale distribution of CHIME/FRB FRBs. The dotted black line shows the joint scattering timescale distribution of Apertif and CHIME/FRB, and the solid black line its fit to a lognormal distribution. The dashed green and orange lines and arrows indicate respectively where Apertif and CHIME/FRB are sensitive to scattering.
%     \textbf{Center:} Joint stacked histogram of observed subcomponent separations. The green histogram represents the Apertif separation distribution normalised by the total number of components in the FRB sample. The orange histogram is the same for CHIME/FRB. The black line shows a fit of the joint histogram to a lognormal distribution.
%     \textbf{Right:} Result from the CHIME/FRB subcomponent separation simulations. The grey histogram shows the simulated fraction of observed multi-component FRBs. The orange vertical region shows the observed fraction of CHIME/FRB multi-component bursts with binomial errors. The green vertical line shows the observed Apertif multi-component fraction.}
%     \label{fig:chime-apertif_tscat}
% \end{figure*}

% ~/Documents/projects/ARTS/scripts/FRB_analysis/multi-component_analysis.ipynb
\begin{figure}[htbp]
    \centering
    \includegraphics[width=\hsize]{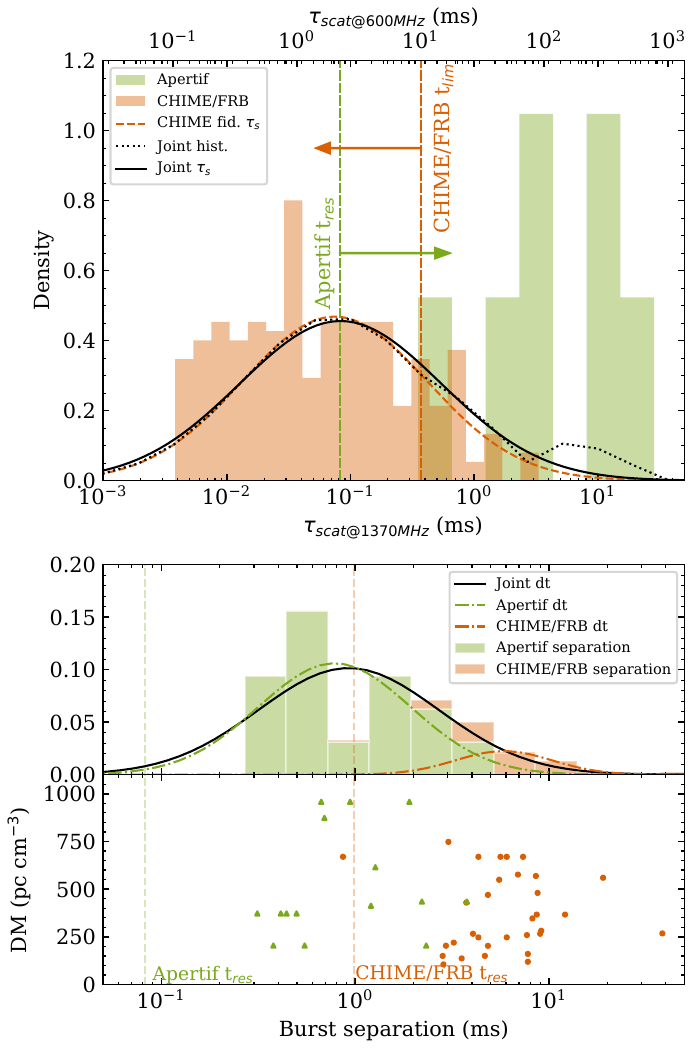}
    \caption{Scattering and subcomponent separation distributions of Apertif and CHIME/FRB.
    \textbf{Top:} Histogram of observed scattering timescales in CHIME/FRB and Apertif FRBs. The dashed orange line
    represents the fiducial scattering timescale distribution of CHIME/FRB bursts. The dotted black line shows the joint
    scattering timescale distribution of Apertif and CHIME/FRB, and the solid black line its fit to a lognormal
    distribution. The dashed green and orange lines and arrows demarcate Apertif and CHIME/FRB sensitivity to
    scattering. The lower and upper horizontal axes show the scattering timescales scaled to the respective band centers
    of Apertif band (1370\,MHz) and  CHIME/FRB (600\,MHz). 
    \textbf{Bottom:} Joint stacked histogram of observed subcomponent separations. The green histogram represents the Apertif separation distribution normalised by the total number of components in the FRB sample. The orange histogram is the same for CHIME/FRB. The black line shows a fit of the joint histogram to a lognormal distribution.}
    \label{fig:chime-apertif_tscat}
\end{figure}

\begin{figure}
    \centering
    \includegraphics[width=\hsize]{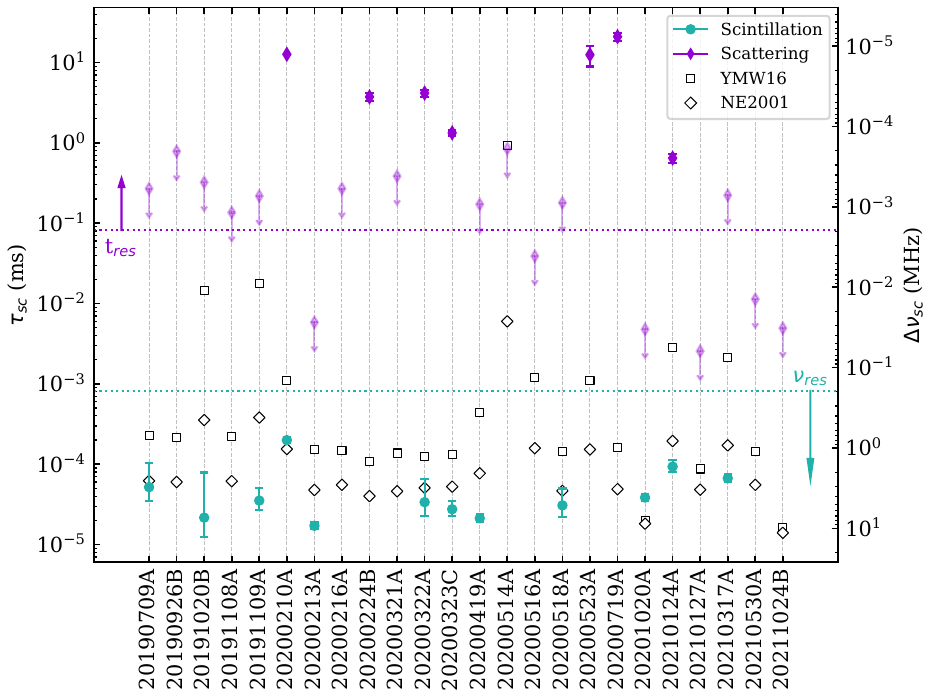}
    \caption{Scattering timescale (left axis) and scintillation bandwidth (right axis, conversion through Eq.~\ref{eq:scint_to_scat}) of Apertif FRBs. 
      % Each vertical line corresponds to a different FRB, arranged by detection date.
%      The left and right axes marks the scattering timescale and corresponding scintillation bandwidth at 1370\,MHz re
      %      using the conversion given by Eq.~\ref{eq:scint_to_scat}.
      Purple diamonds with error bars give the measurable
      scattering timescales, and purple diamonds with arrows the scattering timescale upper limits. Cyan circles with
      error bars give the measured scintillation bandwidth when measurable. The white squares and diamonds give
      respectively the expected Milky Way contribution to scattering from the YMW16 and NE2001 models. All
      values for 1370\,MHz. The horizontal dotted purple and cyan lines mark  the Apertif time and frequency resolution.}
    \label{fig:scintillation}
\end{figure}

Seven of the 24 detected FRBs have measurable scattering, with \tscat\ values ranging between 0.6\,ms and 21\,ms at
%the central frequency of
1370\,MHz (Fig.~\ref{fig:chime-apertif_tscat}, top, in green). 
These timescales are two to four orders of magnitude higher than the MW contribution predicted by NE2001
and YMW16, as visible in Fig.~\ref{fig:scintillation}. 
The scattering is thus likely produced in the host galaxy or the local environment of the FRB, or alternatively in an intervening galaxy in the propagation path of the burst.

Before comparing these values to  surveys at other frequencies,
we note that  % a radio wave propagating through a
\tscat\ evolves with frequency as
\tscat\,${\propto\nu^{-\alpha}}$ with $\alpha$ the scattering index.
% The choice of $\alpha$ is important when comparing FRB surveys observing at different frequencies.
Theoretically, $\alpha=4$ for a simple thin screen model and $\alpha=4.4$ for  propagation  through a turbulent medium
\refbf{with a Kolmogorov spectrum}.
For pulsars, the average $\alpha\sim3.86$  \citep{bhat_multifrequency_2004}.
Estimates of FRB scattering indices, while still scarce, are close to $\alpha=4$ on average, but  compatible with
$\alpha=4.4$ \citep[e.g.][]{thornton_population_2013, ravi_fast_2015, masui_dense_2015, qiu_population_2020}.
We will use a scattering index $\alpha=4$, compatible both with Galactic pulsars and FRB observations.

% cite Cordes & Rickett 1998 for conversion scintillation <-> scattering
% python /home/ines/Documents/projects/ARTS/scripts/FRB_analysis/propagation_properties_distribution.py
% scripts/FRB_analysis/propagation_properties_arts.ipynb

CHIME/FRB, due to its detection biases %quantified with an injection pipeline
\citep[see][]{chimefrb_collaboration_first_2021}, is incomplete for \tscat>10\,ms at 600\,MHz.
This
corresponds to about $>$0.37\,ms at 1370\,MHz. The most scattered CHIME FRBs %detected by CHIME/FRB
thus slightly overlap with the low end of the Apertif sensitivity to scattering. 
This reveals that Apertif can detect a population of highly scattered bursts which are unlikely to be detected by
CHIME/FRB. % due to the scattering reduced S/N at those lower frequencies.
%On the other hand,
Apertif, conversely, cannot resolve scattering timescales below the instrumental broadening, which
%are more likely to be resolved by
CHIME/FRB might.

% In order to roughly estimate how the scattered Apertif bursts modify the
To investigate the intrinsic FRB scattering distribution, we build a joint Apertif-CHIME/FRB scattering distribution.
We add the Apertif scattering histogram, normalised by number of bursts \refbf{and scaled to the same frequency}, to the
CHIME/FRB %fiducial
scattering model \citep{chimefrb_collaboration_first_2021}, and next it by a lognormal model. This method is justified by the small overlap between the different surveys sensitivities to scattering, and it effectively skews the distribution towards higher scattering values. The equation for the lognormal distribution \refbf{measured per unit logarithm of the variable $x$} is given by:
\begin{equation} \label{eq:lognormal}
        p(x) = \frac{m}{\sigma\sqrt{2\pi}} \exp\left(-\frac{\ln{(x/m)^2}}{2\sigma^2}\right),
\end{equation}
where the shape $\sigma$ is a frequency-independent value and $m$ is the frequency-dependent scale in ms. For the joint distribution, we find $\sigma=1.86\pm0.07$ and $m=0.081\pm0.006$\,ms at 1370\,MHz or $m=2.2\pm0.2$\,ms at 600\,MHz (same $\sigma$).
This resulting distribution is a rough estimate; \refbf{we acknowledge that there is a redshift-dependent correction if the scattering originates in the host galaxy. However, since the Apertif and CHIME/FRB redshifts are not known,} correcting for the detailed observational biases and determining the intrinsic FRB scattering distribution is out of the scope of this paper.
%
% This allows us to compare the scattering timescales reported by several surveys observing at different frequencies.
%
% The Apertif FRBs represent a sample of highly scattered bursts.
% The average scattering timescale at 1370\,MHz in the nine FRBs where scattering is resolved is $\sim6$\,ms. 
Overall, we conclude from this joint distribution that
%The CHIME/FRB scattering distribution is the most extensive to date, but it is incomplete above \tscat>10\,ms at 600\,MHz \citep{chimefrb_collaboration_first_2021}. This corresponds to about 0.37\,ms at 1370\,MHz. There is thus a slight overlap between the high end of CHIME and the low end of the Apertif scattering distributions. 
%This means the Apertif FRBs  represent a population of highly scattered bursts. Even though we have not performed a
%burst injection procedure to estimate our biases against detecting FRBs with different properties, the large fraction
%($\sim$29\%) of Apertif FRBs with measurable scattering demonstrate the existence of a
a large fraction of highly scattered FRBs exists in the real, underlying population.

% multicomponent_analysis.ipynb
\begin{figure}
    \centering
    \includegraphics[width=\hsize]{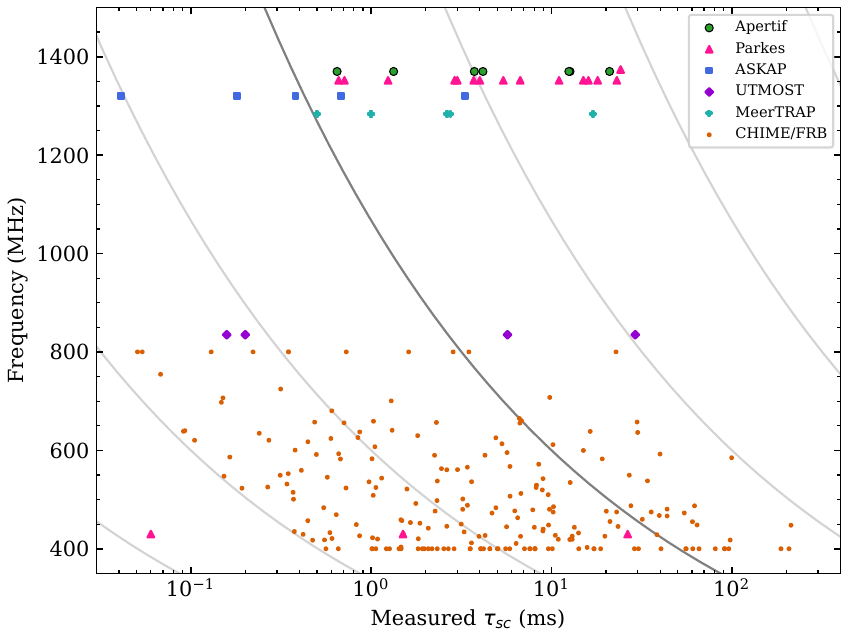}
    \caption{Observed scattering timescales versus  frequency. Green circles represent
      this work, other markers are per the legend.  The values are queried from the TNS database and the First CHIME/FRB Catalog. Grey lines are a reference for the \tscat$\propto\nu^{-4}$ relation. The darkest grey line corresponds to the CHIME/FRB scattering sensitivity limit of 10\,ms at 600\,MHz. Apertif, Parkes, ASKAP, UTMOST and MeerTRAP \tscat\ are shown at their respective central observing frequencies, while the CHIME/FRB \tscat\ are shown at the peak frequency of the burst.}
    \label{fig:scat_multfreq}
\end{figure}

This finding %highly scattered population
is further supported by %the results of
other GHz surveys. % above 1\,GHz.
This is visible when comparing scattering timescales against observing  frequencies, as in  Fig.~\ref{fig:scat_multfreq}.
%, where the scattering timescales measured at different frequencies by different surveys are shown.
Over half of the ASKAP bursts with measurable scattering \citep{day_high_2020, qiu_population_2020}
 exceed the CHIME/FRB  scattering sensitivity limit, once scaled by frequency.
Their voltage capture mode has enabled ASKAP to measure the exponential decay of two FRBs
with a lower scattering timescale than any of the Apertif FRBs, down to \tscat=0.041\,ms for FRB\,20190102C.
The most scattered ASKAP FRB is FRB\,20180130A with \tscat=5.95\,ms, still well less scattered than Apertif
FRB\,20200719A (See Section~\ref{sec:FRB20200719A}). Overall the ASKAP FRBs are less scattered too, with an average
\tscat$\sim2$\,ms versus $\sim6$\,ms for ALERT.

The Parkes FRBs with measurable scattering all fall above the CHIME/FRB sensitivity limit
\citep[e.g.][]{lorimer_bright_2007, petroff_real-time_2015,bhandari_survey_2018, oslowski_commensal_2019}.
Parkes FRBs show marginally larger scattering than Apertif, with an average \tscat$\sim9$\,ms.

Although UTMOST observes at 843\,MHz, considerably lower than Apertif, the measured scattering timescales in four FRBs
do not significantly differ from Apertif and Parkes \citep{farah_five_2019}. This \refbf{is probably explained by the
  UTMOST real-time trigger voltage data and its time resolution of
  $\sim10\,\mu$s}, but it is remarkable nonetheless.
%to note the similarities between UTMOST, Parkes and Apertif both in the measured dispersion measures and scattering timescales.
Once accounted for frequency, though, the UTMOST FRBs represent less scattered bursts than Apertif and Parkes, but similar to ASKAP.
\refbf{We note that we have included no scattering upper limits in this analysis. While the First CHIME/FRB Catalogue reports these upper limits, the TNS does not record such values. A more detailed comparison between the different samples is out of the scope of this paper.} 

\subsubsection{Correlation between dispersion and scattering}

% /Users/user/Documents/projects/ARTS/scripts/FRB_analysis/propagation_properties_distribution.py
% scripts/FRB_analysis/propagation_properties_arts.ipynb
\begin{figure}
    \centering
    \includegraphics[width=\hsize]{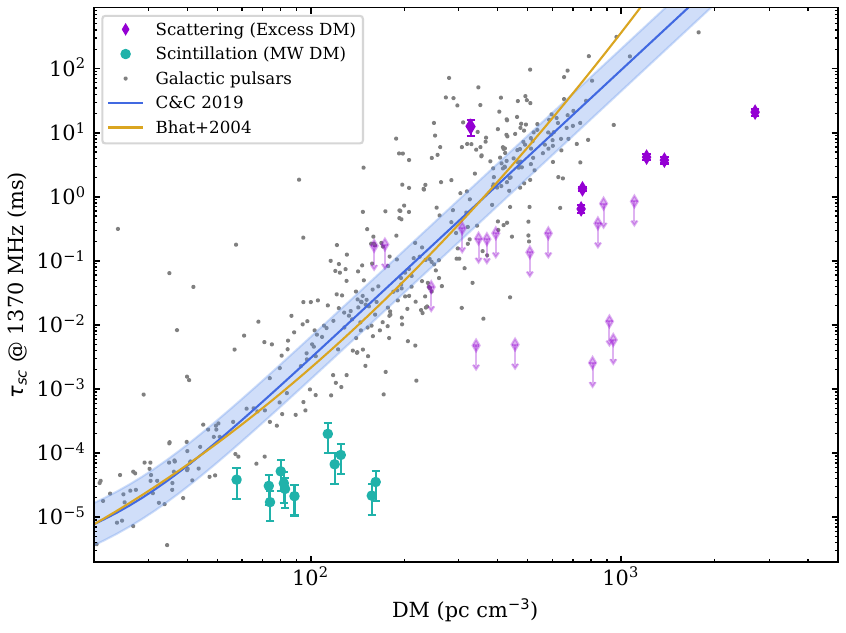}
    \caption{Scattering as a function of DM of Apertif FRBs compared to Galactic pulsars. The measured Apertif \tscat\ as a function of extragalactic DM is shown as purple diamonds (higher transparency for upper limits). The cyan circles correspond to the measured scintillation converted to scattering as a function of the expected Galactic DM. The grey dots are the known pulsar \tscat\ at 1370\,MHz as a function of measured DM.
      Two \tscat--DM relations are shown, from \citet[][yellow line]{bhat_multifrequency_2004}
      and \citet[][blue line with shaded region]{cordes_fast_2019}.
    }
    \label{fig:scat_dm_corr}
\end{figure}

Since the DM and  \tscat\  of Galactic pulsars are correlated,
 we investigated if a similar correlation exists in the Apertif FRBs. Figure~\ref{fig:scat_dm_corr} shows these FRB
 scattering timescales %, including upper limits,
 as a function of excess DM, since we expect the origin of scattering to be \refbf{extra-Galactic}.
 Also plotted for comparison are the pulsar data, and
 %measured DMs and scattering timescales of Galactic pulsars are further plotted for comparison, as well as the
 the \tscat--DM relations from \cite{cordes_fast_2019} and \cite{bhat_multifrequency_2004}. We also plot the scintillation bandwidth of the FRBs converted to scattering timescales as a function of the expected DM$_{\text{MW}}$, since the measured scintillation bandwidths match the expected Galactic contribution.

As the scattering timescales of the FRBs in the top-right of Fig.~\ref{fig:scat_dm_corr} appear to increase with DM, %and thus we carried out further correlation analyses.
%
%In order to
we determined the correlation coefficient between the excess DM and the Apertif \tscat\ values and upper limits.
%of the Apertif FRBs (the purple diamonds in Fig.~\ref{fig:scat_dm_corr}), w
We compute the Kendall  correlation coefficient $\tau$ using the \texttt{cenken} function of the CRAN \texttt{NADA} package\footnote{CRAN NADA package: \url{https://cran.r-project.org/web/packages/NADA/index.html}}, following \citet[][Chapter 10.8.3 and references therein]{feigelson_modern_2012}. This coefficient is a non-parametric correlation test, robust on small sample sizes with censored data \citep{helsel_nondetects_2004, oakes_concordance_1982}, and thus applicable to our case. The function also computes a p-value whose null hypothesis is the absence of correlation.
We find only a weak correlation with  $\tau=0.17$, and a p-value of $0.24$, exceeding the conventional 0.01,
indicating no evidence for a correlation between the excess DM and \tscat\ of Apertif FRBs.

This absence %lack of evidence for
of correlation
is in agreement with previous FRB observations \citep{qiu_population_2020, petroff_fast_2019, cordes_fast_2019}, and supports earlier claims that the IGM does not significantly contribute to scattering \citep{cordes_radio_2016, xu_origin_2016, zhu_dispersion_2021}.

\subsubsection{Origin of scattering} \label{sec:discussion_scattering_origin}

% /users/user/Documents/projects/ARTS/scripts/simulate_scattering.py
\begin{figure}
    \centering
    \includegraphics[width=\hsize]{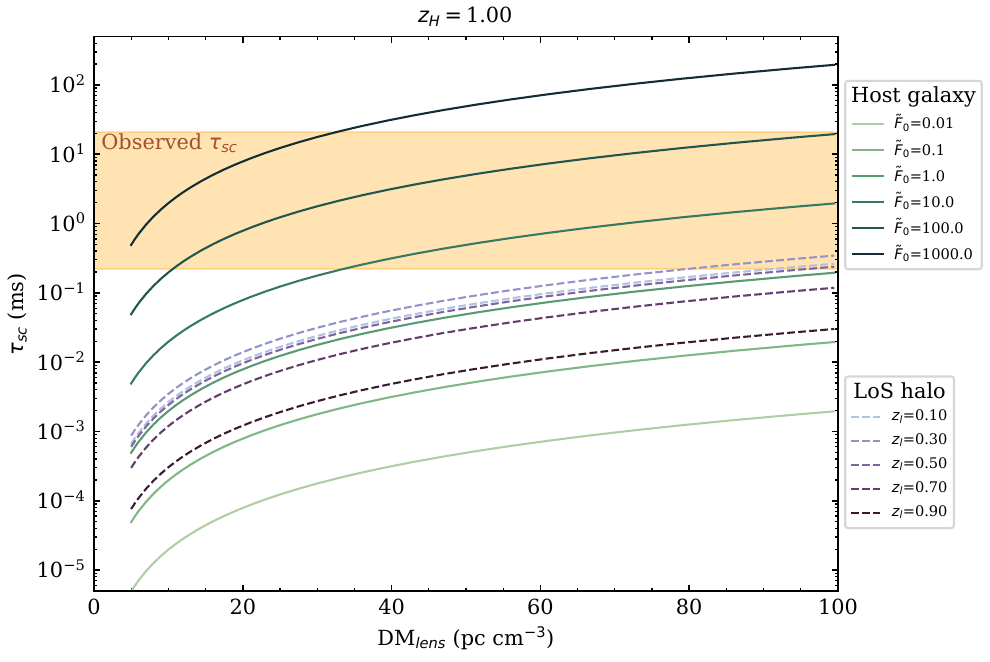}
    \caption{Expected contribution to scattering at 1370\,MHz for different lens DMs from an FRB host galaxy at $z_H=1$ and an intervening galaxy halo within the line of sight. The solid lines show the expected host galaxy contribution for different electron density variations and a geometric boost factor $G_{\text{sc}}=1$. The dashed lines show the expected contribution from an intervening galaxy halo with a thickness $L=30$\,kpc and $\tilde{F}_0=10^{-4}$\pckm\ located at different redshifts. The orange shaded region represents the observed Apertif \tscat\ range. }
    \label{fig:sim_scat}
\end{figure}

Although only FRBs presenting both scattering and scintillation allow us to set upper limits on the distance between the source and the scattering screen, we can determine that the scattering of Apertif FRBs, when present, is much more likely to have been produced at the host galaxy and not in the halo of an intervening galaxy within the LoS.
In their Eq.~2, \citet[]{ocker_constraining_2021} determine a relationship between scattering, the electron density fluctuations of the medium, and a geometric factor that depends on the distances between FRB, scattering medium and observer:
\begin{equation}
    \tau_{\text{sc}} (\text{DM}, \nu, z) \simeq 48.03\,\text{ns} \times \dfrac{A_{\tau}\tilde{F}(z_l)\text{DM}_l^2}{(1+z_l)^3\nu^4} \left[ \dfrac{2d_{\text{sl}}d_{\text{lo}}}{Ld_{\text{so}}} \right], 
\end{equation}
where $A_{\tau}\lesssim1$ is a scaling factor to convert the mean delay to the $1/e$ delay obtained from the pulse shape fit, $\tilde{F}(z)$ in \pckm\ quantifies the electron density variations of the scattering lens, DM$_{l}$ is the DM contribution from the scattering lens, $z_l$ is the redshift of the scattering lens and $\nu$ is the observing frequency in GHz. $d_{\text{sl}}$, $d_{\text{lo}}$, and $d_{\text{so}}$ are angular diameter distances in Gpc, with $d_{\text{sl}}$ the source to lens distance, $d_{\text{lo}}$ lens to observer, $d_{\text{so}}$ source to observer. $L$ is the thickness of the layer in Gpc, and the geometric boost factor $G_{\text{sc}} \simeq d_{\text{sl}}d_{\text{lo}}/Ld_{\text{so}}$. Note that, since all distances are measured in Gpc, $\tau_{\text{sc}}$ is measured in ns. When either the source or the observer are embedded in the scattering medium and the source to observer distance $d$ is much larger than the medium's thickness $L$, $G_{\text{sc}}=1$.
The electron density fluctuations vary with redshift following the cosmic star formation rate (CSFR) as follows \citep[Eq.~22 in][]{ocker_radio_2022}:

\begin{equation}
    \tilde{F}(z) \simeq \tilde{F}_0 \times \dfrac{(1+z)^{2.7}}{1+1[(1+z)/2.9]^{5.6}}.    
\end{equation}

If the scattering lens is located in the host galaxy ($z_l=z_{\text{host}}$), $\tilde{F}_0$ can vary from $10^{-2}$ to $10^3$\,\pckm\, by extrapolation from observations of MW pulsars in the Galactic 
plane. In this case, the FRB would be embedded in the scattering medium and $d_\text{so}$ is much larger than the medium's thickness $L$, thus $G_{\text{sc}}=1$.
If, on the other hand,  the scattering lens is a galactic halo falling within the FRB LoS, the electron density
fluctuations are much lower. Galactic pulsars located in the Milky Way thick disk (at 10--20\,kpc distances) exhibit an $\tilde{F}_0\sim10^{-3}$\,\pckm. An FRB traversing an intervening galactic halo would encounter a much more homogeneous medium, with $\tilde{F}_0\sim10^{-4}-10^{-5}$\,\pckm. Although the geometric boost factor is much larger for an intervening galaxy halo than for the medium where the source or the observer are embedded, the turbulence of the intervening halo is very small unless the FRB passes with a small impact parameter with respect to the galaxy centre.

In Fig.~\ref{fig:sim_scat}, we compare the expected \tscat\ produced by the host galaxy medium and by an intervening
galactic halo as a function of the lens DM contribution. We compute these values at 1.37\,GHz for an FRB located at
$z=1$, the average redshift upper limit of scattered Apertif FRBs. For the host galaxy, we test a range of electron
density fluctuations from $10^{-2}$ to $10^3$. For a scattering lens within the LoS of the FRB with a thickness of
$L=30$\,kpc, we assume $\tilde{F}_0=10^{-4}$\,\pckm\ and test the \tscat\ contribution at different distances varying
from 10\% to 90\% of the host galaxy redshift.

We find the observed Apertif scattering timescales to be much more easily produced in the host galaxy  than by an
intervening galaxy halo located in the burst LoS. This is in agreement with \citet{cordes_redshift_2022}, and it is further supported by the lack of observable scattering in other FRBs which are known to travel through the halos of foreground galaxies. This is the case for FRB\,20191108A, which passes through the halos of M33 and M31  with an impact parameter of 18\,kpc from M33 \citep{connor_bright_2020} and FRB\,20190709A passing $\gtrsim25$\,kpc away from the M33 centre \citep{van_leeuwen_apertif_2023}, as well as the localised FRB\,20181112A \citep{prochaska_low_2019, cho_spectropolarimetric_2020}, and FRB\,20190608B \citep{simha_disentangling_2020}.

By modelling the dispersion and scattering produced throughout the travel path of an FRB, \cite{chawla_modeling_2022} find that circumburst environments with strong scattering properties are required in order to reproduce the FRBs from the first CHIME/FRB Catalog with \tscat>10\,ms at 600\,MHz. This corresponds to $\sim$0.37\,ms at 1370\,MHz, which roughly matches the lowest measured Apertif scattering timescale. Although \cite{chawla_modeling_2022} suggest intervening galaxies within the burst LoS as an alternative explanation, we have determined above that this scenario is more unlikely given the low fluctuation parameter observed in the MW halo \citep{ocker_constraining_2021}. 
A very low impact parameter with respect to the centre of the intervening galaxy ($\lesssim10$\,kpc for a MW-like galaxy, or roughly $\lesssim15$\% of the virial radius as assumed in Section~\ref{sec:FRB20200719A}) would be required in order to produce significant scattering.

A correlation between \tscat\ and DM might have been an indication of a significant contribution to scattering from the
IGM or intervening host galaxies. Meanwhile, scattering in the host galaxy would be highly dependent on the type of
galaxy and its inclination, hence no correlation would be expected.
% \jvl{perhaps take that out; and replace by a paragraph on  the tau $\sim$ DM_h$^2$ discussion from Jim and slack, once we have extended Table A.2} 
The lack of a significant \tscat--DM correlation supports an origin of scattering in the FRB host galaxies.

Beyond the timescales discussed above,
valuable insights on the intervening media for each Apertif FRB are also provided by the
individual scattering indices $\alpha$.
%can provide valuable insights into the properties of the media the FRBs have travelled through.
%In the most simple scenario, scattering produced by a thin screen would result in a scattering index of $\alpha=4$,
%while propagation through a uniform Kolmogorov turbulent medium would produce a scattering index of $\alpha=4.4$.
Of the seven Apertif FRBs with measurable scattering timescales and indices, four are compatible with $\alpha=4$
(for a thin screen scenario), while three are compatible with $\alpha=4.4$ (for a uniform Kolmogorov turbulent medium).
Three bursts present larger scattering indices, incompatible with both %a thin screen and a Kolmogorov turbulence
models (Fig.~\ref{fig:scat_index}). FRB\,20200523A and FRB\,20200719A have low S/N and the scattering timescale can only
be measured in two subbands, so we decide not to draw conclusions from them. However, FRB\,20200210A was bright enough
to divide the bandwidth into 16 subbands and measure the scattering in the top six ones, obtaining a scattering
index of $\alpha=13.8\pm0.9$. \refbf{We obtain consistent indices when measuring the scattering timescale in the top five bands out of eight, or when testing different S/N thresholds, and our measurement is thus robust.} %{\bf ADDRESS REF COMMENT HERE?}
The highest scattering index ever reported in pulsars is $\alpha=9.8$ for B1834--04 \citep{lewandowski_pulse_2013},
although the authors caution %consider this measurement doubtful since
the scattering was only measured at two frequencies.
Scattering index measurements in FRBs %, on the other hand,
are still scarce.
\citet{sammons_two-screen_2023} measure a steep scattering index $\alpha=7.3\pm0.9$ within a narrow observing bandwidth for the repeater FRB\,20201124A, but an index of $\alpha\sim4$ is required to make their scattering measurement consistent with previous observations.
The simplicity of the models that predict scattering indices of $\alpha=4$ or $\alpha=4.4$ is unlikely to apply to more complicated geometries that could arise, for instance, in a clumpy medium, or throughout the inhomogeneities that FRBs are likely to encounter along their travel paths. \citet{walker_extreme_2017} showed that molecular clumps around hot stars \refbf{could potentially} be responsible for extreme radio scintillation events of background radio sources. In
the case of FRB\,20200210A, such a configuration % around the source
could explain both the large scattering timescale and the steep scattering index. 
\refbf{Alternatively, the use of a different kernel modelling a thick screen might produce a scattering index compatible with the theory and other radio pulse observations. The pulse might instead intrinsically broaden at lower frequencies.}
Further study is clearly required to understand the origin of such anomalous scattering profiles.

We thus conclude the scattered Apertif FRBs are likely to be embedded
in extreme environments. These could be star-forming regions or
supernova remnants, as  previously suggested in e.g. \citet{masui_dense_2015, connor_non-cosmological_2016} and  \citet{xu_origin_2016}.

\subsubsection{Dispersion} \label{sec:discussion_dm} \label{sec:result_dm}

% Changing plot position for better placement

In our sample, DMs range from 246 to 2778~\pccm, with a  $\sim$800\,\pccm\ average  and
 $\sim$625\,\pccm\ median. The expected MW and halo contribution in the sample varies between 70 and 300\,\pccm. We
compare the Apertif DMs to other instruments listed in the TNS with samples exceeding 10, namely those reported by CHIME/FRB \citep{chimefrb_collaboration_first_2021}, ASKAP \citep{bannister_detection_2017, shannon_dispersionbrightness_2018, macquart_census_2020}, Parkes \citep{lorimer_bright_2007, keane_origin_2012, keane_fast_2016, thornton_population_2013, burke-spolaor_galactic_2014, petroff_real-time_2015, petroff_polarized_2017, ravi_fast_2015, ravi_magnetic_2016, champion_five_2016}, UTMOST \citep{caleb_first_2017, farah_frb_2018, farah_five_2019}, DSA-110 \citep{law_deep_2024}, and MeerTRAP \citep{jankowski_sample_2023}.
For these surveys, 
Fig.~\ref{fig:dms} visualises the  observed (DM$_{\text{obs}}$) and excess (DM--DM$_{\text{MW}}$)
dispersion measures. Table~\ref{tab:dms} lists the distribution median per survey.
To investigate if the observed Apertif DM distribution agrees with the other surveys, we
performed a Kolmogorov-Smirnov two sample test and obtain the p-values $p_{KS}$ (Table~\ref{tab:dms}).
% If $p_{KS}<0.01$, we cannot reject the
% hypothesis that the observed DMs from two surveys were  drawn from different distributions.

We find the Apertif DM distribution to be compatible with most surveys, \refbf{except ASKAP.} Compared to that survey, with the lowest median DM, $p_{KS}<0.01$ on both the observed and extragalactic
DMs, rejecting that the two surveys drew from the same distribution. 
\refbf{While the ASKAP FRB sample we consider here includes both the fly's eye \citep{shannon_dispersionbrightness_2018} and incoherent-sum survey \citep{macquart_census_2020}, we find the average and median excess DM of the selected bursts from both surveys to be very similar (Average: 379\,\pccm\ for fly's eye, 401\,\pccm\ for incoherent. Median: 361\,\pccm\ for fly's eye, 385\,\pccm\ for incoherent). Although the resulting $p_{KS}$ are larger, which are 0.02 and 0.06 respectively for the fly's eye and incoherent surveys, the smaller sample sizes considered here would not allow for a significant difference to be perceived. Similarly, the MeerTRAP FRB sample consists of the FRBs detected in the incoherent and coherent beams \citep{jankowski_sample_2023}. These have different fields of view and sensitivities, but with the current sample size it is not possible to reveal any significant differences.}
This shows Apertif is sensitive to a more dispersed, more distant population of FRBs than ASKAP.
Hints of a similar difference emerge for DSA-110 ($p_{KS}=0.061$). These differences are also discernible in
Fig.~\ref{fig:dms}.

\begin{table}[]
    \centering
    \caption{Comparison of Apertif DMs to other instrument samples.}
    \begin{tabular}{l cc cc}
    \hline
    \multirow{2}{*}{Instrument} & \multicolumn{2}{c}{Observed DMs} & \multicolumn{2}{c}{Extragalactic DMs}\\ 
    \cline{2-5}
     & Median & $p_{KS}$ & Median & $p_{KS}$\\ 
    \hline
    Apertif & 625 & & 578 &  \\
    ASKAP & 431 & 0.009 & 361 & 0.005 \\ 
    CHIME & 562 & 0.269 & 485 & 0.194 \\ 
    DSA-110 & 452 & 0.061 & 365 & 0.061 \\
    MeerTRAP & 675 & 0.482 & 622 & 0.945 \\
    Parkes & 815 & 0.415 & 694 & 0.774 \\ 
    UTMOST & 647 & 0.701 & 484 & 0.592 \\ 
    \hline
    \end{tabular}
    \tablefoot{Medians for the observed and extragalactic DM values of the main FRB surveys, and their KS two-sample p-value ($p_{KS}$) against Apertif.}
    \label{tab:dms}
\end{table}

\begin{figure}
    \centering
    \includegraphics[width=\hsize]{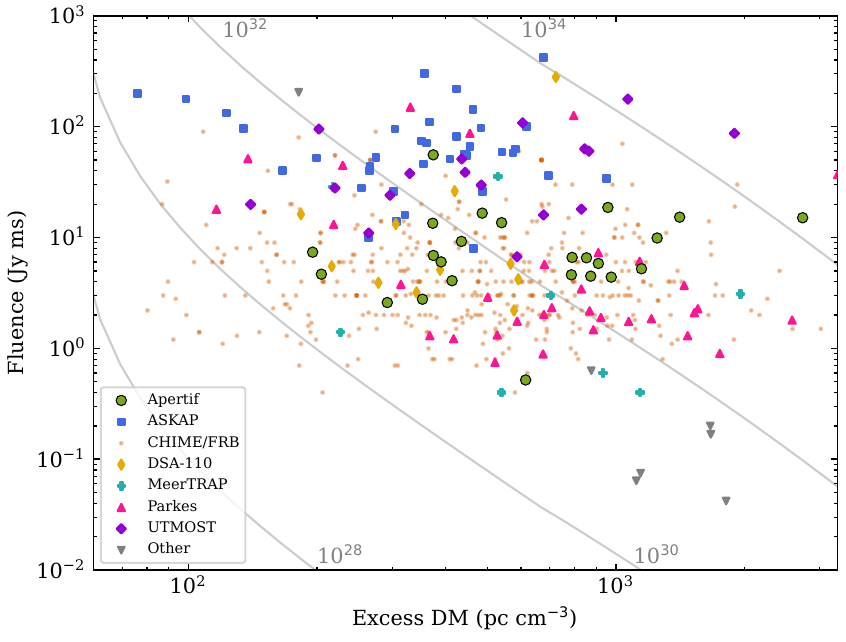}
    \caption{Fluence--excess DM diagram for Apertif and all other TNS FRBs. Surveys are marked per the legend.
      %Apertif is shown as green circles, ASKAP as blue squares, DSA-110 as yellow diamonds, MeerTRAP as turquoise
      %crosses, Parkes as pink upward triangles, UTMOST as purple diamonds, CHIME/FRB as orange dots with a high
      %transparency for better visualisation, and all other FRBs as grey downward triangles.
      Gray lines of constant spectral energy density are shown for reference, labeled in units of  erg\,Hz$^{-1}$. }
    \label{fig:dispersion-fluence}
\end{figure}

%In their analysis, e
Earlier surveys
compared FRB fluences against their extragalactic DMs,
to show that bright, nearby FRBs have energies   comparable to  the more distant, fainter FRBs \citep{shannon_dispersionbrightness_2018, farah_five_2019, niu_crafts_2021}.
The Apertif sample agrees, as visible in 
Fig.~\ref{fig:dispersion-fluence}. %, where we  display this fluence-DM plane.
For guidance, the grey lines represent the equivalent isotropic energy density that FRBs would have assuming the IGM DM contribution from \cite{zhang_fast_2018} and the cosmological parameters from \cite{planck_collaboration_planck_2020}. We adopt an observed host galaxy contribution to the DM of $50/(1+z)$\,\pccm for consistency with \cite{shannon_dispersionbrightness_2018} and \cite{petroff_fast_2019}. 
The Apertif FRBs are located between the Parkes and the UTMOST FRB samples within the fluence-excess DM plane, while overlapping the FRBs from the First CHIME/FRB Catalog. This is in agreement with the DM distributions shown in Fig.~\ref{fig:dms} and the fluence-dependent rates presented in Fig.~\ref{fig:fluence}.

% /Users/user/Documents/projects/ARTS/scripts/FRB_analysis/propagation_properties_distribution.py
% scripts/FRB_analysis/population_properties.ipynb

The Apertif sample contains the FRB with the third largest DM known to date,
 \refbf{at an energy comparable to the largest observed by other surveys (see
   Fig.~\ref{fig:freq_energy})}.
FRB\,20200719A originated at \zmacquart$\sim3.26^{+0.62}_{-1.35}$ (Sect.~\ref{sec:FRB20200719A}).
The maximum cosmic star formation rate (CSFR) took place at $z\sim2$ \citep{madau_cosmic_2014}; FRB\,20200719A could thus have been emitted beyond the CSFR peak. 
Previous works have compared the CHIME/FRB DM distribution to cosmic evolution models, 
and have found it does not appear to track the star formation history of the Universe \citep{qiang_fast_2022, zhang_chime_2022}.
As most FRB progenitor models are based on or related to stellar populations, 
and given that the CSFR decreases by over an order of magnitude from its peak to the current era, we find those results
surprising.
Recently, however, \citet{frboppy_wang_2024} combined for the first time
a  full multi-dimensional Markov chain Monte Carlo search around the  
population synthesis code of \cite{gardenier_synthesising_2019}, 
with the inclusion of the full set of CHIME one-off FRBs. 
They find strong evidence that 
the FRB number density
follows the CSFR.
Detecting additional FRBs like FRB\,20200719A at very large DMs and thus redshifts will better constrain the FRB rate evolution with redshift.
Comparing the FRB host galaxy properties to that of other astrophysical transients has proved useful in constraining the
potential FRB progenitors \citep{heintz_host_2020, safarzadeh_confronting_2020, mannings_high-resolution_2021,
  bhandari_characterizing_2022}; the redshift distribution, \refbf{combined with the highest FRB luminosities,} could
provide with additional information to rule out some of the current FRB progenitor models.
Detecting transients associated with such high-$z$ FRBs in other wavelenghts is currently  challenging through.
%. Long Gamma-ray bursts (LGRBs) have been identified with redshifts up to $z\sim9$ mainly
% with the Swift satellite \citep{cucchiara_photometric_2011, lan_revisiting_2021}, while
The most distant short Gamma-ray burst (SGRB) was found at a redshift of $z\sim2.2$ \citep{selsing_host_2018} while
%This can be explained by a lower Swift sensitivity to high redshift SGRBs compared to LGRBs \citep{guetta_luminosity_2005}.
%The detection of distant optical transients is also limited; the most distant type Ia supernova has a redshift of only
%$z\sim1.9$ \citep{jones_discovery_2013},
 core-collapse supernovae (CCSN) have been found up to $z\sim2.4$
\citep{cooke_type_2009}. % and superluminous supernovae (SLSN) up to $z\sim3.9$ \citep{cooke_superluminous_2012}.
If a link is established between FRBs and any such transients, high DM bursts might help establishing the cosmic evolution of their progenitors. 

\subsubsection{Scintillation} \label{sec:discussion_scintillation}
\label{sec:result_scintillation}

Scintillation is measurable in 11 out of the 24 FRBs, with  bandwidths \scbw\ ranging from 0.8 to 9.3\,MHz, as shown in Fig.~\ref{fig:scintillation}.
While generally falling within the error of 40\% that we assume on the  YMW16 and NE2001 models, the  measured bandwidths  exceed the MW contribution predictions in most cases. % at 1370\,MHz
%
%shows the measured scintillation bandwidths and their comparison to the expected Milky Way contribution.
% The scintillation bandwidth is converted to scattering timescales using Eq.~\ref{eq:scint_to_scat}. %% explained below
% These results will be further discussed in Section~\ref{sec:discussion_scintillation}.
%The scintillation bandwidths of the FRBs in our sample, when measurable,
%fall within the error range of the expected NE2001 contribution, but in most cases they are larger than the predictions.
% HIER WAS IK
This might indicate the electron density along the LoS fluctuates less than predicted by these models. 
The YMW16 predicted scintillation is generally also lower than the Apertif measured values. This may be explained by the way the model predicts the scattering/scintillation values; it uses the \tscat--DM relation instead of modelling the electron density fluctuations along the LoS \citep{yao_new_2017}.
\refbf{Alternatively, this might be related to the difference between the plane wave approximation, appropriate for distant FRBs, and the spherical wave approximation used for galactic pulsars. The plane wave approximation, which assumes weaker scattering, results in a broader scintillation bandwidth, by a factor of $\sim2$, compared to the stronger scattering of spherical waves \citep{lambert_theory_1999}.
In any case,} the scintillation measurements of Apertif FRBs can prove useful in better constraining the Milky Way electron distribution, especially at high Galactic latitudes where there is a dearth of pulsars.

In Fig.~\ref{fig:scat_dm_corr}, we plot the Apertif scintillation bandwidths converted to scattering timescales through
Eq.~\ref{eq:scint_to_scat}, as a function of the expected Galactic contribution to DM (since we expect the scintillation to be produced in the MW).
The measured values do not follow the {\tscat} -- DM relation from \citet{bhat_multifrequency_2004} and
\citet{cordes_fast_2019} for  Galactic pulsars. Since pulsars generally lie in the Galactic plane, they probe a more inhomogeneous medium than the Galactic halo encountered by most FRBs \citep{ocker_constraining_2021}. It is thus understandable that FRBs do not follow the same relation as pulsars.

\subsection{Multi-component bursts} \label{sec:discussion_multcomp}
 \label{sec:multi-component}

Seven out of the 24 detected bursts display more than one discernible component. Our system thus finds $\sim$$30$\,\%  of the bursts at 1370\,MHz are multi-component.
Of those, three can be well fitted by two components (FRB\,20190709A, FRB\,20191109A, and FRB\,20200321A, see Section~\ref{sec:bicomponent_frbs}); FRB\,20200216A is the only one displaying three distinct components; FRB\,20200518A and FRB\,20210530A are well fitted by four components, and finally FRB\,20201020A has five distinct components \citep{pastor-marazuela_fast_2023}. 

In the morphology study of the First CHIME/FRB Catalog bursts, \cite{pleunis_fast_2021} find that only about 5\% of one-off bursts contain multiple components. However, \citet{chimefrb_collaboration_updating_2023} recently published the baseband data for 140 of the bursts from their first catalog, including both one-offs and repeaters. Although no updates on the morphological fractions have been reported yet, by visually examining the published dynamic spectra we determined that $\sim37\pm3$\footnote{Since we determined the multi-component FRBs by eye, the uncertainty comes from the bursts where we could not determine whether a component was real or noise, or when a pulse profile had a single component that did not fit a simple (scattered) Gaussian.} of the bursts from the 130 one-off FRBs in the sample show complex, multi-component morphologies. This would imply a $\sim28$\% fraction of multi-component FRBs, in agreement with the Apertif multi-component fraction. 

We can nevertheless compare the reported multi-component bursts from the First CHIME/FRB Catalog to the Apertif bursts.
While the CHIME/FRB mean subcomponent separation is $\sim7.3$\,ms, with Apertif we observe a mean separation between burst subcomponents of 1.2\,ms. This difference can be explained by the time resolution used in the FRB searches for the First CHIME/FRB Catalog of 0.983\,ms \citep{chimefrb_collaboration_first_2021}. 
In order to better assess the observed subcomponent separation distribution for one-off FRBs, we built histograms of the subcomponent separations and fitted them to a lognormal distribution, as given by Eq.~\ref{eq:lognormal}.
When estimating the separation distribution parameters from the Apertif bursts only, we find a scale parameter $m=0.8\pm0.2$\,ms and a shape parameter $\sigma=0.9\pm0.2$, while for CHIME/FRB we find $m=5.8\pm0.2$\,ms and $\sigma=0.57\pm0.03$. 
To obtain the joint distribution, we normalise each of the instruments histograms; we divide the Apertif histogram by
the total number of subcomponents in the Apertif sample (39 subcomponents), and we do the same for the CHIME/FRB sample
(506 subcomponents). We add the two normalised histograms and fit them to a lognormal distribution, where we find the
parameters $m=0.9\pm0.2$\,ms and $\sigma=1.1\pm0.3$, as shown in the lower panel of Fig.~\ref{fig:chime-apertif_tscat}.

%Since CHIME/FRB observes at
At lower frequency,
%than Apertif did, their detected bursts should be more smeared by scattering. Additionally, we have shown earlier that
%there is a population of highly scattered bursts to which CHIME/FRB is not sensitive. S
scattering is likely to blur together closely spaced subcomponents, and thus one would expect the observed fraction of
multi-component FRBs to decrease with frequency, if the intrinsic subcomponent fraction remains the same. We must consider however the selection effects that the CHIME/FRB baseband pipeline could have on the observed population, since only bursts with S/N>10-12 have a baseband trigger, compared to the S/N>8 trigger threshold to store the intensity data. These different thresholds could be producing a bias in the multi-component fraction of their observed population.

%Studies comparing multi-component fractions at different frequencies are still scarce, but current and upcoming FRB
%surveys with access to high spectro-temporal resolution data, such as MeerTRAP \citep{rajwade_meertrap_2021}, the Deep
%Synoptic Array 110 \citep[DSA-110;][]{kocz_dsa-10_2019}, or the CRAFT COherent upgrade at ASKAP \citep[CRACO;
%  e.g.][]{scott_celebi_2023}, will soon produce large samples of FRBs in the L-band.
Comparing, now and in the near future, the fraction of multi-component FRBs at different frequencies might provide with important clues about the FRB emission mechanism.
% If high time resolution studies at lower or higher frequencies
Proving this fraction changes 
at a level inconsistent with smearing due to scattering or instrumental effects,
would support the existence of frequency-dependent morphology for the population of one-off FRBs. % that we determined through our simulations. 
Such frequency-dependent effects are seen in  radio pulsar profiles,
and are there thought to reflect the change in emission
height  %of different radio frequencies
of the pulsar beam, the so-called radius-to-frequency mapping \citep[RFM; see, e.g.,][]{cordes_observational_1978}. 
This RFM has been suggested to explain the drift rate evolution with frequency observed in repeating FRBs \citep{lyutikov_radius--frequency_2020, tong_three_2022, 2022A&A...658A.143B}.
Observing a similar frequency-dependent relationship in one-off FRB pulse components would  provide further evidence for a neutron-star origin for FRB emission. 

The Apertif FRB components have an average  width of  $\sim$0.5\,ms, while for CHIME/FRB it is $\sim$1\,ms.
In pulsars, the observed profile (and component) width over this same frequency range of 1.4 to 0.4\,GHz evolves too, by an increase of $\sim$20\% (Table~1 in \citealt{kijak_radio_1997}).
If this difference persists when CHIME/FRB collects a larger sample of FRBs with baseband data, this would indicate the evolution of FRB component width versus frequencies is of at least the same scale, if not more, as that seen in pulsars. 

In \citet{karastergiou_empirical_2007}, the profile classification of more than 250 pulsars revealed that 60\% of young, fast-spinning, and highly energetic pulsars show single component average profiles while 40\% show double component profiles, while for older, slower, and less energetic pulsars 45\% are single profile, and the rest are either double or multi-component. 
Although we did not find any current estimates of the radio burst morphology distributions from the magnetar population, observations reveal complex radio bursts to be prevalent \citep{maan_magnetar_2022, caleb_discovery_2022, kramer_quasi-periodic_2023}.
The morphological similarities between radio-loud neutron stars and the observed FRBs suggest a link between the emission process of these phenomena.

We note that the Apertif sample consists entirely of one-off FRBs that have not been seen to repeat. While many previous morphological studies have focused on samples of bursts from known repeaters \citep[e.g.,][]{hewitt_arecibo_2022, platts_analysis_2021, sand_multiband_2022}, the large fraction of multi-component bursts in a self-consistent sample of one-off FRBs is still
\refbf{relatively new \citep[although a smaller sample of ASKAP FRBs already hinted at this;][]{cho_spectropolarimetric_2020, day_high_2020}.}
Ultimately, larger samples of high time resolution one-off FRBs detected by other facilities will be needed to fully contextualise the Apertif sample, as no further studies can be done with Apertif itself. However, we encourage other surveys to further explore the morphological properties of one-off FRBs, as has been done by \cite{faber_morphologies_2023}.
For large numbers, finding the patterns or groups that underlie the burst shapes could be done with unsupervised methods such as those proposed in \citet{2023arXiv231109201V}.

%%%%%%%%%%%%%%%%%%%%%%%%%%%%%%%%%%%%%%%%%

\subsection{Polarisation} \label{sec:discussion_polarisation}
\label{sec:results_polarisation}

Of the 24 ALERT FRBs, 16 triggered a dump of the full Stokes parameters (Fig.~\ref{fig:stokes}).
Of these, eight have measurable RMs, ranging from $\sim120$\,\radsqm\ to ${\sim2050}$\,\radsqm\ in absolute value (see Table~\ref{tab:stokes}). 
The sample includes the one-off FRB with the second largest |RM| ever reported \citep[after FRB\,20221101A from][]{sherman_deep_2024}; FRB\,20200216A, with RM$=-2051.0\pm5.8$ (See Fig.~\ref{fig:pol_FRB20200216A}). 
When converting the RM to the expected redshift range, \zmacquart$=0.44^{+0.12}_{-0.24}$, and after removing the expected MW contribution, this becomes \rmhost$=-4200^{+1300}_{-800}$\,\radsqm. 
FRB\,20200514A, with an RM of $966.1\pm20.5$\,\radsqm, \refbf{although ambiguous}, has the second largest |RM| in our sample, but it is expected to have the largest RM in the reference frame at \zmacquart$=1.35^{+0.30}_{-0.66}$; \rmhost$=6500^{+2000}_{-3200}$\,\radsqm. 
% Two further FRBs have observed RMs between 300 and 500\,\radsqm, which becomes 1300--1500\,\radsqm\ in the reference frame, while four additional FRBs have observed RMs between 100 and 300\,\radsqm, or 200--600\,\radsqm\ in the reference frame. 

The bursts have an average linear polarisation fraction $L/I=43\pm28\%$, while the average circular polarisation fraction is $V/I=9\pm8\%$, with the errors giving the standard deviation, \refbf{if we consider all Apertif FRBs with IQUV data. If we consider only the bursts where we did not force the circular polarisation to be zero, we get $L/I=35\pm27$\% and $V/I=10\pm6$\%, which is compatible with the whole sample.}
Two FRBs have linear polarisation fractions $>80\%$; these are FRB\,20191108A \citep{connor_bright_2020} and FRB\,20210124A. While the first one has a large RM value of $\sim473$\,\radsqm, the latter has no measurable RM within the observing bandwidth.
The two FRBs with the lowest linear polarisation fractions are also the two most scattered bursts with full Stokes data, namely FRB\,20200322A with \tscat$=4.2\pm0.4$\,ms and $L/I=3\pm6\%$, and FRB\,20200323C with \tscat$=1.3\pm0.1$\,ms and $L/I=6\pm3\%$. This could be explained, for instance, by the propagation through a magnetised inhomogeneous plasma screen (see the discussion in Section~\ref{sec:discussion_polarisation}).
None of the bursts have \refbf{large levels of circular polarisation}, with the highest fractions being $V/I=21\pm9\%$ for FRB\,20200514A, and $29\pm17\%$ for FRB\,20200518A. We note, however, that we have calibrated some of the bursts by assuming their circular polarisation fraction to be zero in cases where we saw oscillations in the sign of the $V$ intensity, which could bias our results. 

The PPAs of the bursts with sufficient $L$ signal primarily appear to be flat, and in the case of FRB\,20200518A even within the two subcomponent groups.
In some of the bursts, one or two of the time samples seem to jump up and down in PPA, although this could be produced by noise; see FRB\,20200419A, FRB\,20200514A, and FRB\,20210530A.
FRB\,20200216A presents a pronounced decrease in PPA of ${\sim20^{\circ}}$, while the PPA of FRB\,20210127A appears to decrease by ${\sim5^{\circ}}$ and that of FRB\,20210317A increases by ${\sim6^{\circ}}$. FRB\,20200213A shows an erratic PPA behaviour, which might be explained by the low linear polarisation fraction.

We can classify the polarisation fractions of these 16 FRBs into the four subgroups from \cite{sherman_deep_2024}. These subgroups were determined empirically based on 25 DSA-110 bursts, and they are the following: linearly polarised if $L/I>70\%$ and $V/I<30\%$, circularly polarised if $V/I>30\%$, partially depolarised if $35\%<L/I<70\%$ and $V/I<30\%$, and unpolarised if $L/I<35\%$ and $V/I<30\%$. 
Our sample contains FRBs in three categories: $4/16$ are linearly polarised, $6/16$ are partially depolarised, and
$6/16$ are depolarised. 
The fraction of Apertif FRBs with (partial) linear polarisation of $62.5\%$ is consistent with the $68\%$ fraction found
by DSA-110 \citep{sherman_deep_2024}, as well as with the $\sim70\%$ fraction found in 128 bursts from CHIME/FRB \citep{pandhi_polarization_2024}.
% The average linear polarisation fraction we observe in our sample, $<L/I>=43\pm28\%$

The two FRBs with the lowest polarisation fractions, FRB\,20200322A and FRB\,20200323C, also have the largest scattering timescales of the bursts with full Stokes data. This could be explained if linearly polarised bursts travelled through an inhomogeneous magneto-ionic environment in the vicinity of the source, which would produce depolarisation as well as scattering. This effect has been proposed to explain the depolarisation of repeaters at low frequencies by \citet{feng_frequency-dependent_2022}. They determine an empirical linear correlation between the RM scatter $\sigma_{\text{RM}}$ and the scattering timescale \tscat, where $\sigma_{\text{RM}}$ can be obtained from the depolarisation fraction $f_{\text{RM}}$ with the following equivalence:
\begin{equation}
    f_{\text{RM}} \equiv 1- e^{-2\lambda^4\sigma_{\text{RM}}^2}.
\end{equation}
The Apertif bursts with Stokes data and scattering timescale appear to follow the same relation. However, the upper limit on the scattering timescales for some of these bursts would place them below the limits of the expected relation. This could support the suggestion from \citet{pandhi_polarization_2024} that one-off FRBs do not always have an intrinsic $100\%$ linear polarisation fraction.
\refbf{None of the Apertif FRBs show evidence for depolarisation at lower frequencies within the available bandwidth, so we cannot test for the $\sigma_{\text{RM}}$ and $|\text{RM}|$ observed in \citet{feng_frequency-dependent_2022}.}
% Overall, our observations support the idea of an FRB emission mechanism with intrinsic linear polarisation, and depolarisation occurring along their propagation path.

It is noteworthy that the polarisation fraction of the Apertif FRBs overlap with what has been reported for a sample of
35 young and energetic pulsars observed for the thousand-pulsar-array (TPA) program with MeerKAT
\citep{serylak_thousand-pulsar-array_2021}. The linear polarisation fraction of the latter have a median and standard
deviation of 49\% and 27\% respectively, while for ALERT FRBs it is 43\% and 27\%.
The circular polarisation fraction, on
the other hand, have a mean and standard deviation of 9\% for TPA and 9\% and 8\% respectively for ALERT FRBs. This is
further evidence for the potential link between young, high-energy pulsars and one-off FRBs.

\begin{figure}
    \centering
    \includegraphics[width=\hsize]{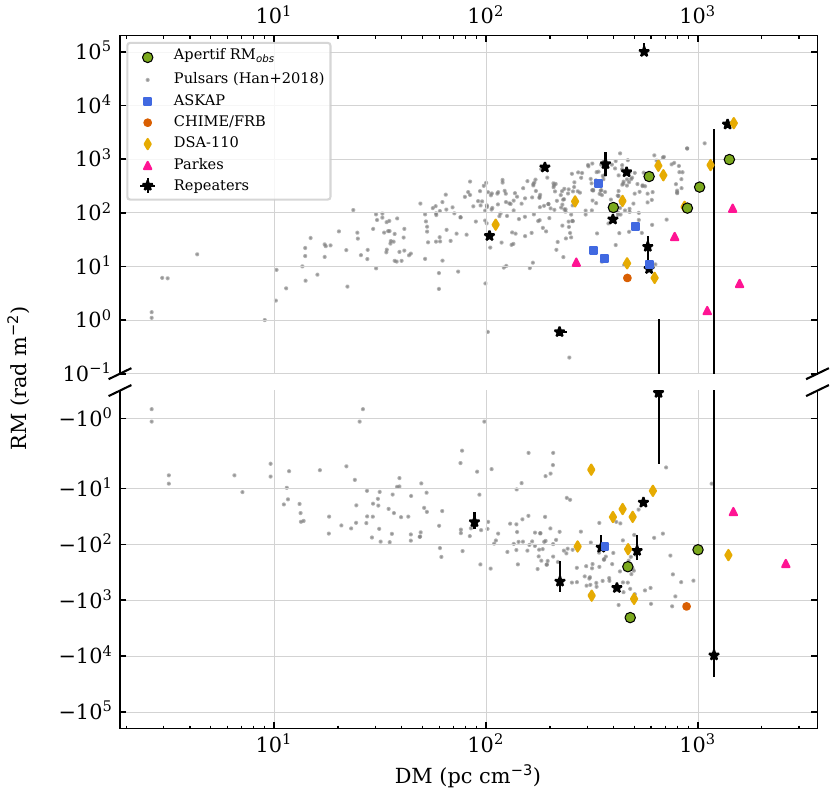}
    \caption{Observed RM of one-off and repeating FRBs compared to galactic pulsars as a function of DM. The green circles show the observed Apertif RMs, the grey dots the measured pulsar RMs from \cite{han_pulsar_2018}, the blue squares are from ASKAP one-off FRBs, the orange circles from CHIME/FRB, the yellow diamonds from DSA-110, and the pink triangles from Parkes FRBs. The mean measured repeater RMs are shown as black stars with error bars showing the observed RM ranges of each source.}
    \label{fig:rms}
\end{figure}

The Apertif FRB sample contains some of the highest RMs ever observed in one-off FRBs, including FRB\,20200216A with
RM$=-2051$\,\radsqm, as can be visualised in Fig.~\ref{fig:rms}\footnote{The CHIME/FRB RMs from
\citet{pandhi_polarization_2024} have not been included since the paper is currently under review and their tables are
not available digitally yet.}. Although no one-off FRB has been yet found to have an RM as extreme as that observed for
the repeaters FRB\,20121102A \citep{hilmarsson_rotation_2021} or FRB\,20190520B \citep{anna-thomas_magnetic_2023}, we
compared the RMs of repeaters and one-offs using a Kolmogorov-Smirnov (KS) test. We find a p-value of 0.51 if we compare
the observed RMs, while the p-value is 0.61 if we correct the RMs for redshift with Equation~\ref{eq:rm_redshift}. This
indicates that the current sample of one-off and repeater RMs could have been drawn from the same distribution, implying
that both FRB classes could be produced in environments with similar magneto-ionic
properties.
\citet{pandhi_polarization_2024} reach a similar conclusion using the CHIME/FRB  polarisation sample,
in line with the more general finding in \citet{2021A&A...647A..30G},
based on a wider range of burst characteristics, that all FRBs originate from a single and mostly uniform population.

If we compare the observed RMs of one-off FRBs from Apertif, \citet{sherman_deep_2024}, and references therein to galactic pulsars, we obtain a KS p-value=0.48, suggesting that the distributions are compatible too. However, if instead we compare the redshift-corrected RMs, the p-value is 0.003, indicating that the RMs have been drawn from different distributions. This reveals that one-off FRBs could originate from environments more extreme than those where pulsars are usually located.
Although RMs of the order of $|\text{RM}|\sim100$\,\radsqm\ might be expected from FRBs originating from MW-like host galaxies with \HII\ regions, reaching $10^3-10^5$\,\radsqm\ would require the FRB to be emitted from a supernova remnant (SNR) or from a \HII\ region \citep{hackstein_fast_2019}. Hence, a significant fraction of one-offs must be produced from within these environments to explain the observed Apertif RMs. Detecting further one-off FRBs with such extreme RMs might be easier in high frequency observations, since the RM oscillations in $Q$/$U$ become stronger at lower frequencies, and might not be resolvable at the instrument's frequency resolution.

% After performing a Fisher exact test \citep{fisher_interpretation_1922, fisher_statistical_1934}\footnote{Applied with the function \texttt{stats.fisher\_exact} from the python package \texttt{scipy}}, we conclude that these subgroup fractions are consistent with what the DSA-110 report. The Fisher exact test returns the probability that we would observe this or an even more imbalanced ratio of multi-component bursts by chance, and it is valid for small sample sizes. For all subgroups, we obtain p-values$>0.28$.

\subsection{All-sky FRB rates and fluence distribution} \label{sec:discussion_rate} \label{sec:event_rate}

In ALERT, we discovered 24 new one-off FRBs in 5259\,h of observing time, an average of one FRB every 9.1\,days. 
Given the effective Apertif FoV of 8.2\,deg$^2$ \citep{van_leeuwen_apertif_2023},
% which already accounts for the sensitivity of the CBs. The
the all-sky rate for $N$ detected FRBs is: % expressed as follows:

\begin{equation} \label{eq:rate}
    R \text{ (sky}^{-1}\text{day}^{-1})= N \times \dfrac{24\text{ h day}^{-1} \times 41253 \text{ deg}^2\text{sky}^{-1}}{5259 \text{ h} \times 8.2 \text{ deg}^2}
\end{equation}

With  $N=24$\,FRBs, $R=550^{+220}_{-170}$\,\skyday, with 90\% Poisson errors \citep{gehrels_confidence_1986}. The rate is consistent with our estimate from \citet{van_leeuwen_apertif_2023}, based on the first 5 Apertif FRB detections, of $700^{+800}_{-400}$\,\skyday.
% $551^{+218}_{-171}$\,\skyday

% python /home/ines/Documents/projects/ARTS/scripts/FRB_analysis/propagation_properties_distribution.py
\begin{figure}[h]
    \centering
    \includegraphics[width=\hsize]{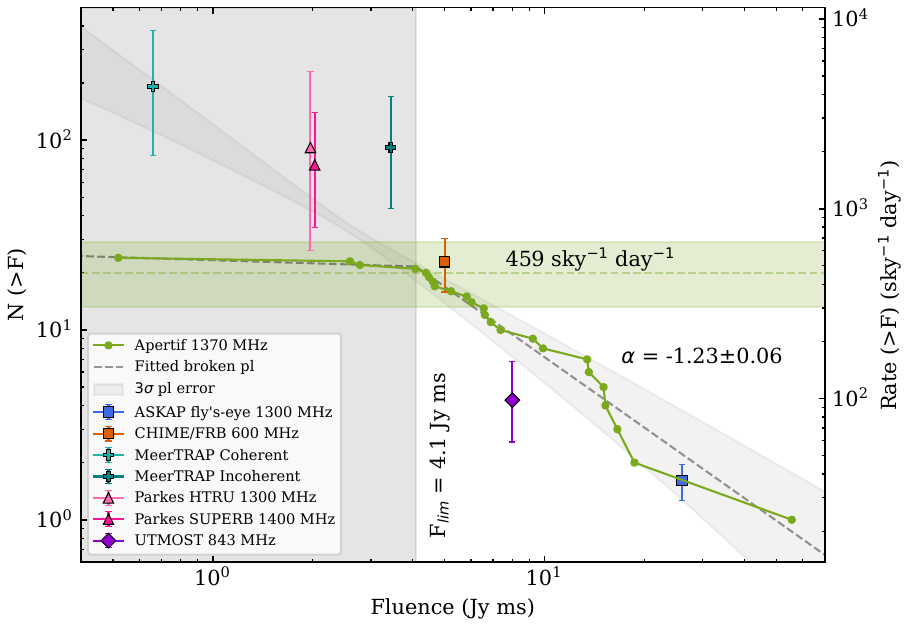}
    \caption{Cumulative fluence distribution of the Apertif FRBs. The left and right ordinate are related through
      %axis       gives the number of FRBs, and the right y axis its conversion to an all sky rate using
      Eq.~\ref{eq:rate}. The green dots give the measured FRB fluences, and the grey dashed line their fit to a broken
      power law. The diagonal grey shaded region gives the $3\sigma$ confidence interval of the power law above the
      fluence limit extrapolated to other fluences. The vertical grey shaded region is below the fluence completeness
      threshold of 4.1\,\jyms, while the horizontal green shaded region gives the all-sky FRB rate above that fluence,
      with Poissonian 95\% confidence limits.
      % The power law above the completeness threshold has an index of $\gamma=-1.23$.
      The markers denote all-sky rates estimated by the surveys in the legend and main text. 
%    blue square for ASKAP fly's-eye \citep{shannon_dispersionbrightness_2018}, 
%    orange square for CHIME/FRB \citep{chimefrb_collaboration_first_2021}, 
%    turquoise cross for MeerTRAP Coherent, and teal cross for MeerTRAP Incoherent \citep{jankowski_sample_2023}
%    light pink triangle for Parkes HTRU \citep{champion_five_2016}, 
%    dark pink triangle for Parkes SUPERB \citep{bhandari_survey_2018}, 
%    and purple diamond for UTMOST \citep{farah_five_2019}.
    }
    \label{fig:fluence}
\end{figure}

To be meaningful, burst rates must be accompanied by a fluence completeness threshold.
% of the instrument, which can be highly variable.
In \citet{van_leeuwen_apertif_2023}, we determined this threshold from the typical Apertif SEFD.
Here we fit a broken power law
to our cumulative fluence distribution (See Fig.~\ref{fig:fluence}), and we assume the  power-law break corresponds to
the completeness threshold $F_{lim}$: 
\begin{equation}
    N(>F) = 
    \begin{cases}
        C (F/F_{lim})^{\gamma_0}, & \text{if } F<F_b\\
        C (F/F_{lim})^{\gamma}, & \text{otherwise,}
    \end{cases}
\end{equation}
with $C$  a constant, and $\gamma_0$ and $\gamma$ the power law indices below and above the threshold.
% This way, we recompute the all-sky rate for the FRBs above the completeness threshold and determine the power-law index of the fluence distribution. 
We find a fluence completeness threshold of $F_{\text{lim}}=4.1\pm0.2$\,\jyms, with $N=20$ FRBs above the
threshold. Using Eq.~\ref{eq:rate}, this yields an FRB all-sky rate at 1370\,MHz of ${R_{1370}(F\geq 4.1 \text{ Jy ms})=
  459^{+208}_{-155}}$\,\skyday. %, with 90\% Poisson errors.
Furthermore, we determine a fluence distribution power law index of $\gamma=-1.23\pm0.06$, where we quote the $1\sigma$ statistical error from the fit. We estimate a systematic error of 0.2 on $\gamma$.

We use the resulting power law to compare our subsequent all-sky rate to the estimates made by other surveys at their respective fluence completeness thresholds. 
In Fig.~\ref{fig:fluence} we plot the all-sky rates from the ASKAP fly's-eye search \citep[$F>26$\,\jyms;][]{shannon_dispersionbrightness_2018}, the Parkes HTRU \citep[$F>2$\,\jyms;][]{champion_five_2016} and SUPERB \citep[$F>2$\,\jyms;][]{bhandari_survey_2018} surveys, the UTMOST survey \citep[$F>8$\,\jyms;][]{farah_five_2019}, MeerTRAP coherent and incoherent \citep[$F>0.66$\,\jyms\ and $F>3.44$\,\jyms, respectively;][]{jankowski_sample_2023}, and the First CHIME/FRB Catalog \citep[$F>5$\,\jyms;][]{chimefrb_collaboration_first_2021}. 

%From the ALERT survey, we compute an all-sky FRB rate at 1370\,MHz of $R_{1370}(F\geq 4.1 \text{ Jy ms})=459^{+208}_{-155}$\,\skyday, with 90\% Poisson errors.
%The power-law fit to the fluence cumulative distribution above this 4.1\,\jyms\ completeness threshold next allows us to  compare our rate to that of other surveys, even though they have different sensitivity limits (See Fig.~\ref{fig:fluence}). The resulting power law index for bursts above the threshold is $\gamma=-1.23\pm0.06\pm0.2$.

\begin{figure}
    \centering
    \includegraphics[width=\hsize]{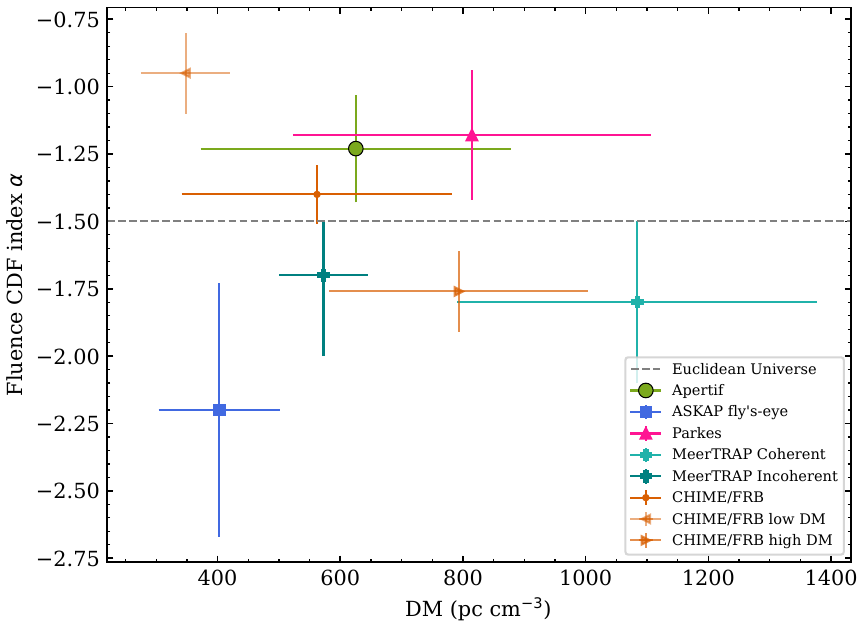}
    \caption{Fluence power-law index $\gamma$ as a function of the median DM, for the surveys marked in the legend.
      % Each
      % data point gives the measured $\gamma$ as a function of the
      Errors on the DM are the median absolute deviation.
      % ASKAP Fly's-Eye survey is shown as a blue square \citep[$\gamma$ values from][]{james_slope_2019}, Parkes is
      % shown as a pink triangle, MeerTRAP Coherent and Incoherent as turquoise and teal crosses respectively, CHIME/FRB
      % as an orange square, the
      The CHIME/FRB subsamples are divided at 500\,{\pccm}.
      %The predicted $\gamma=-1.5$ for a Euclidean Universe is shown by a dashed grey line.
    }
    \label{fig:alpha_dm}
\end{figure}

The Apertif all-sky rate is comparable to most of these surveys when accounting for  their fluence sensitivity
thresholds;
%in the L-band, ASKAP in fly's-eye mode %at 1300\,MHz
%\citep{shannon_dispersionbrightness_2018}, Parkes HTRU % rate at 1300\,MHz
%\citep{champion_five_2016}, SUPERB %rate at 1400\,MHz
%\citep{bhandari_survey_2018}, MeerTRAP Coherent \citep{jankowski_sample_2023}, as well as  CHIME/FRB at 600\,MHz \citep{chimefrb_collaboration_first_2021}, all report rates that agree with ours.
The UTMOST rate at 843\,MHz % \citep{farah_five_2019}
agrees within $3\sigma$, but this rate is obtained from a smaller burst sample. We thus see no evidence for an evolution of the FRB rate with frequency.
In a non-evolving, constant density Euclidean Universe, the expected power law index of the fluence distribution
observed with a perfect telescope is $\gamma=-1.5$. Although our observed power law index appears to be flatter, we cannot rule out that it is consistent with the Euclidean prediction within systematic errors. 
We can compare the Apertif ${\gamma=1.23\pm0.26}$ to what has been reported by other FRB surveys, as shown in Fig.~\ref{fig:alpha_dm}. \cite{bhandari_survey_2018} reported $\gamma=-2.2^{+0.6}_{-1.2}$ for the Parkes burst sample, while \cite{shannon_dispersionbrightness_2018} determined $\gamma=-2.1^{+0.6}_{-0.5}$ for the ASKAP sample. \cite{james_slope_2019} later reanalysed these two FRB samples and determined the Parkes index to be $\gamma=-1.18\pm0.24$ and ASKAP to be $\gamma=-2.2\pm0.47$. While the combined power law index of both surveys is $\gamma=-1.55\pm0.23$, consistent with the Euclidean Universe, they are inconsistent with each other at $2.6\sigma$. This discrepancy was interpreted as a difference in the cosmological population observed by each of these surveys, with ASKAP seeing nearby sources and Parkes more distant ones, following the average DM of each burst sample (See Fig.~\ref{fig:dms} and Table~\ref{tab:dms}).
Meanwhile, the index determined from the First CHIME/FRB Catalog is $\gamma=-1.40\pm0.11^{+0.060}_{-0.085}$, in agreement with the Euclidean prediction and in between the Parkes and ASKAP values \citep{chimefrb_collaboration_first_2021}. This appears to concur with the average DM of the CHIME/FRB sample compared to the other two surveys. 
These studies seemed to reveal an apparent increasing median-DM/power-law-index trend observed in the other surveys, which the Apertif FRBs also follow; the fluence cumulative distribution appeared to be flatter for a sample of FRBs with larger DMs and thus redshifts.
However, recent results from MeerTRAP find power law exponents of $-1.7^{+0.2}_{-0.3}$ and $-1.8^{+0.3}_{-0.3}$ respectively for the  incoherent (median DM$\sim570$\,\pccm) and coherent (median DM$\sim1080$\,\pccm) surveys \citep{jankowski_sample_2023}, which is at odds with the aforementioned trend.
Furthermore, when subdividing the CHIME/FRB catalog into high and low DM FRBs (above and below 500\,\pccm), these sub-samples also follow the opposite trend; the high DM sample has $\gamma=-1.75\pm0.15$ and the low DM sample $\gamma=-0.95\pm0.15$. 
We note that overall,
a number of selection effects modify the single intrinsic $\gamma$ into the observed one,
which may then differ among  surveys.
At lower fluences,
surveys become varyingly incomplete while FRBs at high fluences might be
misidentified as RFI by certain processing
choices. 
Pulses may be more easily detectable if they are intrinsically wide and bright,
but harder to find if their width is result  from dispersion smearing and scattering.
These effects, discussed in e.g. \citet{2019MNRAS.487.5753C}, are simulated in e.g. \citet{frboppy_wang_2024}.
On average, the FRB sample observations mentioned in this section seem to be in agreement with the Euclidean Universe prediction. Future FRB detections and power law index measurements will provide with important information about how the FRB population evolves with redshift. 

% /Users/user/Documents/projects/ARTS/scripts/FRB_analysis/propagation_properties_distribution.py
% scripts/FRB_analysis/population_properties.ipynb

\subsection{Higher-frequency emission} \label{sec:high_f_lum}

% /Users/user/Documents/projects/ARTS/scripts/FRB_analysis/propagation_properties_distribution.py
% scripts/FRB_analysis/population_properties.ipynb
\begin{figure}
    \centering
    \includegraphics[width=\hsize]{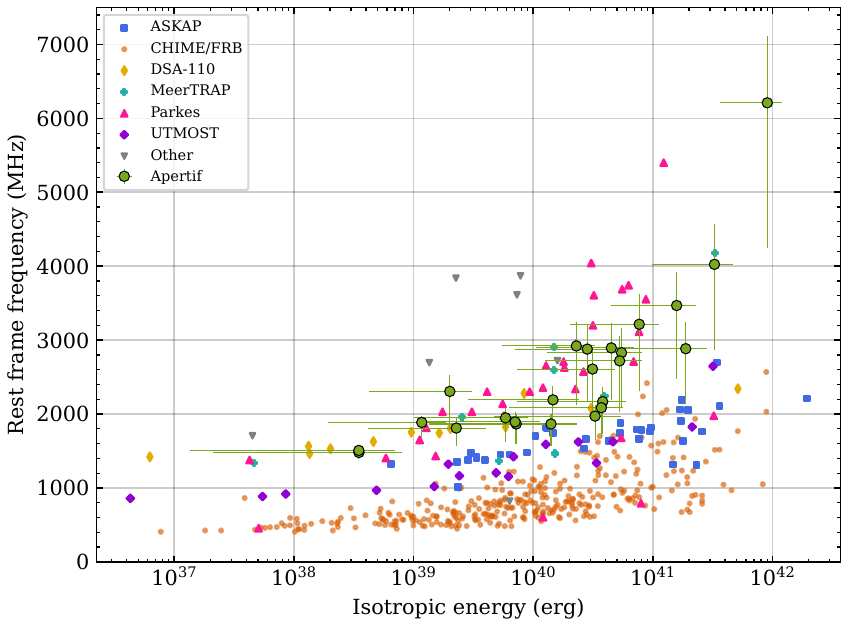}
    \caption{Rest frame frequency as a function of the isotropic energy of all FRBs in the TNS and Apertif. 
      Apertif FRBs are shown as green circles. % ASKAP FRBs as blue squares, DSA-110 as yellow diamonds, MeerTRAP as
                                % turquoise crosses, Parkes FRBs as pink upwards triangles, UTMOST as purple diamonds,
      % CHIME/FRB as orange circles, and all other FRBs as grey downwards triangles.
      All other FRBs, unless localised,  have large error bars for both frequency and energy  that are not shown to
      improve  visibility.}
    \label{fig:freq_energy}
\end{figure}

Although repeating FRBs have been observed to shine at frequencies as high as 8\,GHz
\citep{gajjar_highest-frequency_2018}, no one-off FRBs have been reported above the L-band \citep{petroff_fast_2022}. 
Since the observed FRB radio frequency $\nu_{\text{obs}}$ gets redshifted with distance, we can estimate the intrinsic, rest-frame frequencies at which these FRBs were emitted, $\nu_{0,\text{max}}$, from their expected redshifts:
$\nu_{0,\text{max}} = \nu_{\text{obs}} (1+$\zmacquart). For localised FRBs, we use the host galaxy redshift instead. We
plot the rest frame frequencies as a function of the estimated FRB isotropic energy \citep[Eq.~9
  from][]{zhang_fast_2018} for all FRBs in the TNS database in Fig.~\ref{fig:freq_energy}, and we find 16 FRBs for which
the estimated intrinsic emission frequency is $>3$\,GHz; four from Apertif, eight from Parkes, three from FAST
\citep{zhu_fast_2020, niu_crafts_2021}, and one from MeerTRAP \citep{rajwade_first_2022}. For the FRB\,20200719A
presented in this work (Sect.~\ref{sec:FRB20200719A} and Table~\ref{tab:frb_table2}),
and for the Parkes FRB\,20160102A \citep{bhandari_survey_2018}, the estimated emission frequency
is $>5$\,GHz.
The detection of high-DM one-off FRBs in the L-band thus indicates that bursts are commonly emitted at higher frequencies. % in their host galaxies, and give 

Interestingly, a significant number of individual radio-loud magnetars
are also observable at high frequencies \citep[see e.g.,~][]{levin_1622}, more than normal radio pulsars.
There is also population-based evidence of overlap in high-frequency behaviour between FRBs and magnetars
\citep{gardenier_multi-dimensional_2021}.
Our detection of emission at an inferred frequency > 5\,GHz strengthens the case that FRBs are
emitted by magnetar-like sources.

The strong correlation visible in Fig.~\ref{fig:freq_energy}, between the rest-frame frequency and the FRB isotropic energy, can be well explained by the selection effect caused by the equivalence between the redshift and distance, that together determine
in which frequency the FRBs are observed. \refbf{The correlation is also related to the fluence threshold of each telescope.}
Less energetic high-frequency FRBs most likely exist too, and fill the top-left part of the Fig.~\ref{fig:freq_energy} parameter space, but the current low- and mid-frequency surveys cannot easily detect those.
Nevertheless, if FRBs had steep, declining spectral indices, a source like 
FRB\,20200719A could not exist. 
Our finding that an FRB can produce $E_{\text{iso}} = 9.0\times10^{41}$\,erg above 5\,GHz
means the spectral index of FRBs cannot be too steeply negative.
A quantitative assessment  of the allowed values can only come from multi-survey population modeling
\citep[as discussed in e.g.][]{frboppy_wang_2024}.

\subsection{Motivation for future observations} \label{sec:future}

Our finding that FRBs emit at higher frequencies makes
searches for local FRBs in the S-band (2$-$4\,GHz) and above increasingly  interesting.
Although the smaller FoVs at higher frequency reduce the raw detection rates, the beams shrink equally, enhancing  localisation.  
When enough beams are formed and searched, interferometric S-band searches, using e.g., the MeerKAT  1.75$-$3.5\,GHz system \citep{kramer_meerkat_2018}, could be fruitful.  

While Apertif operations have ceased and the ALERT survey has finished, the science case for continued GHz FRB searching remains strong. From a larger sample of real-time detections, with immediate alerting and repointing of lower-frequency observatories, we can determine the emission bandwidth of one-off FRBs, and understand  their emission mechanism. 
Based on the work presented here,  a larger sample of 1.4\,GHz bursts could be investigated  specifically for  scattering and multi-component bursts.
Such a system could be implemented as a coarse, total-intensity real-time search that preserves baseband data for
detections. While the 1-dimensional nature of the WSRT allowed for full-field beamforming, it did limit the overall
localisation precision. Up to now, 2-D interferometers equipped with PAFs (i.e., ASKAP) have not been able to tile out
the complete primary beam with TABs, reducing the sensitivity (either through incoherent beamforming, or through the
longer integration times in imaging mode). The $\sim$ms integration upgrade to ASKAP for coherent FRB detection over the
entire FoV will increase detection rates while also providing good localisation \citep{2024ATel16468....1B}.
  \ipm{MeerKAT doesn't use PAFs, and the FoV is smaller but also more sensitive with the coherent beams. I think it
    would be worth talking about that and also a bit more about baseband data?} \jvl{Sure, go ahead. I didn't want to
    stray too far, given the PAFs ASKAP is closest and it continues where Apertif stopped. MeerKAT has single pixel dish
    beams, multiple receivers, etc., etc., all very interesting and certainly productive, but a little different from
    Apertif IMO.}

As each WSRT dish has a large collecting area for a "large number$-$small diameter" array, 
improvements to the front ends can be a cost-effective way to increase the system sensitivity.
Cryogenically cooled PAFs combine the strengths of the current system with reduced SEFD \citep{navarrini_design_2018, pingel_commissioning_2021}. 
Such a  successor to Apertif would increase the detection rate by a  factor $\sim$4, and provide better localisation through the higher detection S/N.

%------------------------------------------------------
\section{Conclusions}\label{sec:conclusions}

In this work, we have reported the discovery of 18 new, so-far one-off, FRBs, and analysed the properties of the total of 24 bursts that were detected during the ALERT Apertif FRB survey between July 2019 and February 2022. For each FRB, we determine the localisation region and expected redshift range and perform a flux calibration. We evaluate their morphology, determining the number of components and the spectral properties, and we study the propagation effects by verifying the presence of a resolved scattering tail in time and a scintillation pattern in frequency.

We localise each FRB to a narrow ellipse ($5"-20"$ wide) whose area depends on the detection S/N and the number of CBs where it was
detected. The average localisation area is $\sim5$\,arcmin$^2$. For five new FRBs with a high S/N and a low DM, namely
% FRB\,20200518A, FRB\,20210317A, and FRB\,20200419A, 
FRB\,20200210A, FRB\,20200419A, FRB\,20200518A, FRB\,20210317A, and FRB\,20211024B,
we find a small number of host galaxy candidates in the PanSTARRS DR1 catalogue, $\leq5$, in which one of the galaxies has a \texttt{PATH} probability of being associated with the FRB >50\%. In the case of FRB\,20200419A, we find a single host galaxy candidate with $P(G_1|x)=70\%$.
For the remaining  FRBs we expect too many  galaxies within their comoving volume to uniquely identify the host galaxy.
% the expected number of galaxies contained within their comoving volume is 5 or less. 
% Optical observations of their localisation regions will likely  prove useful in identifying potential host galaxies. In the case of FRB\,20211024B, where the expected number of galaxies within the comoving volume is between three and nine, we found four galaxies with a photometric redshift below the upper limit. Through a PATH analysis, we found the brightest and lowest redshift galaxy to be the most likely host, with $>90$\% probability. For all other FRBs, we did not find any potential galaxy candidates in the NED and GLADE galaxy databases. 

The dispersion measure of our FRB sample resembles that of the Parkes \citep{champion_five_2016, bhandari_survey_2018},
UTMOST \citep{farah_five_2019}, and MeerTRAP \citep{jankowski_sample_2023} FRBs. The median DM is around 100 to
200\,\pccm\ higher than the CHIME/FRB \citep{chimefrb_collaboration_first_2021}, ASKAP
\citep{shannon_dispersionbrightness_2018, macquart_census_2020}, and DSA-100 \citep{law_deep_2024} samples. For the
ASKAP sample, we cannot reject that the DMs have been drawn from a different intrinsic distribution than the Apertif
FRBs. The larger Apertif DMs allow Apertif to observe a more distant population of FRBs than ASKAP. Furthermore, one of
the Apertif bursts, FRB\,20200719A, has the third largest DM of any FRB published to date, with DM$\sim2778$\,\pccm. Its
derived redshift $z\sim3$ implies that one-off FRBs also emit above 5\,GHz,
a frequency resemblant to magnetar bursts, and could be detected in the S-band. In the future, a large sample of highly dispersed FRBs like this one will help us determine the FRB rate as the Universe evolved.

% One of the bursts, FRB\,20200210A, shows both strong scattering and scintillation. From a joint DM-scattering redshift estimation \citep{cordes_redshift_2022}, we infer its redshift to be $z\sim0.11$. This allows us to determine an upper limit of 2\,kpc on the distance between the FRB and the scattering screen, which is fully consistent with being located within the host galaxy.

We find the observed scintillation bandwidth of most FRBs to be compatible with the expected Milky Way contribution from
the NE2001 model within errors, although in many cases the measured values tend to be larger. Since most FRBs are
detected at high Galactic latitudes, this might be evidence that the MW ISM at high Galactic latitudes is more uniform
than models predict, which is evidence that FRBs could be valuable tools for improving our knowledge on the Galactic electron density distribution.

For 16 out of the 24 FRBs, ARTS triggered the capture of full Stokes data.
The distribution of polarisation characteristics of these FRBs (linear, circular, depolarised)
is similar to that seen in  young, energetic pulsars. \refbf{Several of these FRBs show extreme RM values, which could reach |RM|>1000\,\radsqm\ in the source reference frame, supporting the idea that some one-off FRBs are embedded in extreme magneto-ionic environments.}

A significant fraction of the FRBs display a scattering tail $>0.2$\,ms at 1370\,MHz. Most of these \tscat\ are hence
above the CHIME/FRB scattering sensitivity limit of 10\,ms at 600\,MHz, accounting for the difference in frequency. They
thus reveal a population of highly scattered bursts unlikely to be detected at lower frequencies. Such large scattering
timescales could be produced either in the burst environment or an intervening galaxy within the LoS
\citep{chawla_modeling_2022}. For low redshift FRBs ($z\lesssim1$), the low chances of intersecting a galaxy with a
small impact parameter make a dense circumburst environment the most likely explanation. In the case of FRB\,20200210A,
the large scattering tail allows us to estimate its redshift to be $z\sim0.11$ from a joint scattering-DM  analysis
\citep{cordes_redshift_2022}. From the simultaneous presence of scattering and scintillation, we can put an upper limit
constraint of 2\,kpc between the FRB and a first scattering screen. This confirms the observed scattering was produced
within the host galaxy. Fast Radio Burst surveys at high frequencies, such as ALERT, offer the opportunity of studying
FRBs produced in dense environments that would not be detectable at lower frequencies due to the increased scattering
timescales and thus lower S/N. We hypothesise these dense environments  
to be the star-forming regions or  supernova remnants around FRB-emitting neutron stars. 
This is corroborated by the very high RMs we find in a number of one-off FRBs. 

Roughly $\sim$30\% of the bursts are composed of multiple components. Worth mentioning are the structures of
FRB\,20200216A, FRB\,20200518A and FRB\,20210530A, which display more than two subcomponents each.
None show  evidence
for (quasi-)periodic behaviour similar to that seen in FRB\,20201020A \citep{pastor-marazuela_fast_2023}.
The  $\sim$30\% fraction appears to be consistent with the multi-component fraction observed in the CHIME/FRB baseband
data \citep{chimefrb_collaboration_updating_2023}, which is unexpected since the stronger scattering at lower
frequencies should blur together closely spaced subcomponents.
Interestingly, the multi-component fraction is similar to that seen in pulsars. 
Further morphological studies at different bandwidths will reveal whether an evolution of the multi-component fraction
or separation exists, which will shed light on the emission mechanism of FRBs.

We conclude that high time and frequency resolution such as provided by Apertif, combined with polarisation and localisation capabilities, are essential for making progess in our understanding of FRBs.
Our comprehensive analysis of this set of Apertif discoveries, one of the largest uniform samples at 1.4\,GHz to date, shows striking similarities between FRBs and  young, energetic neutron stars.

% By contrast, only $\sim5$\% of the bursts in the First CHIME/FRB Catalogue \citep{pleunis_fast_2021} are found to have multiple components. We find this difference cannot be explained by the scattering timescales becoming larger at lower frequencies as \tscat$\propto\nu^{-4}$ alone;  the difference might thus be due to the intrinsic FRB properties; either a smaller spacing that cannot be resolved at the CHIME/FRB resolution, or an inherent scarcity of multi-component bursts at lower frequencies. \jvl{Check this par, TBC}

\begin{acknowledgements}
We thank Jim Cordes and Stella Ocker for their useful input; Marten van Kerkwijk and Lucy Oswald for interesting
discussions; Eric Kooistra, Jonathan Hargreaves, Daniel van der Schuur, Jisk Attema,  Wim van Cappellen, Andr{\'e}
Gunst, Tom Oosterloo, Betsey Adams, Stefan Wijnholds and many, many others
for conceiving and building Apertif and ARTS;
Samayra Straal,
Oliver Boersma, Pikky Atri, and Kaustubh Rajwade for help with the observations;
and the referee for their feedback.
This research was supported by 
the European Research Council under the European Union's Seventh Framework Programme (FP/2007-2013)/ERC Grant Agreement No. 617199 (`ALERT'). Further research support was provided by 
Vici research project `ARGO' with project number 639.043.815 
and by 'CORTEX' (NWA.1160.18.316), under the research programme NWA-ORC;
both financed by the Dutch Research Council (NWO). 
Instrumentation development was supported by NWO (grant 614.061.613 `ARTS') and the Netherlands Research School for Astronomy (`NOVA4-ARTS', `NOVA-NW3', and `NOVA5-NW3-10.3.5.14').
PI of aforementioned grants is JvL.
IPM further acknowledges funding from an NWO Rubicon Fellowship, project number 019.221EN.019.
EP further acknowledges funding from an NWO Veni Fellowship.
% SMS acknowledges support from the National Aeronautics and Space Administration (NASA) under grant number NNX17AL74G issued through the NNH16ZDA001N Astrophysics Data Analysis Program (ADAP).
DV acknowledges support from the Netherlands eScience Center (NLeSC) under grant ASDI.15.406.
This work makes use of data from the Apertif system installed at the Westerbork Synthesis Radio Telescope owned by ASTRON. ASTRON, the Netherlands Institute for Radio Astronomy, is an institute of NWO.
\end{acknowledgements}

%------------------------------------------------------

\bibliographystyle{yahapj}
\bibliography{biblio,EP,JVL}

%------------------------------------------------------
\begin{appendix}
\section{Fast Radio Burst properties} \label{app:frb_table}

This appendix contains the Tables~\ref{tab:frb_table} and \ref{tab:frb_table2} where the properties of the 24 FRBs detected during the Apertif survey are summarised.

% python ~/Documents/projects/ARTS/scripts/print_table.py
% \clearpage
\renewcommand{\arraystretch}{1.3}
\begin{sidewaystable}
\centering
\caption{Apertif Fast Radio Burst properties.} \label{tab:frb_table}
\begin{tabular}{@{}
        p{2.4cm} % TNS
        S[table-column-width=2.5cm, table-format=5.8] % MJD
        S[table-column-width=1.5cm, table-format=4.2(3)] % DM
        c % S/N
        c % N
        S[table-column-width=1cm, table-format=1.2(2)] % Flux
        S[table-column-width=1.3cm, table-format=1.3(3)] % Fluence
        S[table-column-width=1.3cm, table-format=2.2(2)] % Width
        S[table-column-width=1.4cm, table-format=2.2(3)] % tscat
        % S[table-column-width=1.4cm, table-format=2.1(4)] % scbw
        % S[table-column-width=1.4cm, table-format=2 \pm 2.0, separate-uncertainty=true] % scbw
        c
        S[table-column-width=1cm, table-format=4(2)] % peak freq
        % S[table-column-width=1cm, table-format=3(2)] % BW
        c % BW
        % S[table-column-width=1.6cm, table-format=2.2(2)]
        % S[table-column-width=1.5cm, table-format=2.2(2)]
        @{}}
\hline\hline
{TNS Name} & {MJD} & {DM} & S/N & {$N$} & {Flux} & {Fluence} & {Width} &  {\tscat} & {\scbw} & {$\nu_{\text{peak}}$$^a$} & {BW$^b$} \\
& {barycentric} & {(\pccm)} & & & {(Jy)} & {(\jyms)} & {(ms)} & {(ms)} & {(MHz)} & {(MHz)} & {(MHz)} \\
\hline

FRB 20190709A & 58673.21877799 & 662.83(8) & 15 &2 & 0.09(2) & 0.52(10) & 2.15(20) &  & $3.1\pm1.5$ & 1370 & >300 \\
FRB 20190926B & 58752.03173103 & 957.1(8) & 13 &1 & 0.11(2) & 5.8(12) & 5.0(10) & 0.78(39) &  & 1370 & >300 \\
FRB 20191020B & 58776.78160336 & 464.8(2) & 13 &1 & 0.10(2) & 2.78(56) & 1.36(9) &  & $7.4\pm5.3$ & 1370 & >300 \\
FRB 20191108A & 58795.83162198 & 587.79(6) & 60 &1 & 0.94(19) & 13.6(27) & 0.63(1) &  &  & 1370 & >300 \\
FRB 20191109A & 58796.54963359 & 531.2(2) & 13 &2 & 0.15(3) & 4.08(82) & 2.28(13) &  & $4.5\pm1.3$ & 1370 & >300 \\
FRB 20200210A & 58889.30929416 & 439.7(5) & 38 &1 & 0.16(3) & 55.7(11) & 29.63(55) & 12.65(26) & $0.8\pm0.1$ & 1449(1) & 169(6) \\
FRB 20200213A & 58892.02247546 & 1017.7(2) & 18 &1 & 0.26(5) & 4.39(88) & 1.02(3) &  & $9.3\pm0.8$ & 1361(2) & 145(7) \\
FRB 20200216A & 58895.44348462 & 478.7(2) & 27 &3 & 0.21(4) & 9.2(18) & 7.22(8) &  &  & 1520 & >300 \\
FRB 20200224B & 58903.65955787 & 1450(1) & 17 &1 & 0.12(2) & 15.3(31) & 10.1(11) & 3.73(42) &  & 1520 & >300 \\
FRB 20200321A & 58929.82902069 & 914.71(4) & 13 &2 & 0.05(1) & 4.48(90) & 1.79(19) &  &  & 1436(7) & 226(27) \\
FRB 20200322A & 58930.28670855 & 1290.3(10) & 19 &1 & 0.07(1) & 9.9(20) & 12.8(10) & 4.16(44) & $4.7\pm2.3$ & 1406(7) & 277(33) \\
FRB 20200323C & 58931.38371175 & 833.4(2) & 28 &1 & 0.10(2) & 6.6(13) & 3.89(25) & 1.33(9) & $5.8\pm1.2$ & 1520 & >300 \\
FRB 20200419A & 58958.07914529 & 248.5(3) & 19 &1 & 0.59(12) & 7.4(15) & 0.58(2) &  & $7.5\pm0.8$ & 1370 & >300 \\
FRB 20200514A & 58983.36070020 & 1406.2(2) & 18 &1 & 0.14(3) & 5.2(10) & 2.18(12) &  &  & 1370 & >300 \\
FRB 20200516A & 58985.97336036 & 361.1(4) & 10 &1 & 0.05(1) & 2.59(52) & 2.24(19) &  &  & 1520 & >300 \\
FRB 20200518A & 58987.73145901 & 246.5(1) & 15 &4 & 0.13(3) & 4.67(93) & 3.95(17) &  & $5.2\pm2.0$ & 1370 & >300 \\
FRB 20200523A & 58992.14519881 & 444(9) & 10 &1 & 0.04(1) & 6.9(14) & 50.1(96) & 12.47(356) &  & 1370 & >300 \\
FRB 20200719A & 59049.57422915 & 2778(6) & 13 &1 & 0.05(1) & 15.1(30) & 51.8(58) & 20.96(253) &  & 1459(11) & 259(49) \\
FRB 20201020A & 59142.50645121 & 398.59(8) & 53 &5 & 0.36(7) & 13.5(27) & 2.13(2) &  & $4.1\pm0.4$ & 1370 & >300 \\
FRB 20210124A & 59238.42689790 & 868.25(7) & 24 &1 & 0.11(2) & 4.61(92) & 2.87(24) & 0.65(9) & $1.7\pm0.3$ & 1370 & >300 \\
FRB 20210127A & 59241.36047289 & 891.2(1) & 35 &1 & 0.43(9) & 6.6(13) & 0.83(2) &  &  & 1370 & >300 \\
FRB 20210317A & 59290.29571036 & 466.5(1) & 33 &1 & 0.27(5) & 6.0(12) & 1.14(5) & 0.22(3) & $2.4\pm0.3$ & 1370 & >300 \\
FRB 20210530A & 59364.16761389 & 1000.27(9) & 17 &4 & 0.29(6) & 18.7(37) & 4.15(14) &  &  & 1370 & >300 \\
FRB 20211024B & 59511.38821806 & 509.4(1) & 80 &1 & 0.45(9) & 16.6(33) & 1.45(1) &  &  & 1370 & >300 \\

\hline
\end{tabular}
\tablefoot{\\
$^a$ Central frequency $\nu_{\text{peak}}$ for broadband bursts assumed to be the central frequency, 1370\,MHz. For bursts with power law spectrum, 1520\,MHz and 1220\,MHz if they peak at the top or the bottom of the band respectively. Other values for narrowband bursts fitted to a Gaussian spectrum.\\
$^b$ $BW$ is the FWTM of the bursts with a Gaussian spectrum or $>300$\,MHz for broadband bursts.\\
}
\vspace{9cm}
\end{sidewaystable}

%--------------------------------
\clearpage
\begin{sidewaystable}
\centering
\vspace{9cm}
\caption{Apertif Fast Radio Burst properties, continued.} \label{tab:frb_table2}
\begin{tabular}{@{}
        p{2.4cm} % TNS
        cc % RA, DEC
        S[table-column-width=0.8cm, table-format=4] % a
        S[table-column-width=0.8cm, table-format=2.1] % b
        S[table-column-width=0.8cm, table-format=3.2] % theta
        S[table-column-width=0.8cm, table-format=2.2] % Loc area
        cc % CB, SB
        S[table-column-width=0.4cm, table-format=3] % DM MW
        S[table-column-width=0.01cm] % +
        S[table-column-width=0.3cm, table-format=3] % DM halo
        c % z_max
        % S[table-column-width=0.6cm, table-format=1.2] % z_max
        c % si
        c % alpha
        c % peak freq z
        c % E_iso
        % S[table-column-width=1.2cm, table-format=2.2(2)]
        % S[table-column-width=1.2cm, table-format=2.2(2)]
        % S[table-column-width=1.5cm, table-format=2.2(2)]
        @{}}
% \begin{tabular}{ccccccccc}
\hline\hline
% {TNS Name} & {RA} & {DEC} & {\scriptsize{Loc. area}} & \multicolumn{3}{c}{DM$_{\text{MW+halo}}$$^a$} & {$z_{\text{max}}$$^b$} & {CB} & {SB} & {$\Gamma$$^c$} & {$\alpha$$^d$} & {$\nu_{0,\text{max}}$$^e$} & {$E_{\text{iso}}$$^f$} \\
% & & & {\scriptsize{(arcmin$^2$)}} & \multicolumn{3}{c}{(\pccm)} & & & & & & {(MHz)} & {(erg)} \\
{TNS Name} & {RA$^a$} & {DEC$^a$} & a$^a$ & b$^a$ & $\theta^a$ & {\scriptsize{Loc. area}} & {CB} & {SB} & \multicolumn{3}{c}{DM$_{\text{MW+halo}}$$^b$} & {\zmacquart$^c$} & {$\Gamma^d$} & {$\alpha^e$} & {$\nu_{0,\text{max}}^f$} & {$E_{\text{iso}}^g$} \\
& & & {\scriptsize{(arcsec)}} & {\scriptsize{(arcsec)}} & {($^{\circ}$)} &{\scriptsize{(arcmin$^2$)}} & & & \multicolumn{3}{c}{(\pccm)} & & & & {(GHz)} & {(erg)} \\
\hline

FRB 20190709A & 01h39m19.7s & +32d03m31.3s & 223 & 3.3 & 96.90 & 0.64 & 10 & 36 & 52&{+}&32 & 0.68$^{+0.16}_{-0.36}$ & -- & -- & 2.3 & $2.0\times10^{39}$ \\
FRB 20190926B & 01h42m06.0s & +30d58m05.0s & 1170 & 13.4 & 92.61 & 13.64 & 7 & 39 & 51&{+}&32 & 1.07$^{+0.24}_{-0.54}$ & -- & -- & 2.8 & $5.5\times10^{40}$ \\
FRB 20191020B & 20h30m39.0s & +62d17m43.0s & 555 & 12.5 & 82.88 & 6.05 & 5 & 14 & 102&{+}&44 & 0.32$^{+0.10}_{-0.18}$ & -- & -- & 1.8 & $2.3\times10^{39}$ \\
FRB 20191108A & 01h33m31.4s & +31d41m33.4s & 226 & 5.9 & 108.69 & 1.16 & 21 & 37 & 51&{+}&32 & 0.58$^{+0.14}_{-0.32}$ & -- & -- & 2.2 & $3.8\times10^{40}$ \\
FRB 20191109A & 20h38m11.1s & +61d43m10.8s & 1084 & 13.6 & 144.33 & 13.25 & 18 & 44 & 105&{+}&44 & 0.42$^{+0.10}_{-0.24}$ & -- & -- & 1.9 & $5.9\times10^{39}$ \\
FRB 20200210A & 18h53m59.4s & +46d18m57.4s & 148 & 6.0 & 109.20 & 0.78 & 9 & 37 & 70&{+}&47 & 0.36$^{+0.10}_{-0.22}$ & -- & 13.8$\pm0.9$ & 2.0 & $3.3\times10^{40}$ \\
FRB 20200213A & 09h24m56.9s & +76d49m31.9s & 144 & 7.5 & 78.65 & 0.94 & 9 & 48 & 45&{+}&31 & 1.15$^{+0.24}_{-0.58}$ & -- & -- & 2.9 & $2.3\times10^{40}$ \\
FRB 20200216A & 22h08m24.7s & +16d35m34.6s & 216 & 12.5 & 95.68 & 2.34 & 5 & 30 & 48&{+}&41 & 0.44$^{+0.12}_{-0.24}$ & 10.6$\pm$2.5 & -- & 2.2 & $1.5\times10^{40}$ \\
FRB 20200224B & 02h20m24.7s & +19d26m22.1s & 1037 & 10.2 & 88.92 & 9.26 & 14 & 33 & 44&{+}&31 & 1.65$^{+0.36}_{-0.76}$ & 4.4$\pm$1.3 & 1.5$\pm2.7$ & 4.0 & $3.3\times10^{41}$ \\
FRB 20200321A & 09h38m42.0s & +76d10m45.4s & 190 & 9.0 & 113.74 & 1.49 & 5 & 37 & 45&{+}&31 & 1.00$^{+0.24}_{-0.50}$ & -- & -- & 2.9 & $2.8\times10^{40}$ \\
FRB 20200322A & 22h13m02.5s & +15d16m50.1s & 424 & 10.3 & 103.79 & 3.83 & 6 & 29 & 46&{+}&40 & 1.47$^{+0.32}_{-0.70}$ & -- & 4.5$\pm2.3$ & 3.5 & $1.6\times10^{41}$ \\
FRB 20200323C & 22h12m01.0s & +15d46m21.9s & 300 & 19.0 & 91.42 & 4.97 & 12 & 18 & 47&{+}&40 & 0.90$^{+0.22}_{-0.46}$ & 5.2$\pm$1.1 & 3.5$\pm3.0$ & 2.9 & $4.5\times10^{40}$ \\
FRB 20200419A & 19h00m34.2s & +81d43m20.5s & 212 & 7.0 & 132.69 & 1.29 & 32 & 56 & 55&{+}&35 & 0.08$^{+0.04}_{-0.06}$ & -- & -- & 1.5 & $3.5\times10^{38}$ \\
FRB 20200514A & 01h47m32.5s & +64d18m37.3 & 2374 & 12.1 & 105.49 & 25.01 & 2 & 40 & 181&{+}&44 & 1.35$^{+0.30}_{-0.66}$ & -- & -- & 3.2 & $7.7\times10^{40}$ \\
FRB 20200516A & 18h56m35.9s & +46d49m37.4 & 1121 & 10.3 & 134.05 & 10.08 & 18 & 22 & 73&{+}&47 & 0.24$^{+0.06}_{-0.14}$ & 6.9$\pm$2.6 & -- & 1.9 & $1.2\times10^{39}$ \\
FRB 20200518A & 09h36m45.3s & +77d22m36.8s & 164 & 11.6 & 87.40 & 1.67 & 25 & 68 & 45&{+}&31 & 0.10$^{+0.04}_{-0.08}$ & -- & -- & 1.5 & $3.5\times10^{38}$ \\
FRB 20200523A & 18h55m37.1s & +47d07m38.1s & 513 & 14.8 & 78.28 & 6.64 & 24 & 19 & 72&{+}&46 & 0.36$^{+0.10}_{-0.20}$ & -- & 12.0$\pm5.8$ & 1.9 & $7.2\times10^{39}$ \\
FRB 20200719A & 09h18m41.1s & +77d23m06.8s & 307 & 14.4 & 78.45 & 3.93 & 22 & 20 & 46&{+}&31 & 3.26$^{+0.62}_{-1.35}$ & -- & 11.1$\pm4.5$ & 6.2 & $9.0\times10^{41}$ \\
FRB 20201020A & 13h51m24.7s & +49d02m03.6s & 115 & 8.3 & 80.88 & 0.84 & 29 & 17 & 29&{+}&32 & 0.36$^{+0.10}_{-0.22}$ & -- & -- & 1.9 & $1.4\times10^{40}$ \\
FRB 20210124A & 19h41m25.3s & +58d55m02.1s & 147 & 7.0 & 99.54 & 0.89 & 11 & 48 & 80&{+}&44 & 0.90$^{+0.22}_{-0.46}$ & -- & 4.4$\pm3.3$ & 2.6 & $3.1\times10^{40}$ \\
FRB 20210127A & 16h49m56.4s & +26d37m30.0s & 155 & 5.5 & 84.89 & 0.75 & 21 & 34 & 41&{+}&44 & 0.98$^{+0.24}_{-0.50}$ & -- & -- & 2.7 & $5.3\times10^{40}$ \\
FRB 20210317A & 19h36m27.4s & +59d51m50.7s & 129 & 4.8 & 97.13 & 0.54 & 23 & 35 & 76&{+}&43 & 0.38$^{+0.10}_{-0.22}$ & -- & -- & 1.9 & $7.0\times10^{39}$ \\
FRB 20210530A & 22h08m56.0s & +16d32m25.4s & 185 & 18.4 & 95.56 & 2.97 & 6 & 36 & 47&{+}&40 & 1.11$^{+0.26}_{-0.54}$ & -- & -- & 2.9 & $1.9\times10^{41}$ \\
FRB 20211024B & 13h20m34.5s & +42d29m20.7s & 123 & 7.2 & 107.23 & 0.77 & 18 & 40 & 25&{+}&31 & 0.52$^{+0.12}_{-0.30}$ & -- & -- & 2.1 & $3.7\times10^{40}$ \\

\hline
\end{tabular}
\tablefoot{\\
$^a$ Properties of the localisation region fitted to an ellipse. RA and DEC are the central coordinates of the ellipse, a and b the semi-major an semi-minor axes, respectively, of the 99\% contour. The angle $\theta$ is the ellipse angle measured from West (lowest RA) through the North, using the same convention as \href{https://sites.google.com/cfa.harvard.edu/saoimageds9}{ds9}.\\
$^b$ Milky Way DM from NE2001 model \citep{cordes_ne2001i_2003} and halo DM from \citep{yamasaki_galactic_2020}.\\
$^c$ \zmax computed from \citep{zhang_fast_2018}.
$^d$ Spectral index $\Gamma$ of bursts with power law spectrum.\\
$^e$ Scattering index $\alpha$ of bursts with scattering tail, defined by \tscat$\propto\nu^{-\alpha}$.\\
$^f$ Rest frame frequency upper limit $\nu_{0,\text{max}}$.\\
$^g$ Isotropic energy upper limit.
}
\end{sidewaystable}

%-------------------------------
\clearpage
\section{Host galaxy candidates} \label{app:host_gal}

In Table~\ref{tab:host_galaxies}, we present the galaxies identified within the error regions for those FRBs with $\leq5$ host galaxy candidates, as well as the resulting association probabilities from the \texttt{PATH} analysis.

\begin{table*}[b] \label{tab:host_galaxies}
    \centering
    \caption{Host galaxy candidates and association probability for FRBs with $\leq5$ galaxies within the error region.}
    \begin{tabular}{c c c c c c c c}
    \hline\hline
    FRB & ID & Galaxy name & $z_{\text{phot}}$ & $r_r$ (") & $m_r$ & $P(O)$ & $P(O|x)$ \\
    \hline
    \multirow{4}{*}{FRB\,20200210A}
    & G1 & PSO J283.5110+46.3322 & $0.11\pm0.01$ & 6.1 & 18.0 & 0.924 & 0.580 \\
    & G2 & PSO J283.5087+46.3348 & $0.40\pm0.08$ & 3.3 & 21.0 & 0.040 & 0.003 \\
    & G3 & PSO J283.5028+46.3205 & $0.46\pm0.06$ & 2.9 & 21.5 & 0.025 & 0.289 \\
    & U &  &  &  &  & 0.01 & 0.129 \\
    \hline
    \multirow{5}{*}{FRB\,20200216A}
    & G1 & PSO J332.0996+16.6087 & $0.49\pm0.07$ & 2.6 & 21.4 & 0.325 & 0.424 \\
    & G2 & PSO J332.0953+16.5341 & $0.51\pm0.20$ & 2.5 & 22.3 & 0.138 & 0.004 \\
    & G3 & PSO J332.1018+16.6091 & $0.52\pm0.09$ & 2.4 & 21.6 & 0.265 & 0.370 \\
    & G4 & PSO J332.0976+16.5926 & $0.56\pm0.32$ & 2.4 & 22.1 & 0.171 & 0.139 \\
    & U &  &  &  &  & 0.1 & 0.064 \\
    \hline
    \multirow{2}{*}{FRB\,20200419A}
    & G1 & PSO J285.2463+81.7361 & $0.15\pm0.03$ & 3.5 & 19.6 & 0.999 & 0.704 \\
    & U &  &  &  &  & 0.001 & 0.296 \\
    \hline
    \multirow{3}{*}{FRB\,20200518A}
    & G1 & PSO J144.1948+77.3261 & $0.08\pm0.03$ & 4.6 & 19.9 & 0.350 & 0.415 \\
    & G2 & PSO J144.2179+77.3481 & $0.08\pm0.02$ & 4.6 & 19.4 & 0.640 & 0.557 \\
    & U &  &  &  &  & 0.01 & 0.028 \\
    \hline
    \multirow{3}{*}{FRB\,20200719A}
    & G1 & PSO J139.6135+77.4537 & $0.81\pm0.96$ & 2.4 & 22.7 & 0.048 & $1.8\times10^{-4}$ \\
    & G2 & PSO J139.6677+77.3898 & $1.45\pm0.51$ & 3.7 & 22.6 & 0.052 & 0.101 \\
    & U &  &  &  &  & 0.9 & 0.899 \\
    \hline
    \multirow{4}{*}{FRB\,20210317A}
    & G1 & PSO J294.1082+59.8453 & $0.15\pm0.05$ & 3.2 & 20.0 & 0.391 & 0.541 \\
    & G2 & PSO J294.1097+59.8285 & $0.15\pm0.06$ & 4.5 & 19.9 & 0.412 & 0.350 \\
    & G3 & PSO J294.1028+59.8412 & $0.30\pm0.05$ & 2.4 & 20.7 & 0.186 & 0.088 \\
    & U &  &  &  &  & 0.01 & 0.021 \\
    \hline
    \multirow{6}{*}{FRB\,20211024B}
    & G1 & PSO J200.1431+42.4925 & $0.24\pm0.07$ & 2.9 & 20.5 & 0.194 & 0.235 \\
    & G2 & PSO J200.1445+42.4911 & $0.24\pm0.02$ & 5.2 & 19.4 & 0.604 & 0.609 \\
    & G3 & PSO J200.1529+42.5097 & $0.34\pm0.35$ & 1.0 & 24.5 & 0.006 & $3.3\times10^{-4}$ \\
    & G4 & PSO J200.1548+42.5192 & $0.56\pm0.27$ & 2.7 & 21.8 & 0.052 & $4.2\times10^{-4}$ \\
    & G5 & PSO J200.1392+42.4859 & $0.58\pm0.18$ & 1.9 & 22.0 & 0.045 & 0.042 \\
    & U &  &  &  &  & 0.1 & 0.113 \\
    \hline
    \end{tabular}
    \tablefoot{The galaxies are sorted by increasing redshift. Their ID is the same identifier as used in the text. The galaxy name is given by the Pan-STARRS source catalogue. The photometric redshift and errors are given by PS1-STRM. $r_r$ and $m_r$ are respectively the Kron radius and the magnitude in the $r$ band as given by Pan-STARRS. $P(O)$ and $P(O|x)$ are respectively the \texttt{PATH} prior and posterior probability that the galaxy is associated with the FRB. For each FRB, U represents the unseen galaxies. \ipm{Add separation from error region.}}
\vspace{55mm}  %%% Moving the Table up on the empty page
\end{table*}

% $6.5\times10^{-4}$

%-------------------------------

\clearpage
\section{Stokes data} \label{app:stokes}
In this appendix, we show the Stokes $I,~Q,~U,$ and $V$ data of the 14 new FRBs with Stokes data dumps, in Fig.~\ref{fig:stokes}. When possible, we have performed an RM measurement, shown in Figures~\ref{fig:RM_FRB200213} to \ref{fig:RM_FRB210530}.

\renewcommand{\sisetup}{separate-uncertainty}
\begin{table*}[bh]
    \centering
    \caption{Apertif Fast Radio Burst polarisation properties.} \label{tab:stokes}
    \begin{tabular}{
        c % TNS Name
        % cccc
        S[table-column-width=1.5cm, table-format=2 \pm 2.0, separate-uncertainty=true] % L
        S[table-column-width=1.5cm, table-format=2 \pm 2.0, separate-uncertainty=true]  % V
        S[table-column-width=2cm, table-format=4.1 \pm 2.1, separate-uncertainty=true]  % RM
        S[table-column-width=2cm, table-format=3 \pm 2.0, separate-uncertainty=true]  % RM_MW
        c  % RM_host
    }
    \hline\hline
    TNS Name & $L/I$ & $|V|/I$ & \rmobs & \rmmw & \rmhost\\
    & \multicolumn{1}{c}{(\%)} & \multicolumn{1}{c}{(\%)} & \multicolumn{1}{c}{(\radsqm)} & \multicolumn{1}{c}{(\radsqm)} & \multicolumn{1}{c}{(\radsqm)} \\
    \hline
FRB 20191020B & 31 \pm 7 & 5 \pm 7 &  & 24 \pm 17 & \\
FRB 20191108A & 86 \pm 2 & 0 \pm 1 & 473.1 \pm 2.2 & -63 \pm 13 & 1400$^{+300}_{-500}$\\
FRB 20200213A & 10 \pm 3 & 8 \pm 6 & 300.3 \pm 2.1 & -17 \pm 7 & 1500$^{+400}_{-500}$\\
FRB 20200216A & 38 \pm 6 & 11 \pm 4 & -2051.0 \pm 5.8 & -36 \pm 10 & $-$4200$^{+1300}_{-800}$\\
FRB 20200321A & 17 \pm 5 & 13 \pm 9 &  & -16 \pm 6 & \\
FRB 20200322A & 3 \pm 6 & 14 \pm 9 &  & -28 \pm 9 & \\
FRB 20200323C & 6 \pm 3 & 0 \pm 3 &  & -30 \pm 6 & \\
FRB 20200419A & 77 \pm 6 & 4 \pm 6 &  & 9 \pm 8 & \\
FRB 20200514A & 51 \pm 5 & 21 \pm 9 & 979.8 \pm 20.5 & -215 \pm 68 & 6500$^{+2000}_{-3200}$\\
FRB 20200516A & 17 \pm 13 & 14 \pm 7 &  & 1 \pm 12 & \\
FRB 20200518A & 72 \pm 12 & 29 \pm 17 &  & -17 \pm 7 & \\
FRB 20201020A & 36 \pm 2 & 8 \pm 5 & 125.1 \pm 37.5 & 17 \pm 3 & 200$^{+160}_{-140}$\\
FRB 20210124A & 86 \pm 8 & 15 \pm 12 &  & 32 \pm 18 & \\
FRB 20210127A & 53 \pm 2 & 4 \pm 5 & 123.5 \pm 0.4 & 35 \pm 9 & 340$^{+130}_{-160}$\\
FRB 20210317A & 50 \pm 5 & 3 \pm 4 & -251.7 \pm 1.3 & 26 \pm 20 & $-$530$^{+160}_{-130}$\\
FRB 20210530A & 52 \pm 4 & 0 \pm 7 & -125.1 \pm 4.6 & -37 \pm 10 & $-$390$^{+190}_{-160}$\\
    \hline
    \end{tabular}
    \tablefoot{Only FRBs with triggered full-Stokes data dumps are listed here. To ensure uniform layout, insignificant
      digits are sometimes also included.
      \ipm{Add columns with linear and unpolarised calibrators?}\\
    % $^a$ Linear polarisation fraction.\\
    % $^b$ Circular polarisation fraction.\\
    % $^c$ Faraday rotation measure obtained through RM synthesis.\\
    % $^d$ Expected Milky Way contribution of the RM and standard deviation in the direction of the FRB according to \cite{hutschenreuter_galactic_2022}.
    }
\vspace{110mm}  %%% Moving the Table up on the empty page
\end{table*}

% arts041:/home/arts/pastor/scripts/plot_iquv.ipynb
\begin{figure*} 
    \centering
    \includegraphics[width=0.24\textwidth]{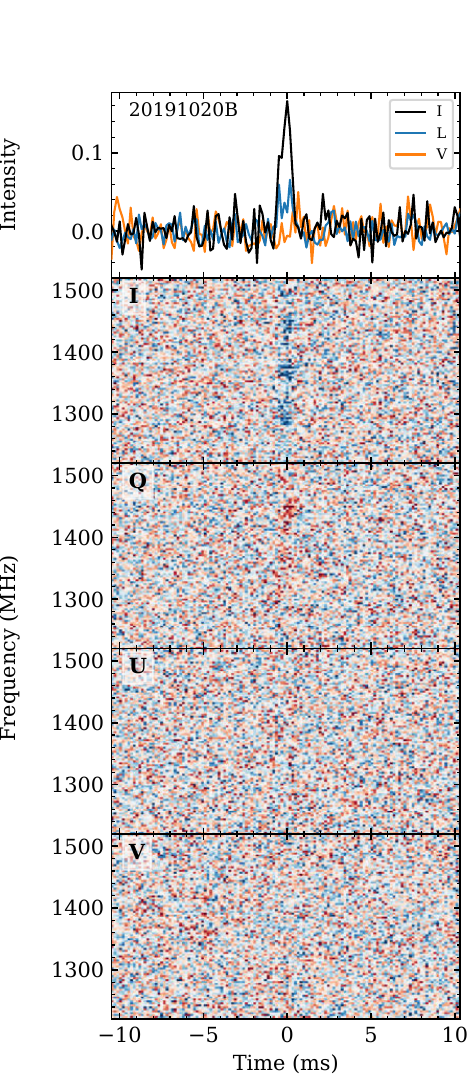}
    \includegraphics[width=0.24\textwidth]{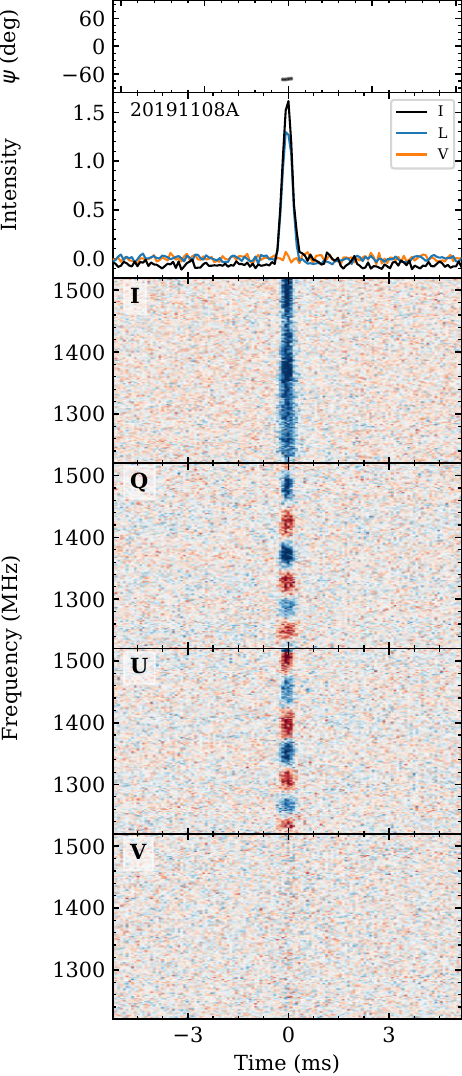}
    \includegraphics[width=0.24\textwidth]{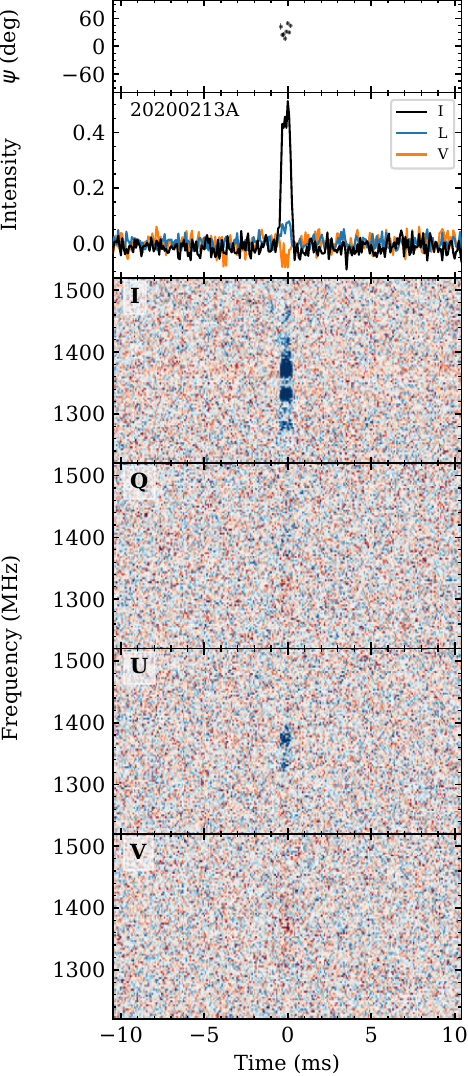}
    \includegraphics[width=0.24\textwidth]{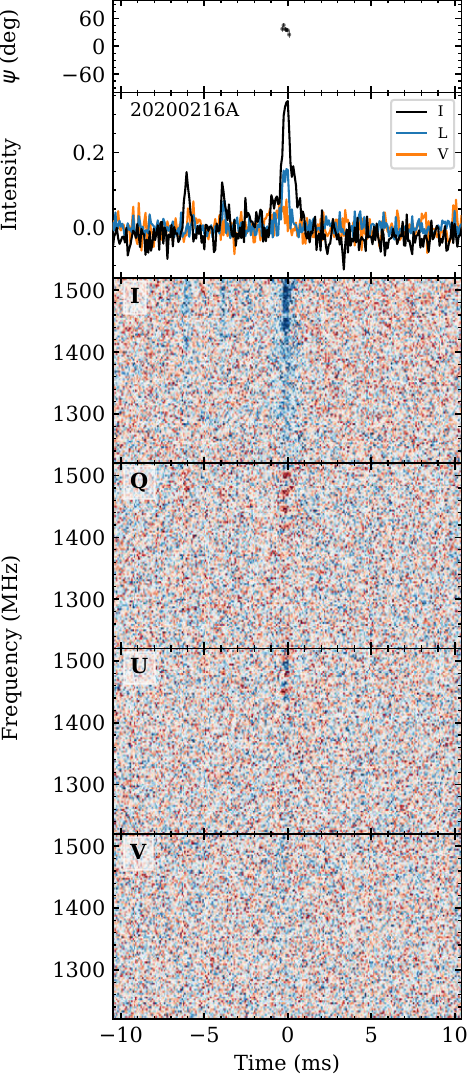}\\
    \includegraphics[width=0.24\textwidth]{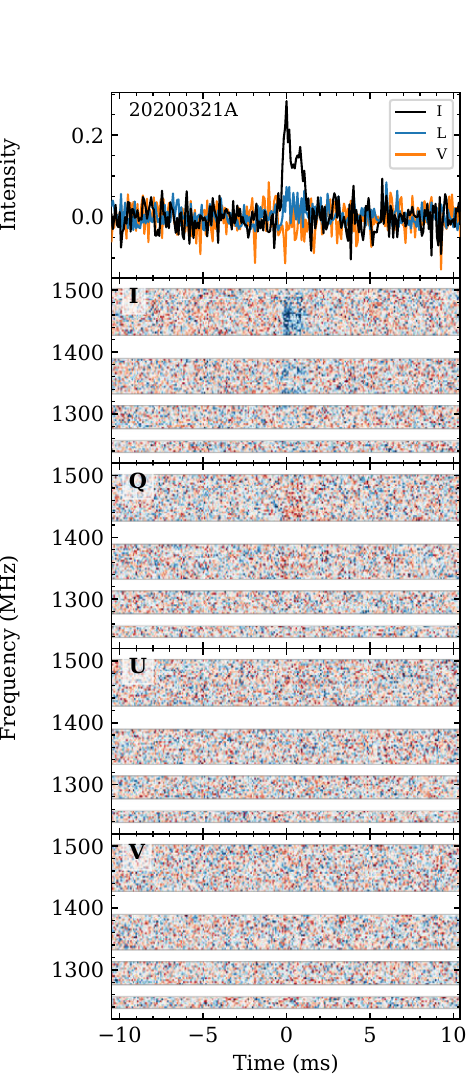}
    \includegraphics[width=0.24\textwidth]{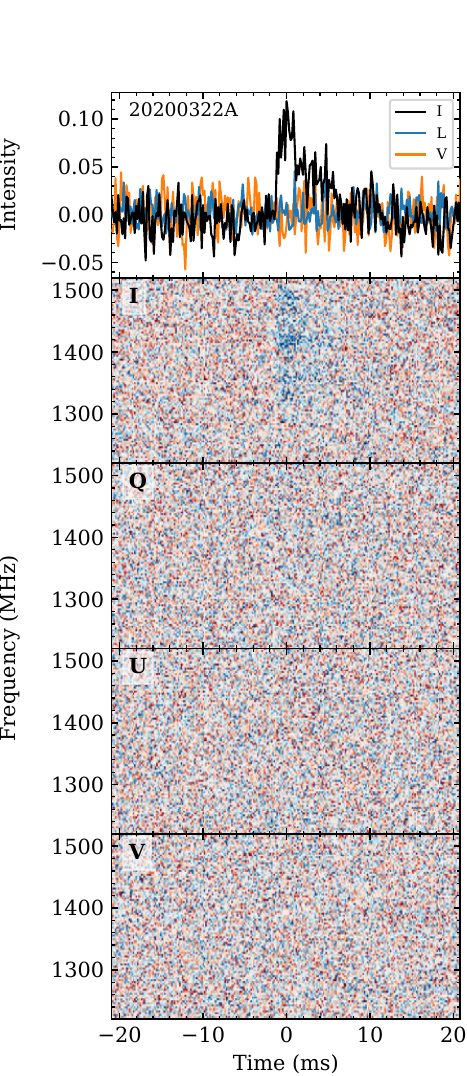}
    \includegraphics[width=0.24\textwidth]{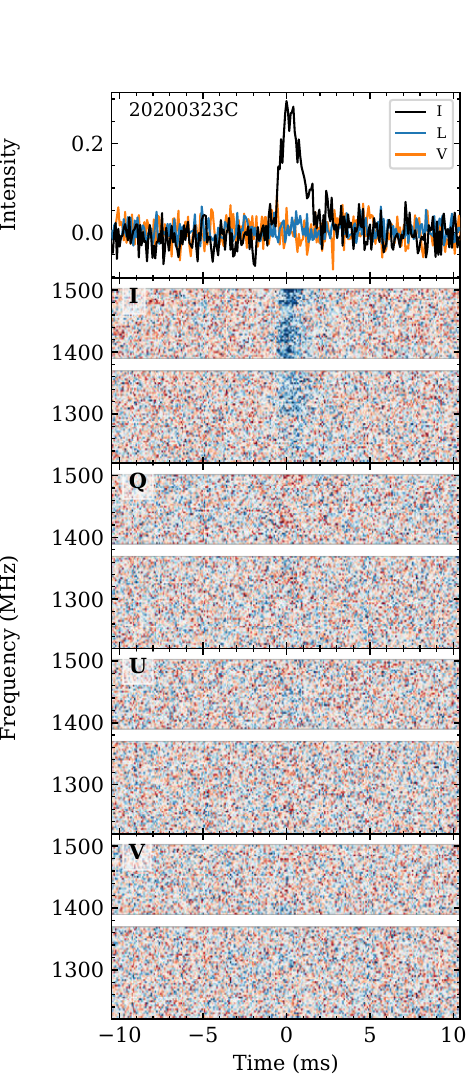}
    \includegraphics[width=0.24\textwidth]{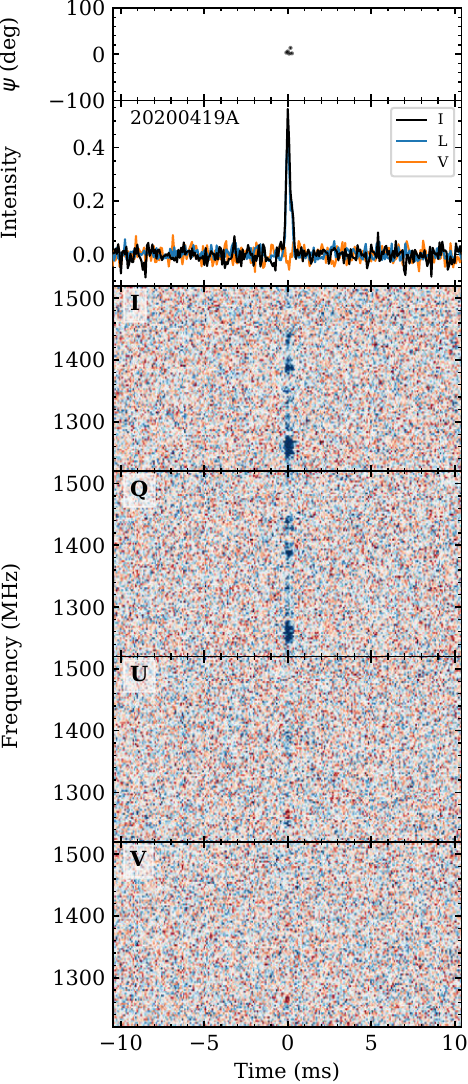}
    \caption{Stokes IQUV data of triggered FRBs. Each top panel shows the pulse profile of total intensity (Stokes I - black), linear polarisation ($\sqrt{Q^2+U^2}$ - blue) after Faraday de-rotation when an RM is known, and circular polarisation (Stokes V - orange). The bottom panels show respectively Stokes I, Q, U, and V, without Faraday de-rotation. Blue indicates a positive value and red a negative value. The FRB name is indicated on the top right corner.}
    \label{fig:stokes}
\end{figure*}

\addtocounter{figure}{-1}
\begin{figure*}
    \centering
    \includegraphics[width=0.24\textwidth]{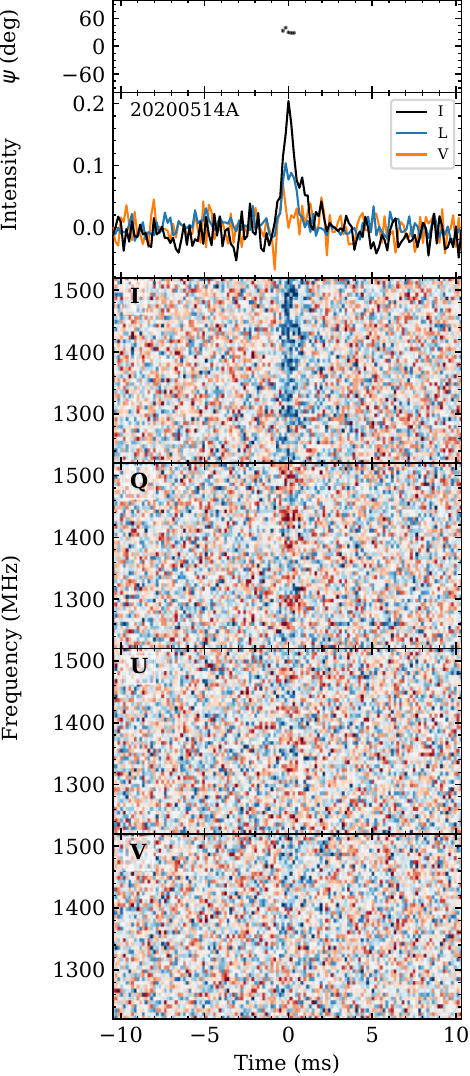}
    \includegraphics[width=0.24\textwidth]{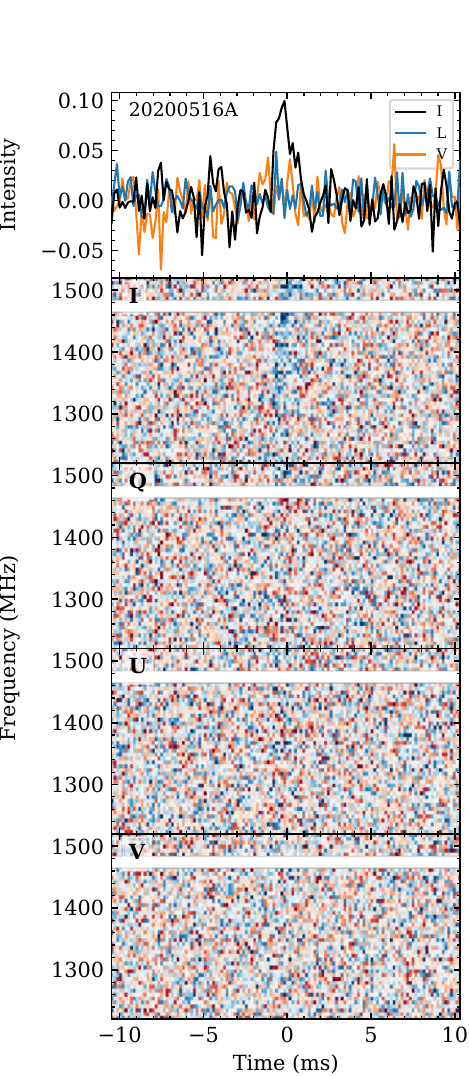}
    \includegraphics[width=0.24\textwidth]{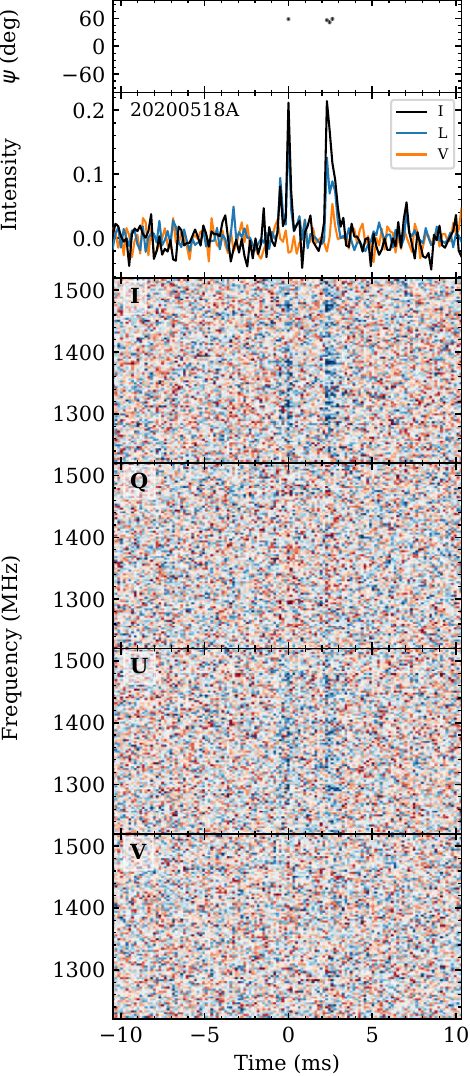}
    \includegraphics[width=0.24\textwidth]{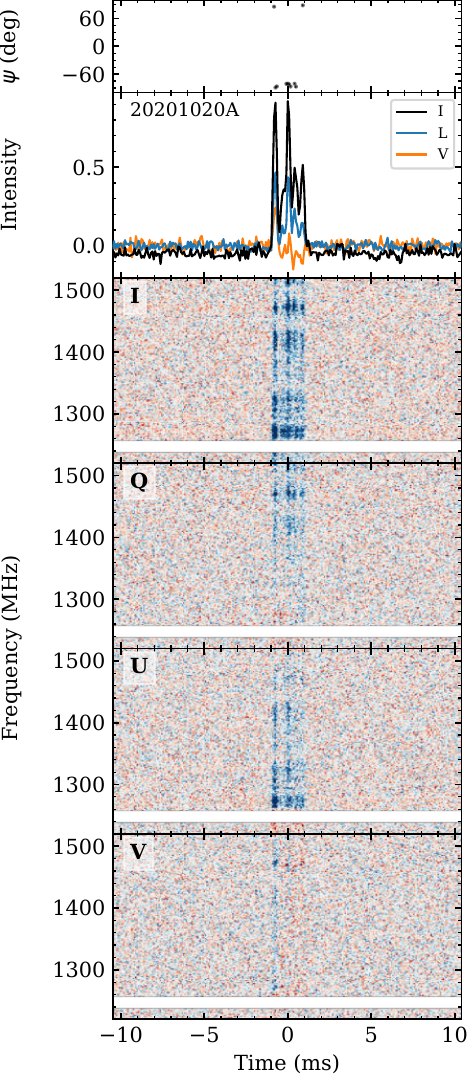}\\
    \includegraphics[width=0.24\textwidth]{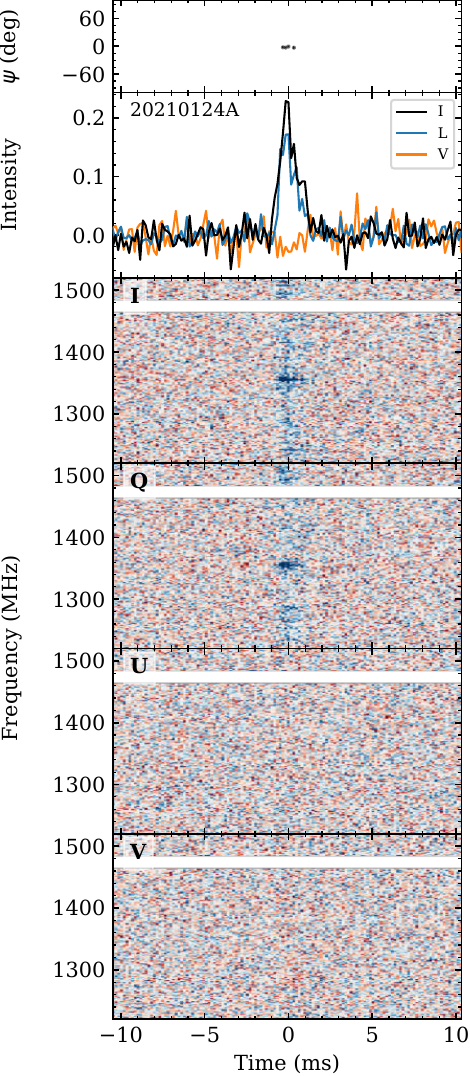}
    \includegraphics[width=0.24\textwidth]{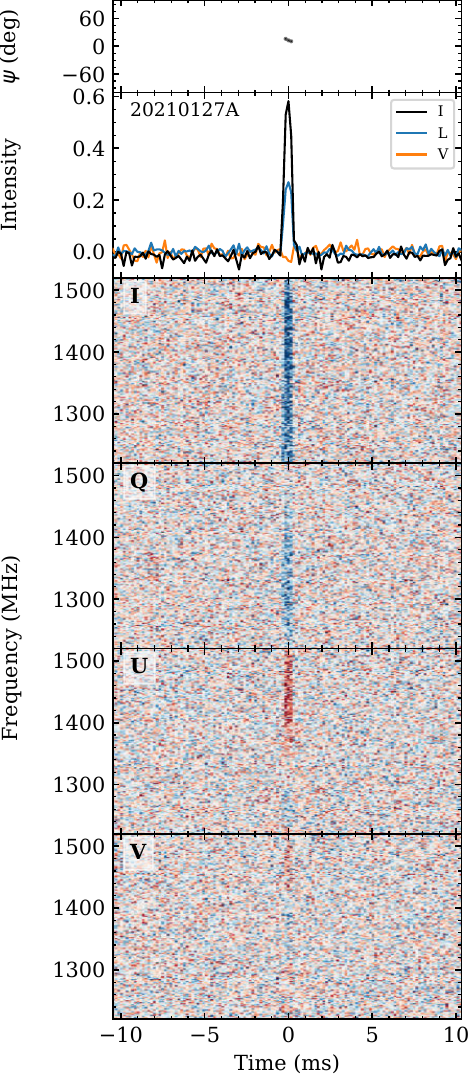}
    \includegraphics[width=0.24\textwidth]{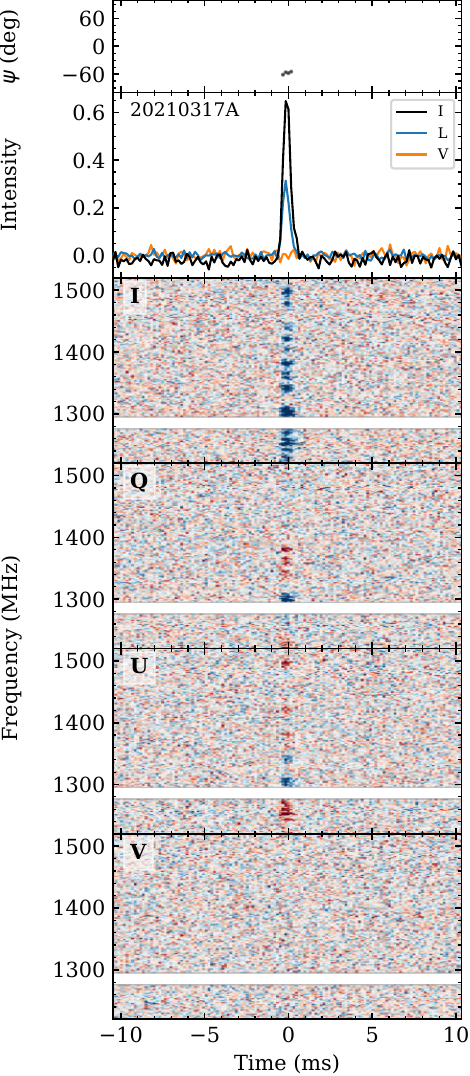}
    \includegraphics[width=0.24\textwidth]{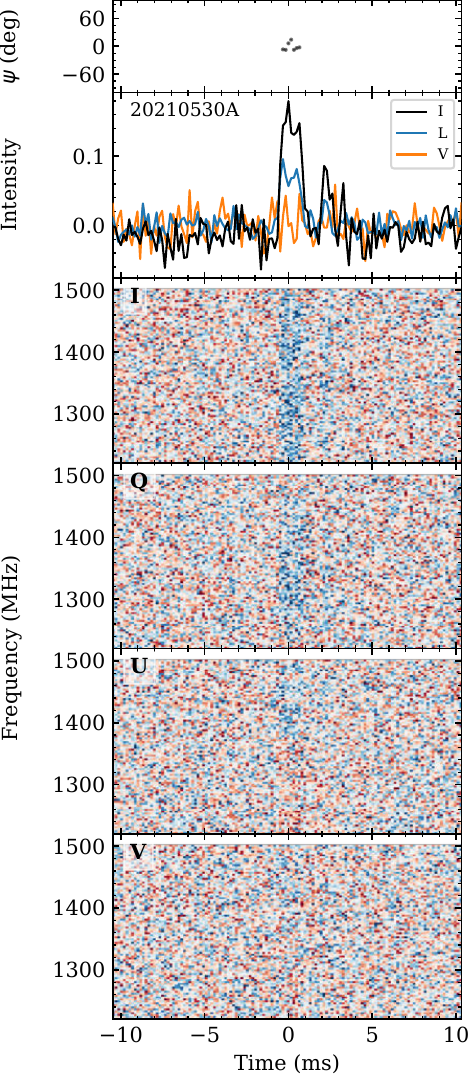} 
    \caption{(Continued)}
\end{figure*}

\begin{figure}
    \centering
    \includegraphics[width=\hsize]{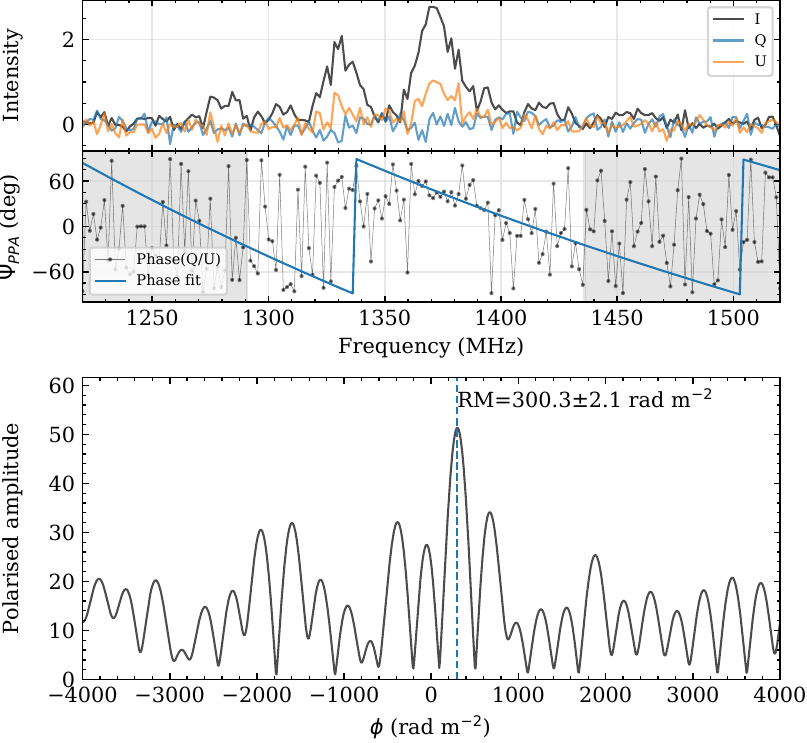}\\
    \caption{Rotation measure analysis of FRB\,20200213A. The frequency channels between 1291 and 1436\,MHz were used for the RM synthesis, since they are contained within the full width at tenth maximum of the spectrum fitted to a Gaussian.}
    \label{fig:RM_FRB200213}
\end{figure}

\begin{figure}
    \centering
    \includegraphics[width=\hsize]{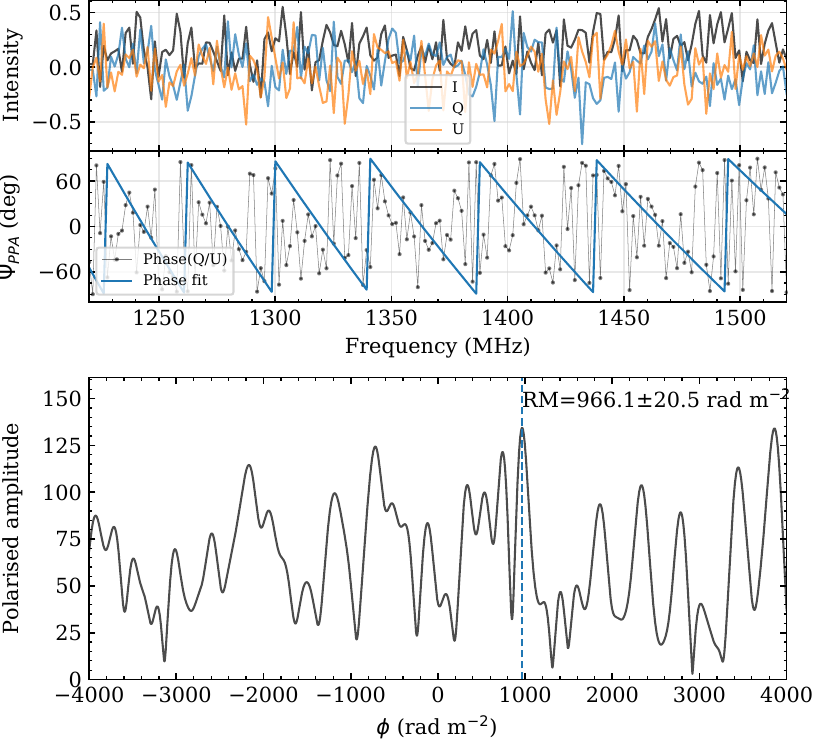}\\
    \caption{Rotation measure analysis of FRB\,20200514A. The whole bandwidth was used. }
    \label{fig:RM_FRB200514}
\end{figure}

\begin{figure}
    \centering
    \includegraphics[width=\hsize]{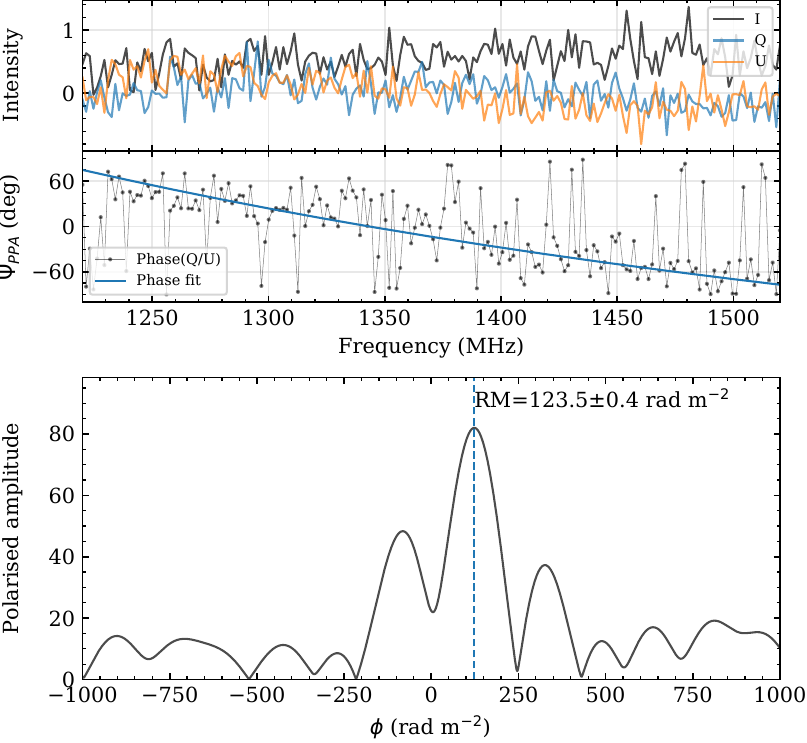}\\
    \caption{Rotation measure analysis of FRB\,20210127A. The whole bandwidth was used.}
    \label{fig:RM_FRB210127}
\vspace{5mm}    
\end{figure}

\begin{figure}
    \centering
    \includegraphics[width=\hsize]{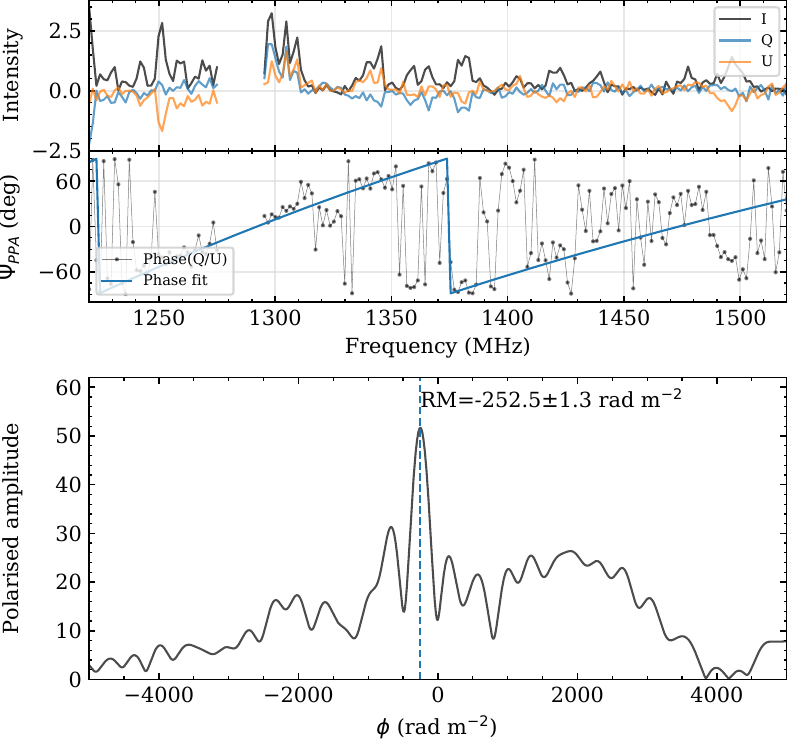}\\
    \caption{Rotation measure analysis of FRB\,20210317A. The whole bandwidth was used.}
    \label{fig:RM_FRB210317}
\end{figure}

\begin{figure}
    \centering
    \includegraphics[width=\hsize]{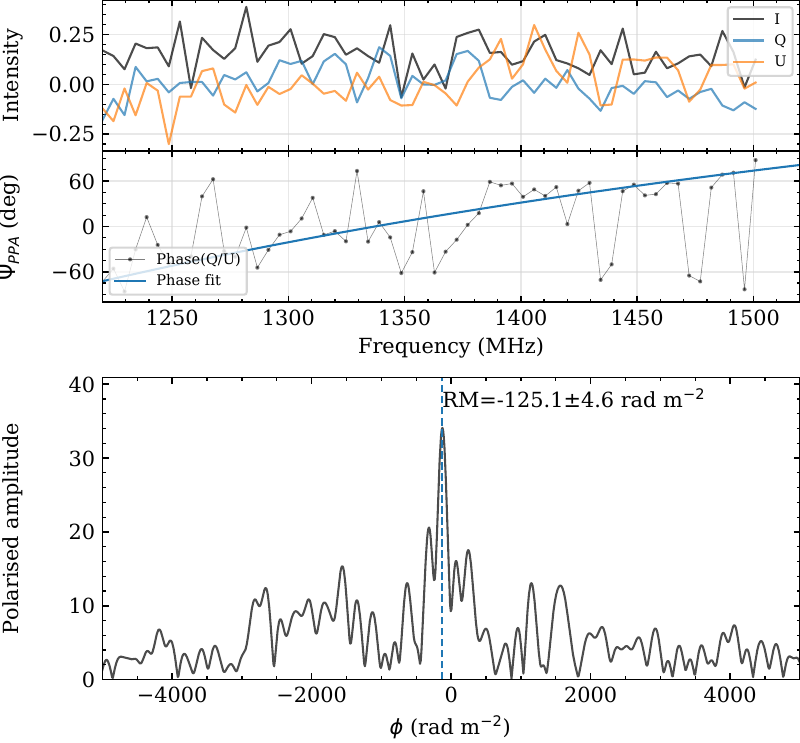}\\
    \caption{Rotation measure analysis of FRB\,20210530A. The whole bandwidth was used.}
    \label{fig:RM_FRB210530}
\end{figure}

\FloatBarrier 
\section{Extended Analysis, Results and Discussion} 
\subsection{Faraday rotation measure analysis} \label{app:RM_synth}

The RM (\radsqm) measures the integrated strength of the magnetic field parallel to the line of sight along the propagation path \citep[e.g.][]{petroff_fast_2019}. In the case of a source located at a redshift $z=z_{\text{src}}$, the RM can be expressed as \citep{mckinven_polarization_2021}:
\begin{equation}
    \text{RM} = -C_R \int_0^{z_{\text{src}}} \dfrac{n_e(z)B_{||}(z)}{(1+z)^2}\dfrac{dl}{dz}dz,
\end{equation}
with $C_R=811.9$\,\radsqm/($\mu$G\,\pccm), $n_e$ the free electron density along the line of sight, $B_{||}$ the magnetic field strength parallel to the propagation path, and $dl(z)$ the distance element along the line of sight at redshift $z$.

The RM is observed as oscillations in the $Q$ and $U$ intensities, periodic on $\lambda^2$, with $\lambda$ the observed wavelength in metres.
For FRBs where we observe oscillations in the $Q$ and $U$ intensities, we apply the RM synthesis method to estimate the Faraday rotation \citep[see][]{burn_depolarization_1966, brentjens_faraday_2005}. This technique applies the equivalent of a Fourier transformation to the complex linear polarisation, $P(\lambda^2)=Q(\lambda^2)+iU(\lambda^2)$, though this expression only has a physical meaning for $\lambda^2\geq0$; 
\begin{equation} \label{eq:rm-synth}
    F(\phi) = \int_{-\infty}^{+\infty} P(\lambda^2) e^{-2i\phi\lambda^2} d\lambda^2,
\end{equation}
where $|F(\phi)|$ is the total linearly polarised intensity within the observed bandwidth after Faraday de-rotating $P(\lambda^2)$ by the Faraday depth $\phi$ (\radsqm).
The complex linear polarisation can also be written as $P(\lambda^2)=|P|e^{2i\psi}$, with the PPA $\psi(\lambda^2) = \psi_0+\text{RM}\lambda^2$, and $\psi_0$ the polarisation position angle at a reference frequency. 
A Faraday dispersion function (FDF) is built by computing $F(\phi)$ at different $\phi$ and $\psi$ trial values, and the RM and $\psi_0$ are given by the ($\phi$, $\psi$) values that maximise the FDF. 
To determine the best RM value and errors, we fit the FDF peak to a parabolic curve. 

Once the (RM, $\psi_0$) solution has been found, the Stokes $Q$ and $U$ can be Faraday de-rotated, and the resulting ${L/I = \sqrt{Q^2+U^2}/I}$ will give us the linear polarisation fraction. The circular polarisation fraction will in turn be given by $V/I$.
The PPA can be expressed as a function of time $t$ and frequency $\nu$ as follows:
\begin{equation}
    \psi(t,\nu)=\dfrac{1}{2}\tan^{-1}\dfrac{U(t,\nu)}{Q(t,\nu)}.
\end{equation}
After Faraday de-rotating Stokes $Q$ and $U$, we can study the PPA evolution with time. This PPA is given by the polarisation angle of the emission at the source, and it is thus a property intrinsic to the FRB. 
% A PPA evolution with time would be expected from a source whose magnetic field orientation is changing (for instance, in the case of a spinning neutron star), and it can thus provide crucial information about the emission mechanism.
It is set by the properties of the magnetic field in the emission regions, since the plane of the linear polarisation coincides with the local plane of curvature of the magnetic field lines. This holds true both for pulsar-like and synchrotron maser emission  mechanisms.

\subsection{Scattering of FRB\,20200210A} \label{app:FRB20200210A_scattering}

\begin{figure}[h]
    \centering
    \includegraphics[width=\hsize]{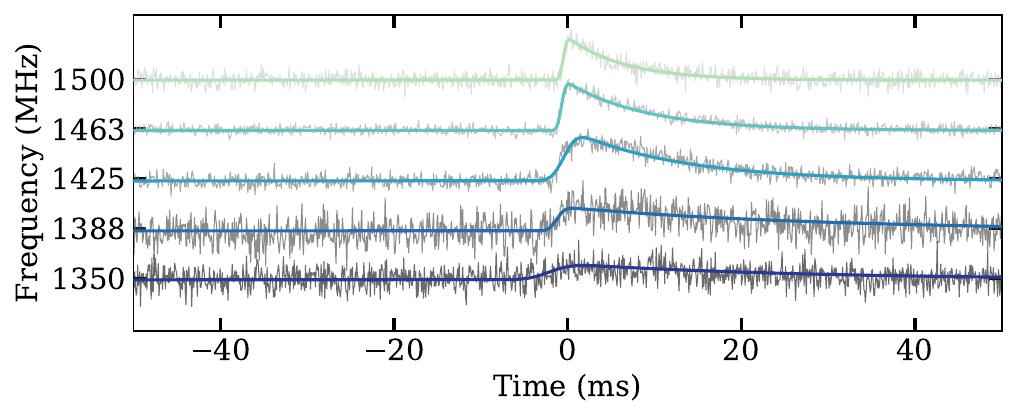}
    \includegraphics[width=\hsize]{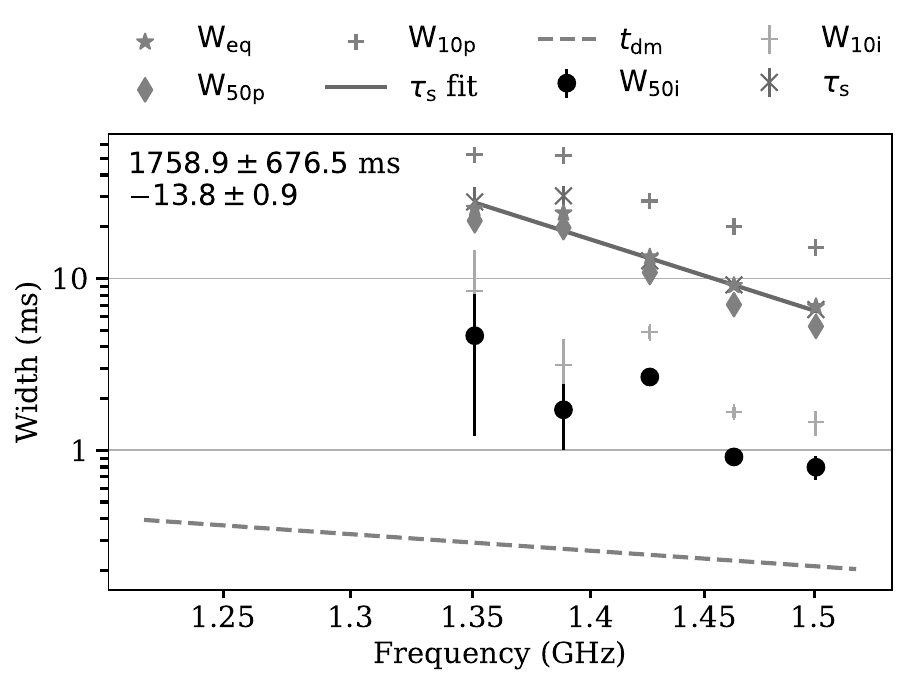}
    \caption{Multi-frequency scattering fit result for FRB\,20200210A. The top panel shows the pulse profile in 6 subbands with S/N>4.5, each one fitted to a Gaussian convolved with an exponential decay. The bottom panel shows the fit result at each subband. The scattering timescales are shown as crosses, while the fit to a scattering index is shown as a solid line. The text at the top right corner gives the expected scattering timescale at 1\,GHz given the spectral index, and the line below the scattering index $\alpha$.}
    \label{fig:FRB20200210A_scattering}
\end{figure}

To better characterise the scattering of FRB\,20200210A, we used \texttt{scatfit} to divide the full bandwidth into 8 subbands, and computed the S/N of the FRB in each one. If the S/N was above a threshold of 4.5, we performed a scattering fit. We used the scattering timescales in each subband to fit them to a power law as a function of frequency, since we expect \tscat\,${\propto\nu^{-\alpha}}$. We find a scattering index $\alpha=13.8\pm0.9$ (Note that \texttt{scatfit} defines the scattering index with an opposite sign). This result is robust, since we find a consistent $\alpha$ value when changing the number of frequency subbands to 16 and use a different S/N threshold. If this scattering index is used to scale the scattering timescale to 1\,GHz, it would result in \tscat$\sim1700\pm700$\,ms. The result is shown in Fig.~\ref{fig:FRB20200210A_scattering}.

% python /opt/anaconda3/lib/python3.9/site-packages/scatfit/apps/fit_frb.py CB09_12.0sec_dm0_t06_sb49_tab-01.fil 440.5 --fitscatindex --smodel scattered_isotropic_bandintegrated --fscrunch 192 --zoom -50 50 --fast

\newpage
\subsection{Foreground galaxy clusters of FRB\,20200719A}  \label{app:foreground_clusters}

\begin{figure}[H]
    \centering
    \includegraphics[width=\hsize]{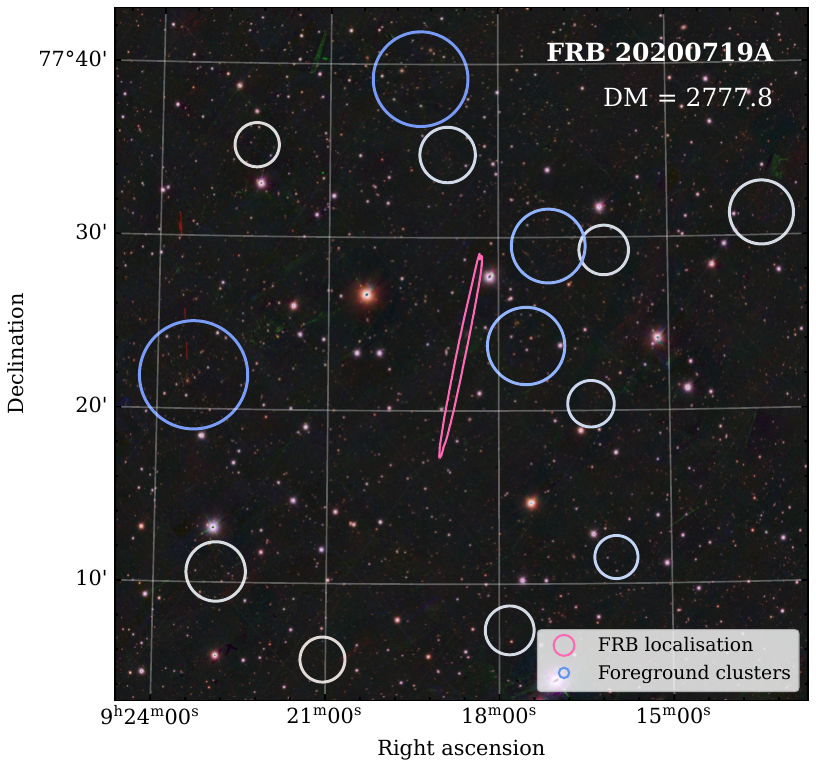}
    \caption{Foreground galaxy clusters around the localisation of FRB\,20200719A. The FRB localisation is indicated by a pink ellipse, while the clusters are represented by circles ranging from blue to white with increasing redshift. The size of the circles shows the $R_{500}$ radius \citep{zou_galaxy_2021}. The background image is from Pan-STARRS.}
    \label{fig:FRB20200719A_gal_cluster}
\end{figure}

\subsection{Scattering}  \label{app:ext_scatt}
\begin{figure}[H]
    \centering
    \includegraphics[width=\hsize]{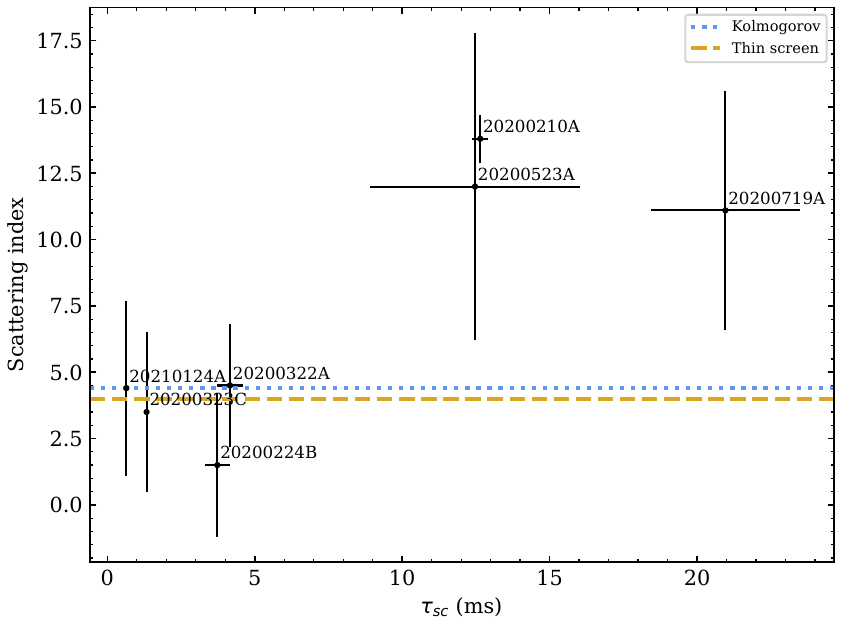}
    \caption{Scattering index as a function of scattering timescale, 
      for the applicable Apertif FRBs (black markers). %with error bars. % where these could be measured.
      %, with the TNS name of each indicated next to the marker.
      The golden and blue dashed horizontal line indicate scattering indices of 4 (as expected for a thin screen), and
      %the blue dotted line the scattering index of
      4.4 (for Kolmogorov turbulence) respectively.}
    \label{fig:scat_index}
\end{figure}

\clearpage
\subsection{Dispersion}  \label{app:ext_DM}
% Putting this plot here for better placement
% python propagation_properties_distribution.py 
% scripts/FRB_analysis/population_properties.ipynb
\begin{figure*}[b]
    \centering
    \includegraphics[width=17cm]{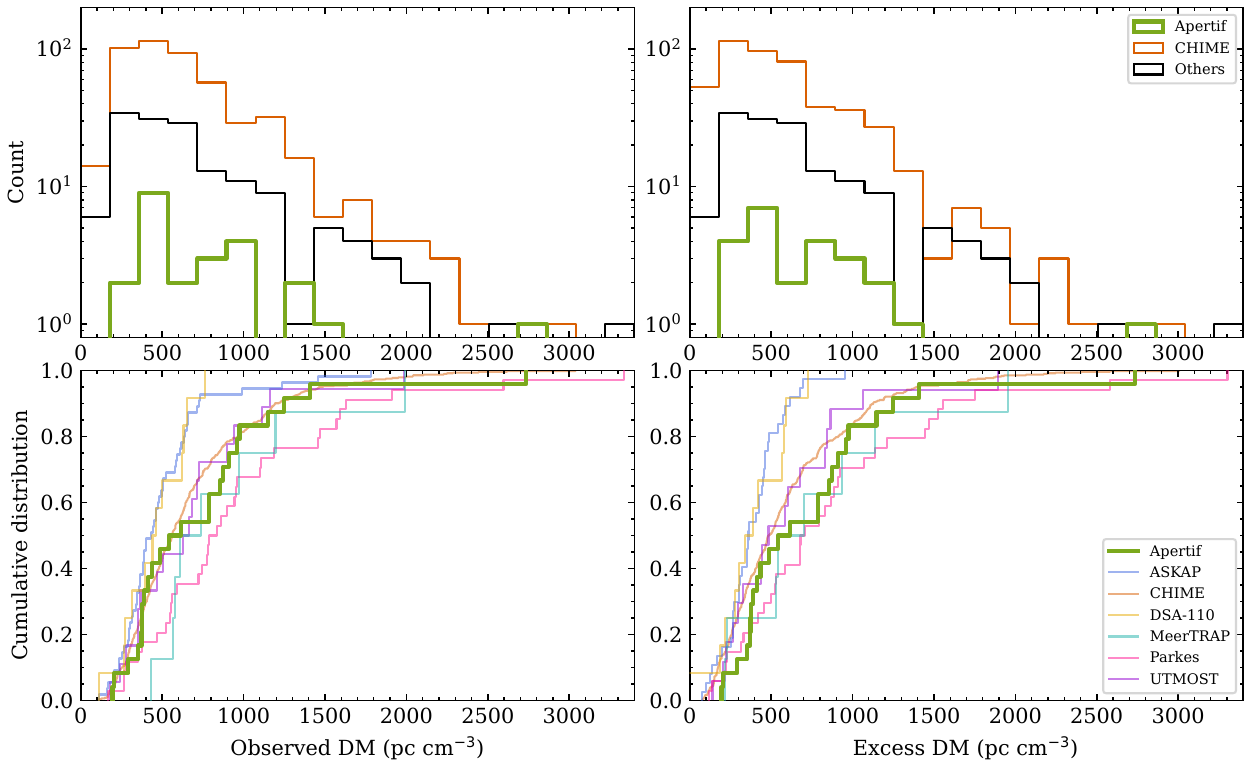}
    \caption{Distribution of Apertif FRB DMs compared to other FRB samples. The top  panels show DMs as observed (left)
      and the excess given the Milky Way contribution (right).  with Apertif in green, CHIME/FRB in orange, and all
      other FRBs in the TNS in black. The bottom  panels show the respective cumulative distributions, now for the
      single-instrument samples marked in the legend.}
    \label{fig:dms}
    \vspace{100mm}  %%% Moving the Fig up on the empty page
\end{figure*}

\end{appendix}

\end{document}